\pdfoutput=1 
\documentclass{JINST}

\usepackage{subfig}
\usepackage{cite}

\usepackage{enumitem}

\title{Monitoring and data quality assessment of the ATLAS liquid argon calorimeter}

\author{The ATLAS Collaboration}

\abstract{
The liquid argon calorimeter is a key component of the ATLAS detector installed at the CERN Large Hadron Collider. 
The primary purpose of this calorimeter is the measurement of electron and photon kinematic properties. It also provides a crucial input for measuring jets and missing transverse momentum. An advanced data monitoring procedure was designed to quickly identify issues that would affect detector performance and ensure that only the best quality data are used for physics analysis.
This article presents the validation procedure developed during the 2011 and 2012 LHC data-taking periods, in which more than 98\% of the proton--proton luminosity recorded by ATLAS at a centre-of-mass energy of 7--8~TeV had calorimeter data quality suitable for physics analysis.
}


\newcommand{\CL}[0]{calibration loop}
\newcommand{\LADIeS}[0]{signoff team}

\newcommand{\streamJet}[0]{JetTauEtmiss stream}
\newcommand{\streamExp}[0]{Express stream}
\newcommand{\streamCosmic}[0]{CosmicCalo stream}
\newcommand{\streamLCE}[0]{LArCellsEmpty stream}
\newcommand{\streamLC}[0]{LArCells stream}
\newcommand{\LB}[0]{luminosity block}
\newcommand{\LBs}[0]{luminosity blocks}

\newcommand{\Std}[0]{Standard}
\newcommand{\Sat}[0]{Saturated}
\newcommand{\ythrees}[0]{$Y_{3\sigma}$}
\newcommand{\emptyBG}[0]{empty bunch group}
\newcommand{\filledBG}[0]{filled bunch group}

\begin{document}
\section{Introduction} \label{sect:intro}

The ATLAS liquid argon calorimeter (LAr calorimeter) was designed to measure accurately electron and photon properties in a wide pseudorapidity ($\eta$) region,\footnote{ATLAS uses a right-handed coordinate system with its origin at the nominal interaction point (IP) in the centre of the detector and the $z$-axis along the beam pipe. The $x$-axis points from the IP to the centre of the LHC ring, and the $y$-axis points upward. Cylindrical coordinates $(r,\phi)$ are used in the transverse plane, $\phi$ being the azimuthal angle around the beam pipe. The pseudorapidity is defined in terms of the polar angle $\theta$ as $\eta=-\ln\tan(\theta/2)$.} $|\eta| < 2.5$. It also significantly contributes to the performance of jet and missing transverse momentum measurements ($E_{\rm{T}}^{\rm{miss}}$) in the extended pseudorapidity range $|\eta| <4.9$. This detector played a key role in the discovery of the Higgs boson \cite{higgs2012}.

\begin{figure}[!ht]
  \center
    \subfloat[]{\label{fig:LArDetector}\includegraphics[width=0.55\textwidth]{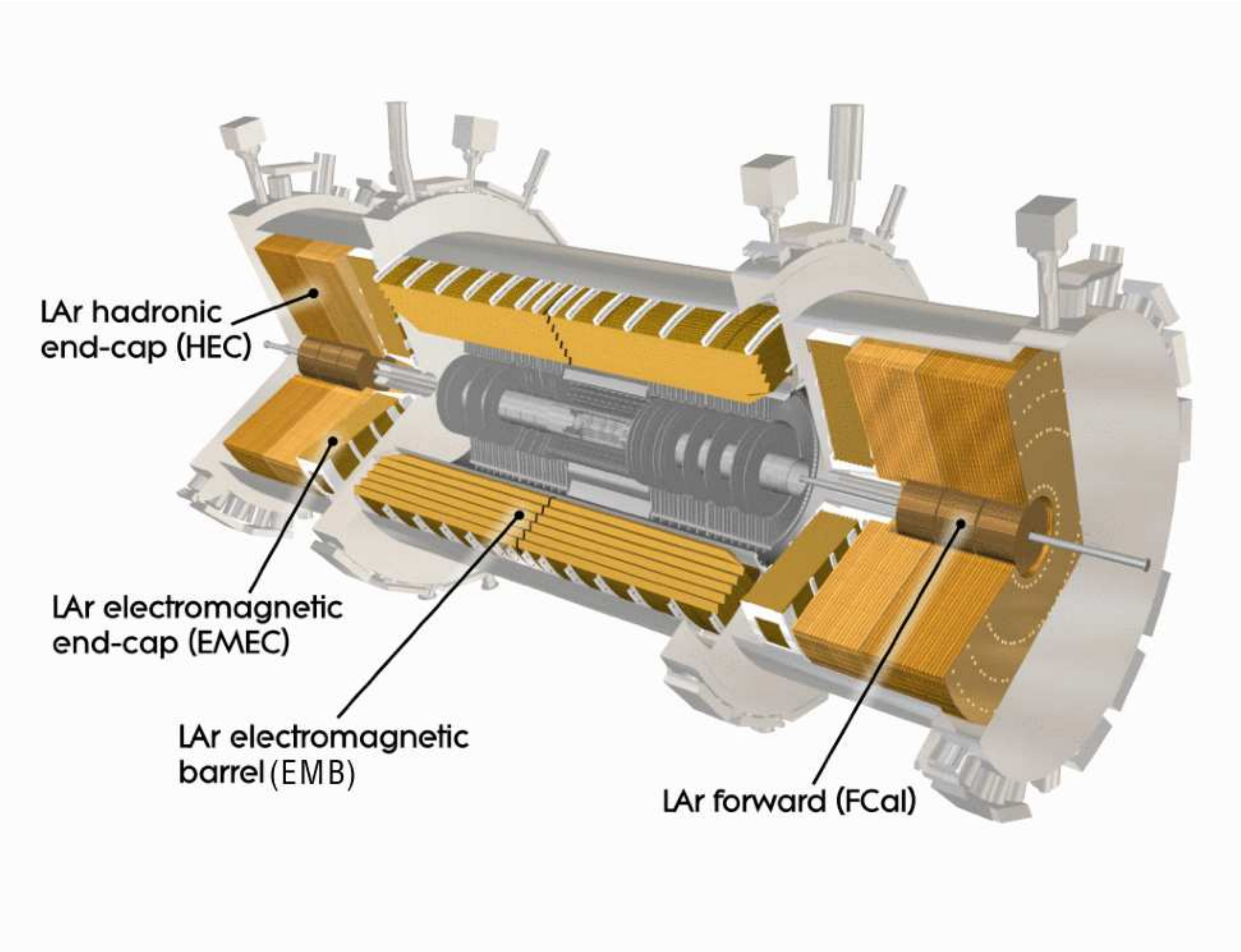}}
    \subfloat[]{\label{fig:PulseShape}\includegraphics[width=0.4\textwidth]{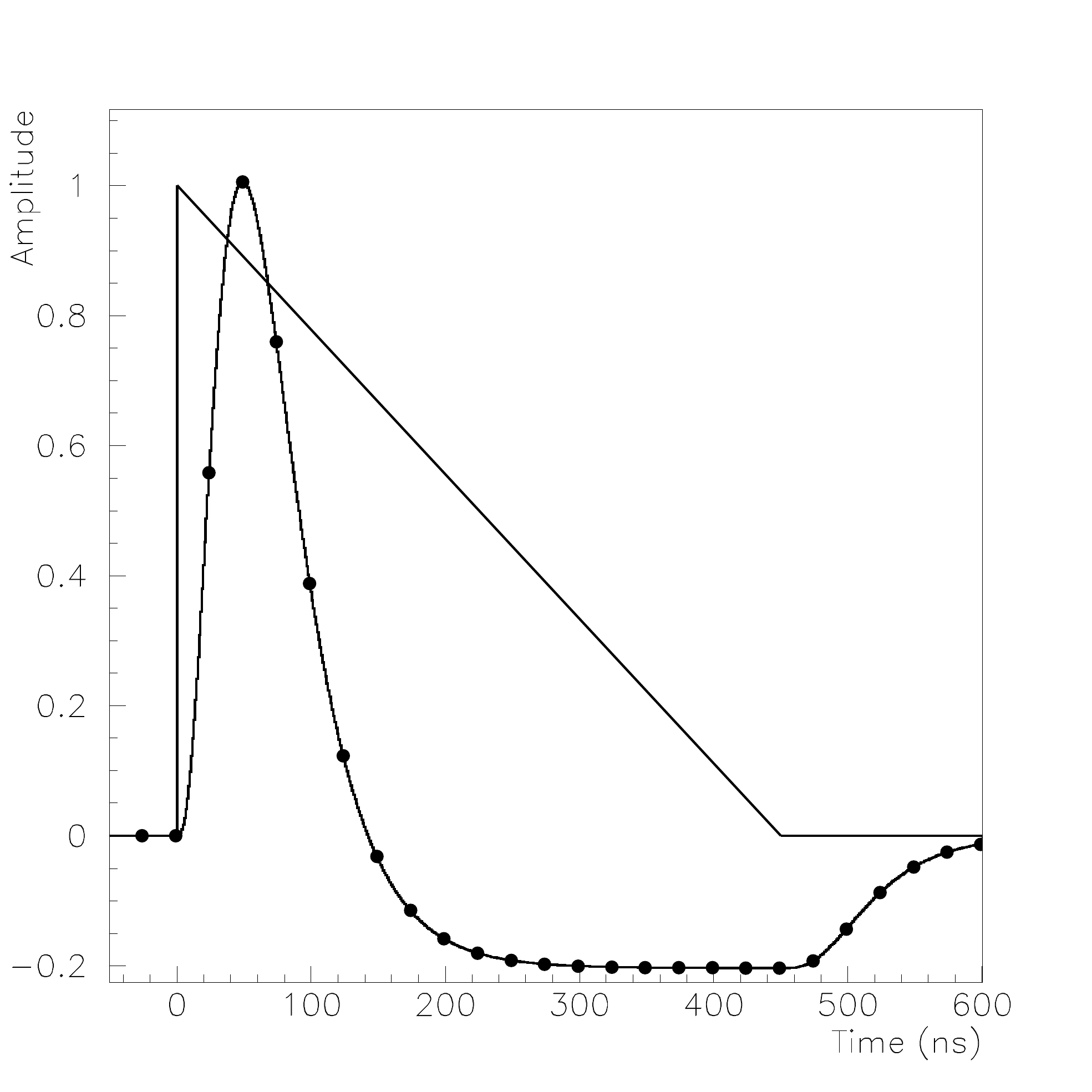}}
   \caption{(a) Cut-away view of the liquid argon calorimeter. (b) Signal shape as produced in the electromagnetic barrel (triangle), and after shaping (curve with dots). The dots represent the time and amplitude of the digitized samples.}
  \label{fig:1}
\end{figure}

Figure~\ref{fig:1}\subref{fig:LArDetector} shows the LAr calorimeter, which consists of four distinct sampling calorimeters \cite{atlasJINST,larTDR}, all using liquid argon as the active medium. The electromagnetic barrel (EMB) and endcaps (EMEC) use lead as the passive material, arranged in an accordion geometry. This detector geometry allows a fast and azimuthally uniform response as well as a coverage without instrumentation gap. The electromagnetic calorimeters cover the pseudorapidity region $|\eta| < 3.2$ and are segmented into layers (three in the range $|\eta| < 2.5$, two elsewhere) to observe the longitudinal development of the shower and determine its direction. Furthermore, in the region $|\eta| < 1.8$ the electromagnetic calorimeters are complemented by a presampler, an instrumented argon layer that provides information on the energy lost in front of the electromagnetic calorimeters. For the hadronic endcaps (HEC) covering the pseudorapidity range $1.5<|\eta|<3.2$, copper was chosen as the passive material and a parallel plate geometry was adopted. For the forward calorimeter (FCal), located at small polar angles where the particle flux is much higher and the radiation damage can be significant, a geometry based on cylindrical electrodes with thin liquid argon gaps was adopted. Copper and tungsten are used as passive material. The hadronic and forward calorimeters are also segmented in depth into four and three layers respectively. The four detectors are housed inside three cryostats (one barrel and two endcaps) filled with liquid argon and kept at a temperature of approximately 88~K. Each detector part is referred to as a {\itshape partition} named EMB, EMEC, HEC and FCal with an additional letter, C or A, to distinguish the negative and positive pseudorapidity regions respectively.\footnote{The barrel is made of two halves housed in the same cryostat.}  Hence, there are eight different partitions.

Although each detector has its own characteristics in terms of passive material and geometry, a special effort was made to design uniform readout, calibration and monitoring systems across the eight partitions. The 182468 calorimeter channels are read out by 1524 front-end boards (FEBs)~\cite{electronicPerf,FEBDesign} hosted in electronics crates located on the three cryostats. These FEBs shape the signal and send the digitized samples via optical links to 192 processing boards (named ``RODs'' for read-out drivers)~\cite{RODDesign} that compute the deposited energies before passing them to the central data-acquisition system. The signal shapes before and after the FEB shaping are shown in Figure~\ref{fig:1}\subref{fig:PulseShape}.

This article describes the data quality assessment procedure applied to ensure optimal calorimeter performance together with low data rejection, emphasizing the performance achieved in 2012, when 21.3~$\rm{fb^{-1}}$ of proton--proton collisions were recorded by the ATLAS experiment. The integrated luminosity is derived, following the same methodology as that detailed in reference ~\cite{lumi2011}, from a preliminary calibration of the luminosity scale derived from beam-separation scans performed in November 2012. This dataset is divided into ten time periods within which data-taking conditions were approximately uniform: the characteristics of these periods are summarized in table \ref{tab:dataPeriod}. The dataset is also divided into {\itshape runs} that correspond to a period of a few hours of data taking (up to 24 hours depending on the LHC beam lifetime and the ATLAS data-taking performance). Each run is divided into one-minute blocks (periods known as {\itshape} {\LB}s).

\begin{table}[!htbp]
\caption{Characteristics of the ten data-taking periods defined in 2012. The F and K periods not considered in this article correspond to data taking without LHC collisions, and are hence not relevant for data quality assessment.}
\begin{center}
\begin{tabular}{|p{3.8cm}|p{0.75cm}|p{0.75cm}|p{0.75cm}|p{0.75cm}|p{0.75cm}|p{0.75cm}|p{0.75cm}|p{0.75cm}|p{0.75cm}|p{0.75cm}|}
  \hline
2012 data-taking periods                             & A    & B   & C   & D   & E   & G   & H   & I   & J   & L\\
  \hline
Start date (day/month)                                         & 4/4  & 1/5 & 1/7 & 24/7 & 23/8 & 26/9 & 13/10 & 26/10 & 2/11 & 30/11\\
  \hline
Integrated luminosity \newline  recorded ($\rm{fb^{-1}}$)     & 0.84  & 5.30 & 1.54 & 3.37 & 2.70 & 1.30 & 1.56 & 1.06 & 2.72 & 0.89\\
  \hline
Peak luminosity \newline ($10^{33}\rm{cm^{-2}s^{-1}})$ & 5.5 & 6.7 & 6.2 & 7.3 & 7.6 & 7.3 & 7.5 & 7.3 & 7.4 & 7.5\\
  \hline
Mean instantaneous luminosity ($10^{33}\rm{cm^{-2}s^{-1}})$ & 2.0 & 3.6 & 2.9 & 3.8 & 4.2 & 4.0 & 4.3 & 4.4 & 4.2 & 4.3\\
  \hline
\end{tabular}
\end{center}
\label{tab:dataPeriod}
\end{table}
The article is organized as follows: the ATLAS data processing organization and data quality assessment infrastructure are described in section~\ref{sect2_dqpolicy}. Sections \ref{sect3_dcsHVTrips}--\ref{sect7_noisyCells} detail the specific LAr calorimeter procedures developed to assess the data quality in all aspects: detector conditions (section~\ref{sect3_dcsHVTrips}), data integrity (section~\ref{sect:dataIntegrity}), synchronization (section~\ref{sect:timing}), large-scale coherent noise (section~\ref{sect:noiseBursts}) and isolated pathological cells (section~\ref{sect7_noisyCells}). For each aspect of the data quality assessment, the amount of rejected data is presented chronologically as a function of data-taking period. For illustration purposes, the ATLAS run 205071 from June 2012 is often used. With 226~$\rm{pb^{-1}}$ accumulated in 18~hours of the LHC collisions period (``fill'') number 2736, it is the ATLAS run with the highest integrated luminosity. Finally, section~\ref{sect:concl} recaps the data quality performance achieved in 2011 and 2012, and provides a projection towards the higher energy and luminosity conditions scheduled for the LHC restart in 2015.

\section{Data quality assessment operations and infrastructure} \label{sect2_dqpolicy}
 
The ATLAS data are monitored at several stages of the acquisition and processing chain to detect as early as possible any problem that could compromise their quality. Most of the monitoring infrastructure is common to both the online and offline environments, but the levels of details in the monitoring procedure evolve with the refinement of the analysis (from online to offline).

\subsection{Online monitoring}

During data taking, a first and very crude quality assessment is performed in real time on a limited data sample by detector personnel called {\itshape shifters} in the ATLAS control room. The shifters focus on problems that would compromise the data quality without any hope of improving it later, such as serious data corruption or a significant desynchronization. During data taking, tracking the calorimeter noise is not considered a priority as long as the trigger rates remain under control. The trigger rates are checked by a dedicated trigger shifter who can decide, if needed, to take appropriate action. This may consist of either simply ignoring the information from a noisy region of typical size $\Delta\phi\times\Delta\eta = 0.1 \times 0.1$  or setting an appropriate prescale factor for the trigger item saturating the bandwidth (see next section for more details of the trigger system).

To assess the data quality of the ongoing run, the ATLAS control room shifters run simple algorithms to check the content of selected histograms, and the results are displayed using appropriate tools~\cite{OnlineDQMF}. Even though the running conditions are constantly logged in a dedicated electronic logbook \cite{ATLOG}, no firm data quality information is logged at this point by the shifters.

\subsection{Relevant aspects of LHC and ATLAS trigger operations} \label{offlinePolicy}

The LHC is designed to contain trains of proton bunches separated by 25~ns \cite{lhcJINST}. The corresponding 25~ns time window, centred at the passage time of the centre of the proton bunch at the interaction point, defines a bunch crossing. The nominal LHC configuration for proton--proton collisions contains 3564 bunch crossings per revolution, each of which is given a unique bunch crossing identifier (BCID). However, not all BCIDs correspond to bunches filled with protons. The filling is done in bunch trains, containing a number of equally spaced bunches. Between the trains, short gaps are left for the injection kicker and a longer gap occurs for the abort kicker. A configuration frequently used in 2012 consists of a mixture of 72- and 144-bunch trains (typically a dozen) with a bunch spacing of 50~ns for a total of 1368 bunches. Each train therefore lasts 3.6--7.2~$\mu$s, and two trains are spaced in time by between 600~ns and 1~$\mu$s. The BCIDs are classified into bunch groups by the ATLAS data-acquisition system\cite{BeamBackground2011}. The bunch groups of interest for this article are
\begin{itemize}[topsep=3pt,itemsep=2pt,parsep=3pt]
\item {\itshape{\filledBG}}: a bunch in both LHC beams;
\item {\itshape{\emptyBG}}: no proton bunch. 
\end{itemize}
In the configuration widely used in 2012, the {\emptyBG} consisted of 390 BCIDs, roughly three times less than the {\filledBG} (1368 BCIDs). As the average electron drift time in the liquid argon (of the order of several hundreds nanoseconds) is longer than the time between two filled bunches, the calorimeter response is sensitive to collision activity in bunch crossings before and after the BCID of interest. These unwanted effects are known as out-of-time pile-up. To limit its impact, the BCIDs near a filled BCID (within six BCIDs) are excluded from the {\emptyBG}.

The ATLAS trigger system consists of three successive levels of decision \cite{triggerTDR_LVL1,triggerTDR_DAQ,trigger2010}. A {\itshape trigger chain} describes the three successive trigger items which trigger the writing of an event on disk storage. The ATLAS data are organized in {\itshape streams}, defined by a trigger menu that is a collection of trigger chains. The streams are divided into two categories: calibration streams and physics streams. The calibration streams are designed to provide detailed information about the run conditions (luminosity, pile-up, electronics noise, vertex position, etc.) and are also used to monitor all the detector components while the physics streams contain events that are potentially interesting for physics analysis. 

In the case of the LAr calorimeter, four main calibration streams are considered for the data quality assessment.
\begin{itemize}[topsep=3pt,itemsep=2pt,parsep=3pt]
\item The {\itshape Express} stream contains a fraction of the data (around 2--3\% of the total in 2012) representative of the most common trigger chains used during collision runs; almost all of these trigger chains are confined to the {\filledBG}.
\item The {\itshape CosmicCalo} stream contains events triggered in the {\emptyBG}, where no collisions are expected.
\item The {\itshape LArCells} stream contains partially built collision events~\cite{partialEventBuilding}, where only a fraction of the LAr data are stored (the cells belonging to a high-energy deposit as identified by the second level of the trigger system). The reduced event size allows looser trigger conditions and  significantly more events in the data sample.
\item The {\itshape LArCellsEmpty} stream benefits from the same ``partial event building'' facility as the \streamLC, and the trigger is restricted to the {\emptyBG}.
\end{itemize}
The CosmicCalo, LArCellsEmpty and LArCells streams mainly contain trigger chains requesting a large energy deposit in the calorimeters.

Several physics streams are also mentioned in this article. The {\itshape JetTauEtmiss} stream is defined to contain collision events with jets of large transverse momentum, $\tau$ lepton candidates or large missing transverse momentum. The {\itshape EGamma} stream is defined to contain collision events with electron or photon candidates.

The LAr calorimeter data quality assessment procedure is meant to identify several sources of potential problems and to address solutions. The calibration streams containing collision events (Express and LArCells streams) are used to identify data corruption issues, timing misalignments and large coherent noise. The CosmicCalo and LArCellsEmpty streams, filled with events triggered in the {\emptyBG}, are used to identify isolated noisy cells. 

The LAr calorimeter data quality assessment procedure is not meant to monitor higher-level objects (such as electron/photon, J/$\psi$ candidates, etc.) and their characteristics (uniformity, calibration, mass, etc.): this task is performed in a different context and is beyond the scope of this article.

\subsection{Practical implementation of the data quality assessment}

\begin{figure}[!b]
  \center
   \includegraphics[width=0.96\textwidth]{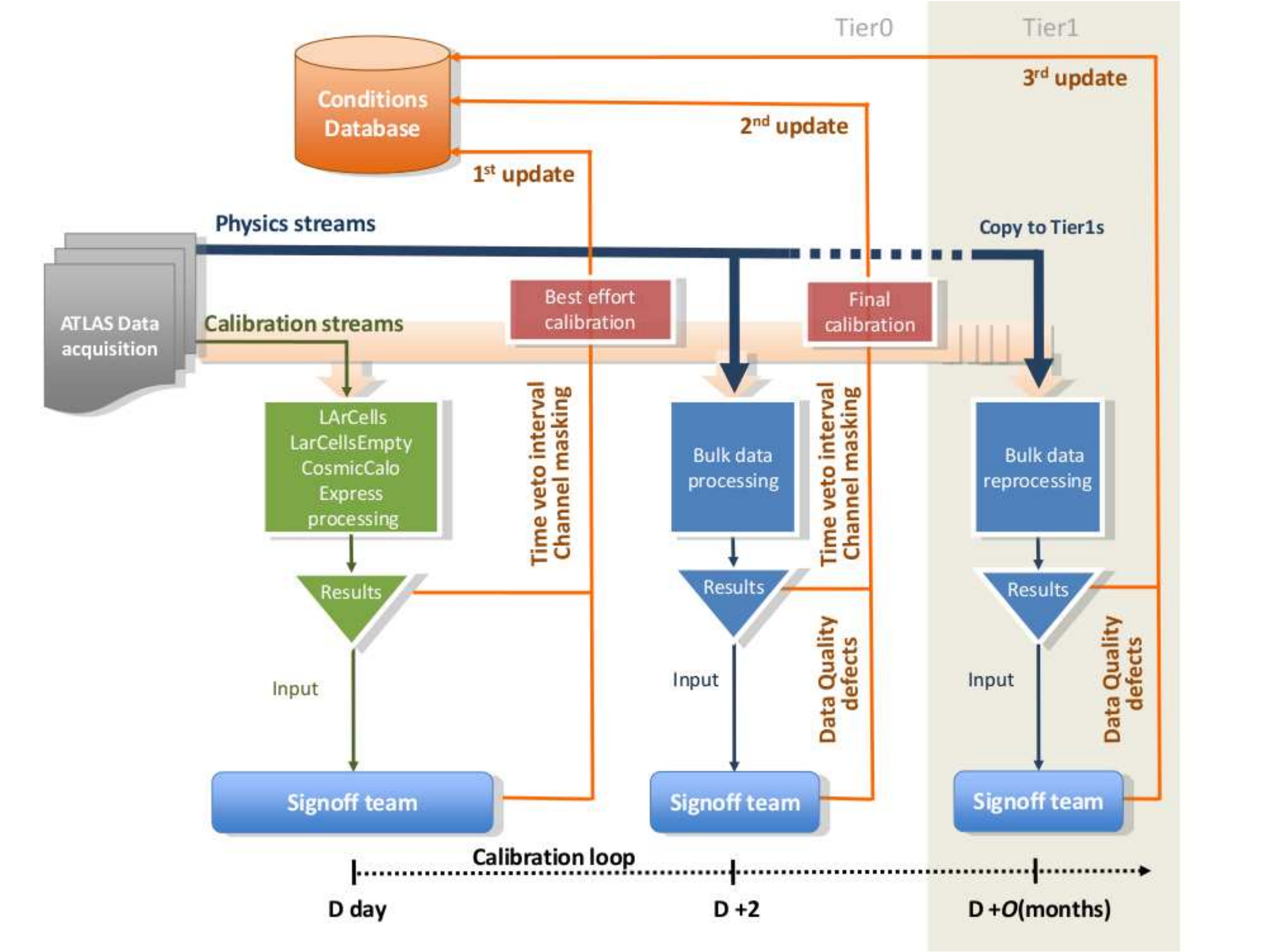}
   \caption{ATLAS data processing and monitoring organization and {\CL} scheme (with focus on the LAr calorimeter case).}
   \label{fig:andreasPlot}
\end{figure}

A graphical view of the ATLAS data processing organization is shown in figure~\ref{fig:andreasPlot}. Since the information provided by the calibration streams is necessary to reconstruct the physics data, the calibration streams are promptly processed during the {\itshape express processing} which is launched shortly after the beginning of a run. The data are processed with the ATLAS Athena software on the CERN computing farms \cite{computingTDR}, either the Grid Tier 0 farm or the Calibration and Alignment Facility (CAF) farm \cite{cernCAF}.  The monitoring histograms are produced at the same time within the Athena monitoring framework and then post-processed with dedicated algorithms to extract data quality information. The data quality results are available through a central ATLAS web site \cite{offlineDQMF} for all the ATLAS subdetectors. A first data quality assessment is performed at this stage. The conditions databases \cite{COOLStatus} which store the complete picture of the detector status and the calibration constants as a function of time are also updated. These tasks are completed within 48 hours after the end of the run, before the start of the physics stream reconstruction. The 48-hour period for this primary data quality review is called the {\itshape {\CL}}. 

Given the complexity of the checks to be completed over the 182468 calorimeter cells, a dedicated web infrastructure was designed. It enables quick extraction and summarization of meaningful information and optimization of data quality actions such as the automated production of database updates. 
Despite the high level of automation of the LAr calorimeter data quality procedure, additional supervision by trained people remains mandatory. In 2011 and 2012, people were assigned during daytime hours, seven days per week, to assess the relevance of the automatically proposed actions. These one or two people are referred to as the {\itshape \LADIeS}. 

Once the database conditions are up-to-date and the 48-hour period completes, the processing of all the physics streams (also called {\itshape the bulk}) is launched. Typically, the complete dataset is available after a couple of days, and a final data quality assessment is performed to check if the problems first observed during the {\CL} were properly fixed by the conditions updates. If the result of the bulk processing is found to be imperfect, further database updates may be needed. However, such new conditions data are not taken into account until the next data reprocessing, which may happen several months later. The final data quality assessment for the bulk processing is done using exactly the same web infrastructure as for the primary data quality assessment with the express processing. 

\subsection{Data quality logging}
\label{dqLogging}

At each stage, any problem affecting the data quality is logged in a dedicated database. The most convenient and flexible way to document the data losses consists of assigning a {\itshape defect} \cite{defectDB} to a {\LB}.
Approximately 150 types of defects were defined to cover all the problems observed in the LAr calorimeter during the 2011 and 2012 data taking. These defects can be either global (i.e. affecting the whole calorimeter) or limited to one of the eight partitions. A defect can either be {\itshape intolerable}, implying a systematic rejection of the affected {\LB}, or {\itshape tolerable}, and mainly set for bookkeeping while the data are still suitable for physics analysis.

The defects are used to produce a list of {\LBs} and runs that are declared as ``good'' for further analysis. This infrastructure is powerful, as it permits precise description and easy monitoring of the sources of data loss; it is also flexible, since a new list of good {\LBs} and runs can be produced immediately after a defect is changed. However, since the smallest time granularity available to reject a sequence of data is the luminosity block, the infrastructure is not optimized to deal with problems much shorter than the average {\LB} length (i.e. one minute). 

To reduce the data losses due to problems lasting much less than a minute, a complementary method that stores a status word in each event's header block allows event-by-event data rejection. In order not to bias the luminosity computation, small time periods are rejected rather than isolated events. This {\itshape time-window veto} procedure allows the vetoed interval to be treated like another source of data loss: the corresponding luminosity loss can be accurately estimated and accounted for in physics analyses. The time periods to be vetoed are defined in a standard ATLAS database before the start of the bulk processing. The database information is read back during the Tier 0 processing, and the status word is filled for all events falling inside the faulty time window. Since this information must be embedded in all the derived analysis files, the database conditions required to fill this status word must be defined prior to the start of bulk reconstruction, i.e. during the {\CL}. In that sense, the status word is less flexible than the defect approach, but it can reject very small periods of data. 

\section{Detector conditions} \label{sect3_dcsHVTrips}

Stable operation in terms of detector safety, powering and readout is essential for ensuring high quality of data. Information about the detector conditions is provided by both the ATLAS Detector Control System (DCS) \cite{pvssReference} and the Tier 0 processing output.

\subsection{Detector control system infrastructure}

The ATLAS DCS system provides a {\itshape state} and a {\itshape status} word per partition: the state reflects the present conditions of a partition (``Ready'', ``Not\_Ready'', ``Unknown'', ``Dead''), while the status is used to flag errors (``OK'', ``Warning'', ``Error'', ``Fatal''). The state/status words are stored in a database and used by the ATLAS DCS data quality calculator \cite{DCSDQ} to derive an overall DCS data quality flag that is specific to the LAr calorimeter for each {\LB} and is represented by a colour. The condition assigned to each {\LB} is based on the worst problem affecting the data during the corresponding time interval, even if the problem lasted for a very short time. Table \ref{DCS:tableFlag} summarises the policy used to derive the LAr calorimeter DCS data quality flags. The colour hierarchy is the following with increasing severity: green -- amber -- grey -- red.

The DCS system allows the masking of known problems to avoid continuous state/status errors, as this would prevent the shifter from spotting new problems during data taking. Therefore, a {\itshape green} flag does not always mean that the LAr calorimeter is in an optimal state. A green flag ensures that the detector conditions from the DCS point of view remain uniform during a run, since no new problem masking is expected during data taking.

\begin{table}[!bhp]
\caption{Assignment policy for LAr calorimeter DCS data quality flag.}
\begin{center}
\begin{tabular}{|p{3.1cm}|p{3.3cm}|p{1.5cm}|p{6.2cm}|}
  \hline
  State & Status & DCS flag & Possible source of problem\\
  \hline
  Ready  & OK & Green & - \\
  \hline
  Unknown Dead & Warning, Error & Amber & Loss of communication \\
  \hline
  Not\_Ready & Warning, Error, Fatal & Red & Power supply trip\\
  \hline
  Anything else & Anything else & Grey & Corrupted/missing data in DCS database \\
  \hline
\end{tabular}
\end{center}
\label{DCS:tableFlag}
\end{table} 
There is no defect automatically derived from the DCS flag. However, the {\LADIeS} is expected to understand any DCS flag differing from green and cross-check with other sources, such as the monitoring algorithm and operation reports. For the period 2010--2012, the main source of abnormal DCS flags was high-voltage power supply trips.

\subsection{Monitoring of high-voltage conditions}
\label{DCS:HV}

The high voltage (HV) -- applied for charge collection on the active liquid argon gaps of the calorimeter~-- is distributed among 3520 sectors of typical size $\Delta\eta\times\Delta\phi=0.2\times 0.2$ (in the three layers of the electromagnetic calorimeters)~\cite{larTDR}. Each sector is supplied by two or four independent HV lines in a redundant scheme. Because the HV conditions impact the amount of signal collected by the electrodes, and therefore are a crucial input for the energy computation, they are constantly monitored online, and stored in a dedicated conditions database. The HV values are written every minute or every time a sizeable variation (greater than 5~V) is observed.

The most common issue encountered during data taking is a trip of one HV line, i.e. a sudden drop of voltage due to a current spike. When a current spike occurs, the HV module automatically reduces the voltage in that sector. The HV line is usually ramped up automatically directly afterwards. If the automatic ramp-up procedure fails (or before automatic ramping was used, e.g. early 2011), the HV line can either be ramped up manually or left at zero voltage until the end of the run; in the latter case, thanks to the redundant HV supply, the affected regions remain functional although with a worse signal/noise ratio. During data acquisition, the calibration factors associated with the HV settings are stored in registers of the ROD boards~\cite{RODDesign} and cannot be changed without a run stop; therefore they remain constant during a run, even if the effective HV value changes. 
As reduced HV settings induce a reduced electron drift speed, the energy computed online is underestimated and impacts the trigger efficiency near the trigger threshold. Given the limited size of a sector and the rare occurrence of such a configuration, this had a negligible impact. As previously described, the HV trips are recorded by the DCS data quality flag, but a dedicated HV database including all the trip characteristics is also filled daily by an automated procedure.

During the offline Tier 0 reconstruction, a correction factor is automatically applied by the reconstruction software based on the HV reading. A variation of HV conditions also requires an update of the expected noise per cell, which has to be corrected in the same way as the energy in order not to bias the clustering mechanism. Due to the data reconstruction model, this update cannot be automated and requires human intervention within the 48-hour calibration loop delay.

The data quality assessment makes use of the three different sources of information (DCS flags, HV database and offline HV correction monitoring) to get a consistent picture of the HV conditions during a run. During a trip, the HV, and therefore the energy scale, vary too quickly to be accurately assessed. In addition, the {\LB} in which the trip happened is usually affected by a large burst of coherent noise (see section~\ref{sect:noiseBursts}) and is hence unusable for physics. Therefore, the luminosity blocks where a HV drop occurred are systematically rejected by marking an intolerable defect. The policy regarding {\LB}s with HV ramp-up has evolved over time. Initially rejected, these periods are now corrected offline with the proper HV values and marked with a tolerable defect, after a careful check of the noise behaviour. The studies performed on data with HV ramping are detailed in section~\ref{sect:rampup}.
\begin{figure}[!htb]
    \subfloat[]{\label{fig:trip1}\includegraphics[width=0.54\textwidth]{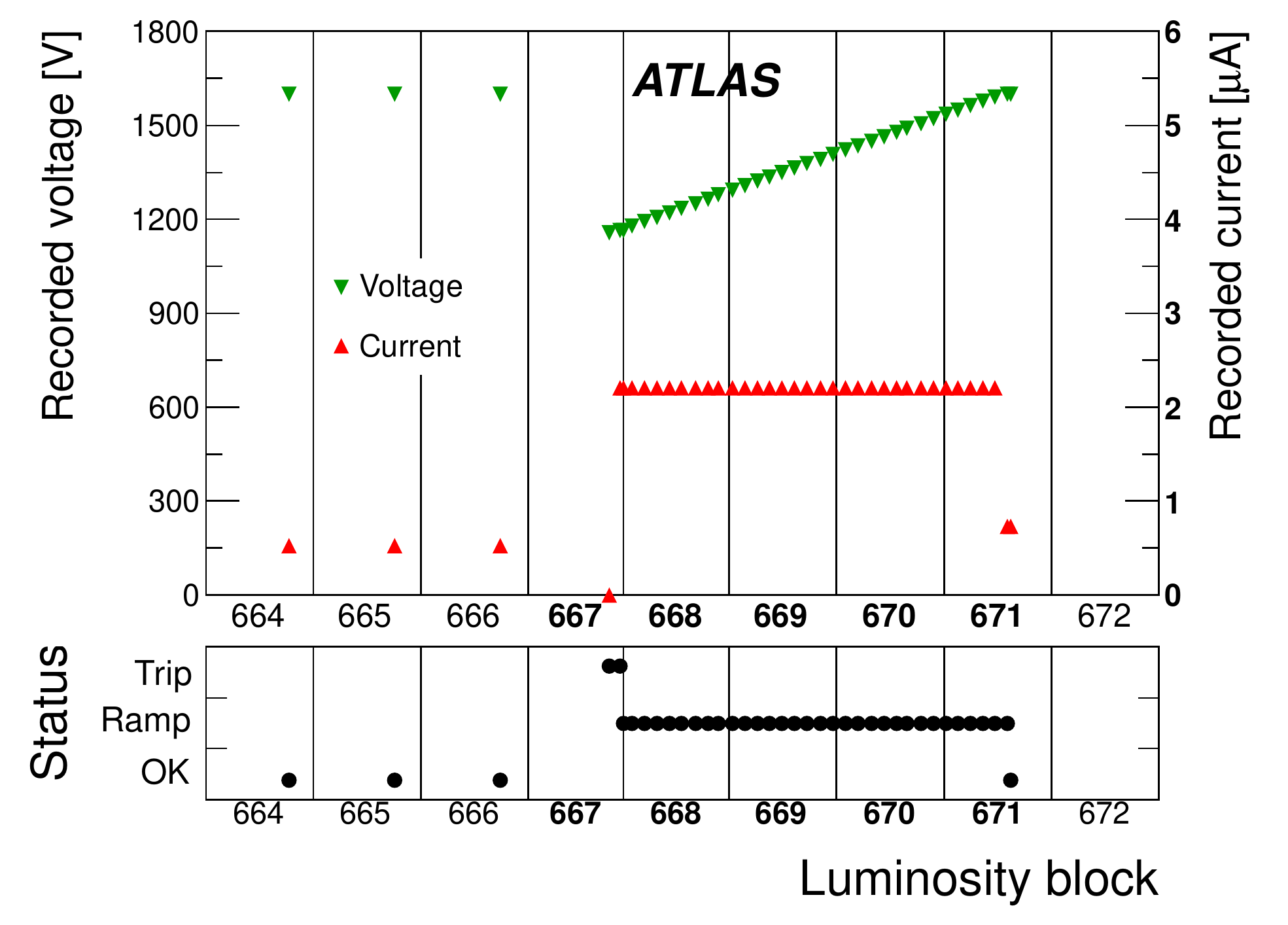}}
    \subfloat[]{\label{fig:HVcorrection}\includegraphics[width=0.42\textwidth]{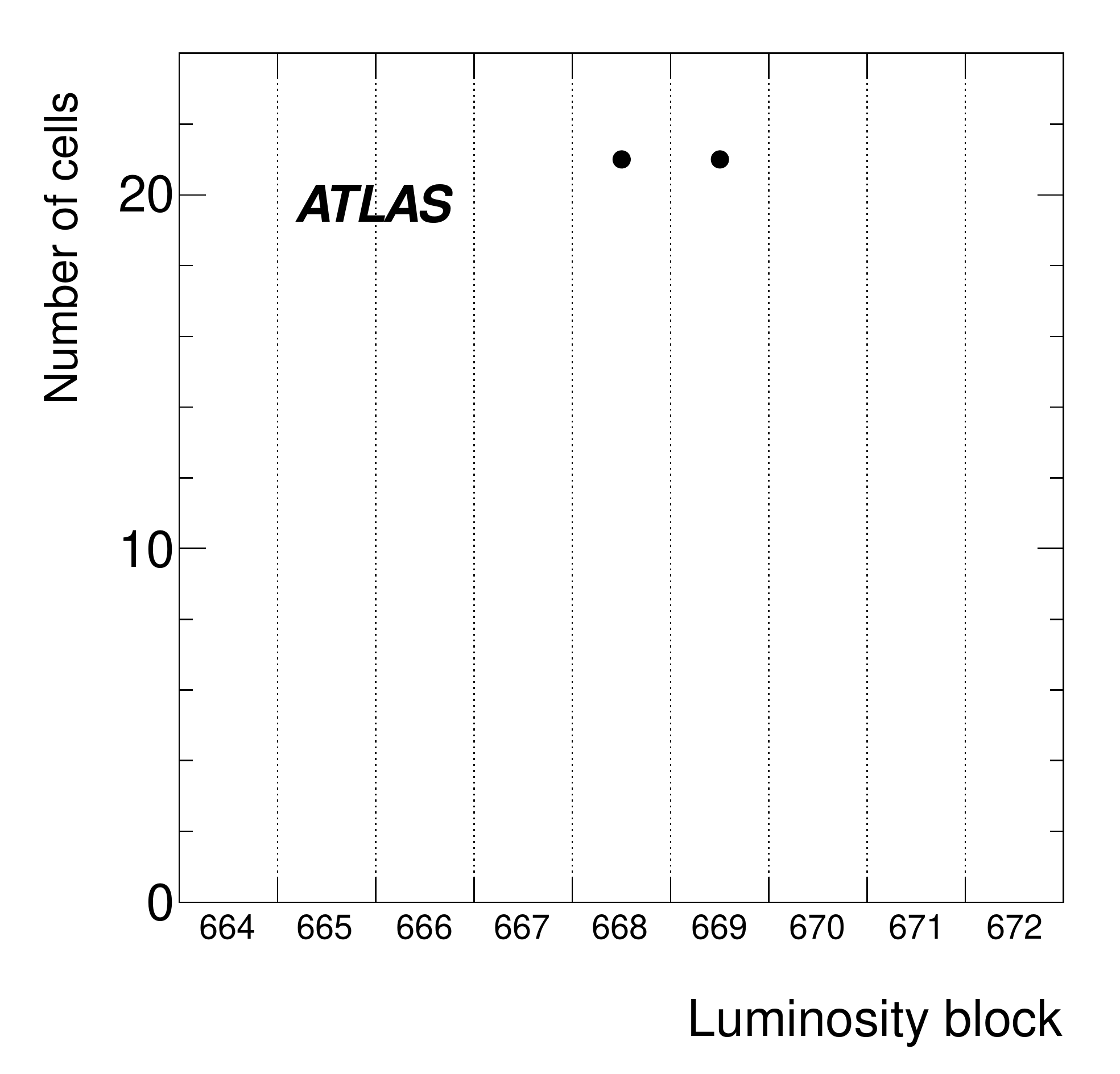}}
   \caption{Example of a typical trip of a HV line supplying one HEC sector. (a) Recorded voltage, current and status evolution. The {\LB} numbers shown in bold indicate a red DCS flag. (b) Number of readout cells with a HV correction greater than 5\% (with respect to the start of run) as a function of the luminosity block number.}
   \label{fig:trip340}
\end{figure}

The DCS information about a typical trip of a HV line supplying one hadronic calorimeter sector is shown in figure~\ref{fig:trip340}\subref{fig:trip1}. A voltage drop of 500~V (from 1600~V down to 1100~V) is observed in {\LB} 667. The high-voltage was then automatically ramped up at a rate of 2~V/s, lasting approximately four minutes. The nominal HV value was recovered during luminosity block~671. The DCS flag is red for five luminosity blocks 667--671, which is consistent with the error status bit also displayed in figure~\ref{fig:trip340}\subref{fig:trip1} for this interval. Figure~\ref{fig:trip340}\subref{fig:HVcorrection} shows the corresponding offline monitoring plot for the same HV trip, displaying how many calorimeter cells have a HV correction factor greater than 5\% at the beginning of the {\LB}. Only two luminosity blocks are identified: 668 and 669.\footnote{The correction factors depend nonlinearly on the voltage and in this case are smaller than the relative voltage change.} Based on this consistent information, the {\LB} 667 was marked with an intolerable defect. The luminosity block range 668--671 when the ramping voltage occurred was marked with a tolerable defect.

\subsection{Validation of data taken during the ramp-up procedure}
\label{sect:rampup}

As already mentioned, the offline software takes into account the effective HV settings to correct the energy. The electronics noise correction is estimated at the beginning of the ramp-up period, and considered constant until the voltage is stable again. As the noise correction factor is maximal at the start of the ramp-up period, this means that during this short time, the electronics noise is slightly overestimated, inducing a negligible bias in the clustering algorithm. The reconstruction software therefore appears to cope well with HV channel variations. However, before declaring the ramping HV data as good for physics, a further check is performed to detect any non-Gaussian noise behaviour that could be induced by the ramping operations.

\begin{figure}[!htb]
    \subfloat[]{\label{fig:rampup1}\includegraphics[width=0.48\textwidth]{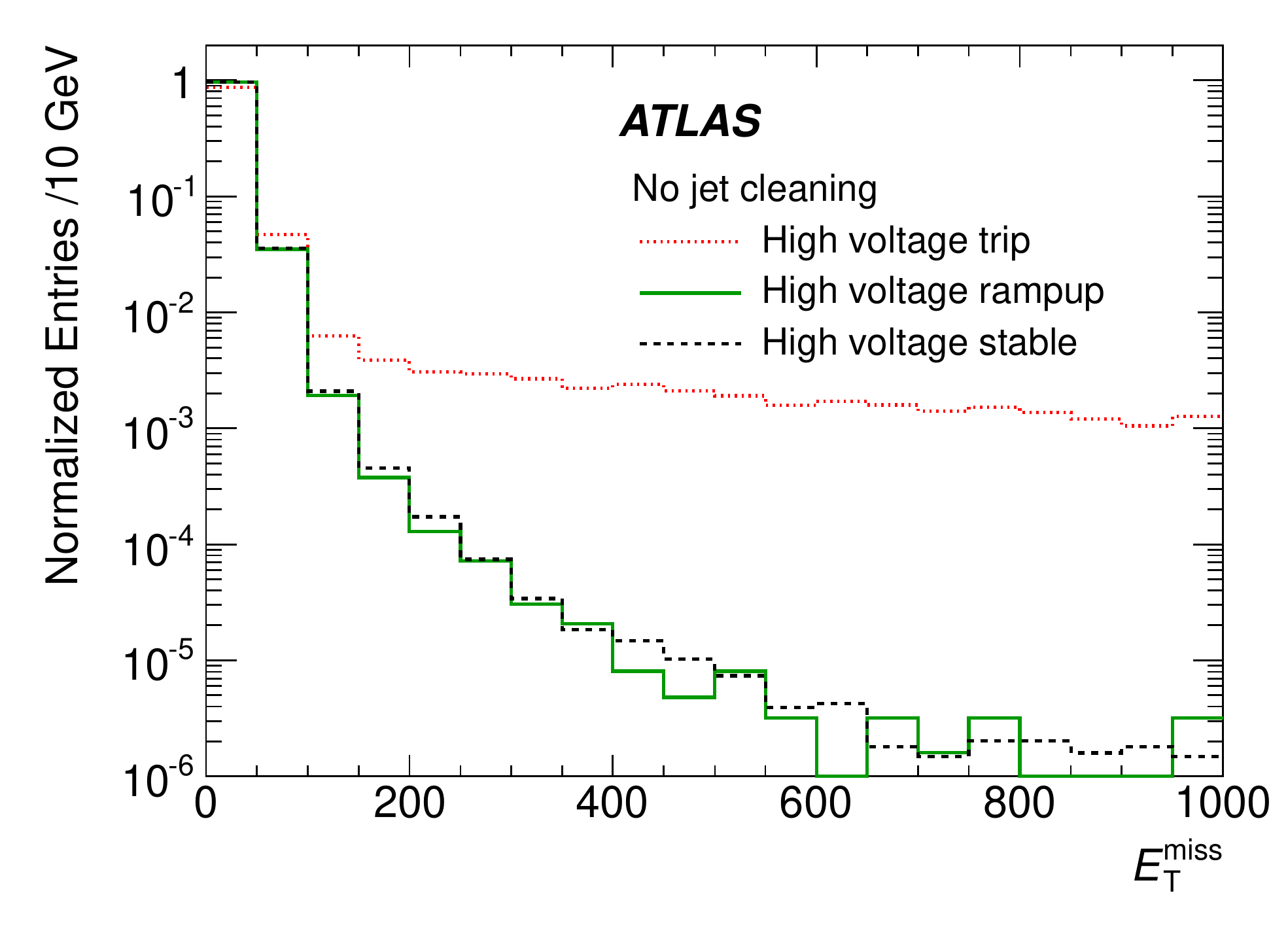}}
    \subfloat[]{\label{fig:rampup2}\includegraphics[width=0.48\textwidth]{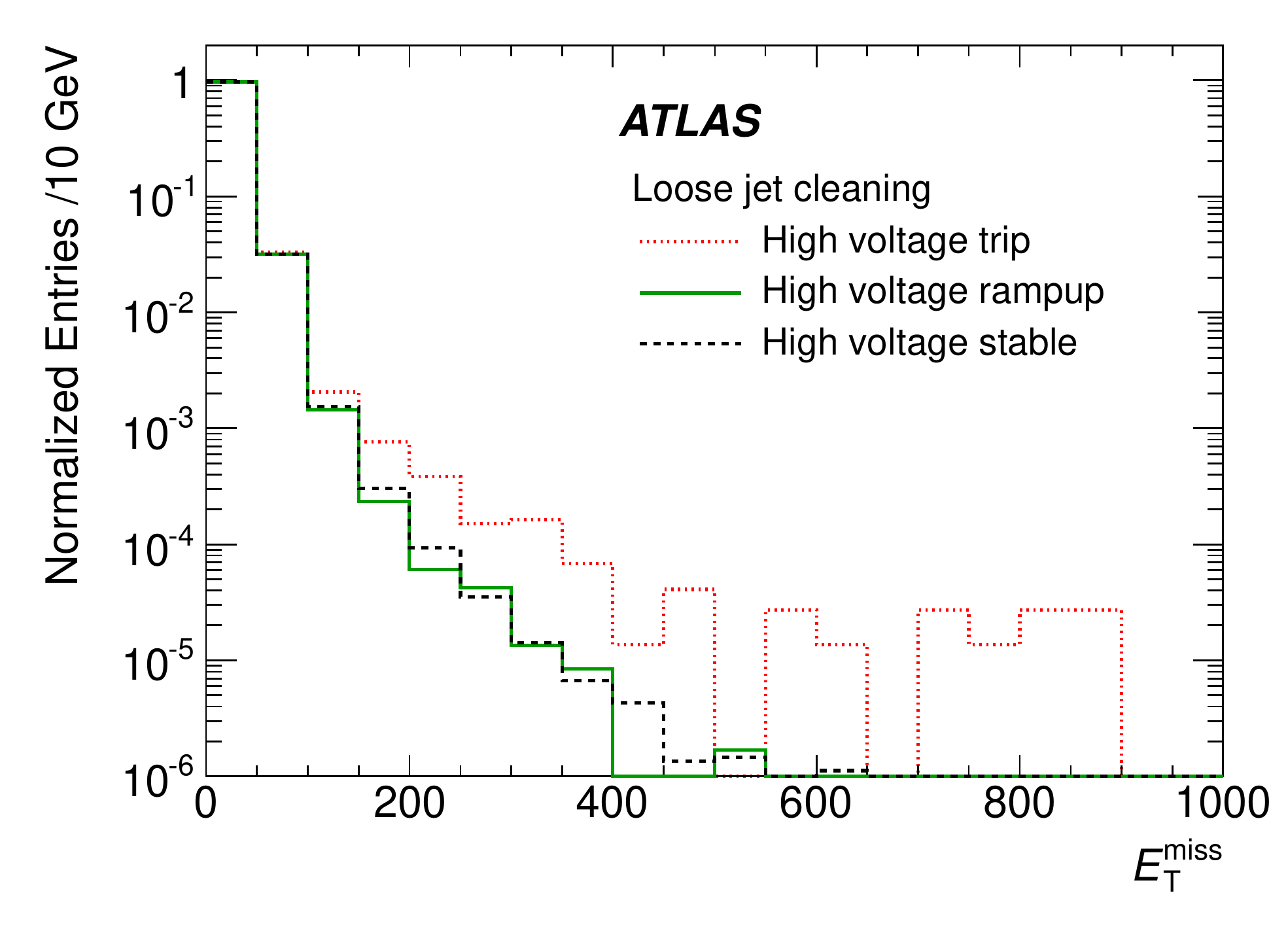}}
   \caption{Distributions of missing transverse momentum, $E_{\rm{T}}^{\rm{miss}}$, measured in 2011 collision data in {\streamJet} for {\LB}s with stable HV conditions (dashed line), a HV trip (dotted line) and a HV line ramping up (full line). Distributions are shown (a) without any jet-cleaning and (b) with a loose jet-cleaning procedure applied.}
   \label{fig:rampup}
\end{figure}

All the 2011 collision data containing {\LB}s affected by a HV trip or a ramp-up were considered for this study. A search for a potential noise excess was performed on the {\streamJet} data by considering the missing transverse momentum distributions computed in {\LBs} with different HV conditions (trip, ramping up, stable). In figure~\ref{fig:rampup}\subref{fig:rampup1}, a clear noise excess is seen in the {\LBs} when a trip occurred. The {\LBs} with a ramping HV line exhibit behaviour very similar to that of the regular {\LBs}. Figure~\ref{fig:rampup}\subref{fig:rampup2} shows the same distributions after applying the ``loose jet-cleaning procedure'' applied routinely to ATLAS physics analyses~\cite{JetCleaning2010,BeamBackground2011}. This cleaning procedure is based on a set of variables related to hadronic shower shapes, characteristics of ionization pulse shapes, etc. and is meant to remove fake jets due to calorimeter noise and out-of-time pile-up. The noise observed in the {\LBs} (systematically rejected) where a trip occurred is largely reduced, whereas the other types of {\LBs} still exhibit very similar behaviours. 

A complementary cross-check was performed by considering the rate of reconstructed jets in the same three types of {\LBs} in the {\streamCosmic} where no collision is expected. Before any jet-cleaning procedure, it appears that the rate of jets in the {\LBs} where a trip occurred is 1.6 times larger than in regular {\LBs}. In the case of {\LBs} with a ramping HV line, no difference from the regular {\LBs} is observed within a statistical error of 10\% on the ratio of the number of jets.

Hence, these studies confirm  that the {\LBs} with a ramping HV line can safely be kept for analysis. Those {\LB}s are, however, marked with a tolerable defect, in order to keep track of this hardware feature and ease the extraction of the corresponding data for detailed studies.

\subsection{Monitoring of coverage}

The LAr calorimeter design nominally provides full hermeticity in azimuth and longitudinal coverage up to $|\eta| = 4.9$. However, when hardware failures (though rare) occur, this coverage may be degraded. The inefficiencies can, for example, be due to a faulty HV sector where all HV lines are down. In this case, the resulting dead area is of typical size $\Delta\eta\times\Delta\phi=0.2\times 0.2$, and usually affects several calorimeter layers at the same time. Since such degraded coverage might significantly affect the physics performance, the corresponding data are systematically rejected by marking them with an intolerable coverage defect.

The detector coverage can also be degraded by a readout system defect. If the inactive region is limited to a single isolated FEB, the impact is usually restricted to a single layer in depth,\footnote{Due to cabling reasons, this statement does not apply to the hadronic calorimeters.} and the data are not systematically rejected. An intolerable defect is set only when four or more FEBs are simultaneously affected. If an important readout problem cannot be immediately fixed and must remain present during a long data-taking period, the intolerable defect policy is not acceptable, since ATLAS cannot afford to reject all the data taken for an extended period. Instead, for such incidents the inactive region is included in the Monte Carlo simulation of the detector response to automatically account for the acceptance loss in physics analysis. Such a situation happened once in 2011: six FEBs remained inactive for several months due to a hardware problem that prevented the distribution of trigger and clock signals. The problem was traced to a blown fuse in the controller board housed in the same front-end crate as the affected FEBs. Given the impossibility of swapping out boards while the ATLAS detector is closed, the problem was remedied only during the 2011--2012 technical stop. However, a spare clock and trigger distribution board was installed in summer 2011, allowing the recovery of four FEBs out of six for the last months of 2011 data taking. Also, three FEBs had to be switched off for approximately two weeks in 2012 due to a problem with the cooling circuit.

\subsection{Associated data rejection in 2012}\label{subs:dcs-dataRejection}

\begin{figure}[!htb]
    \subfloat[]{\label{fig:2012HVTRIP}\includegraphics[width=0.48\textwidth]{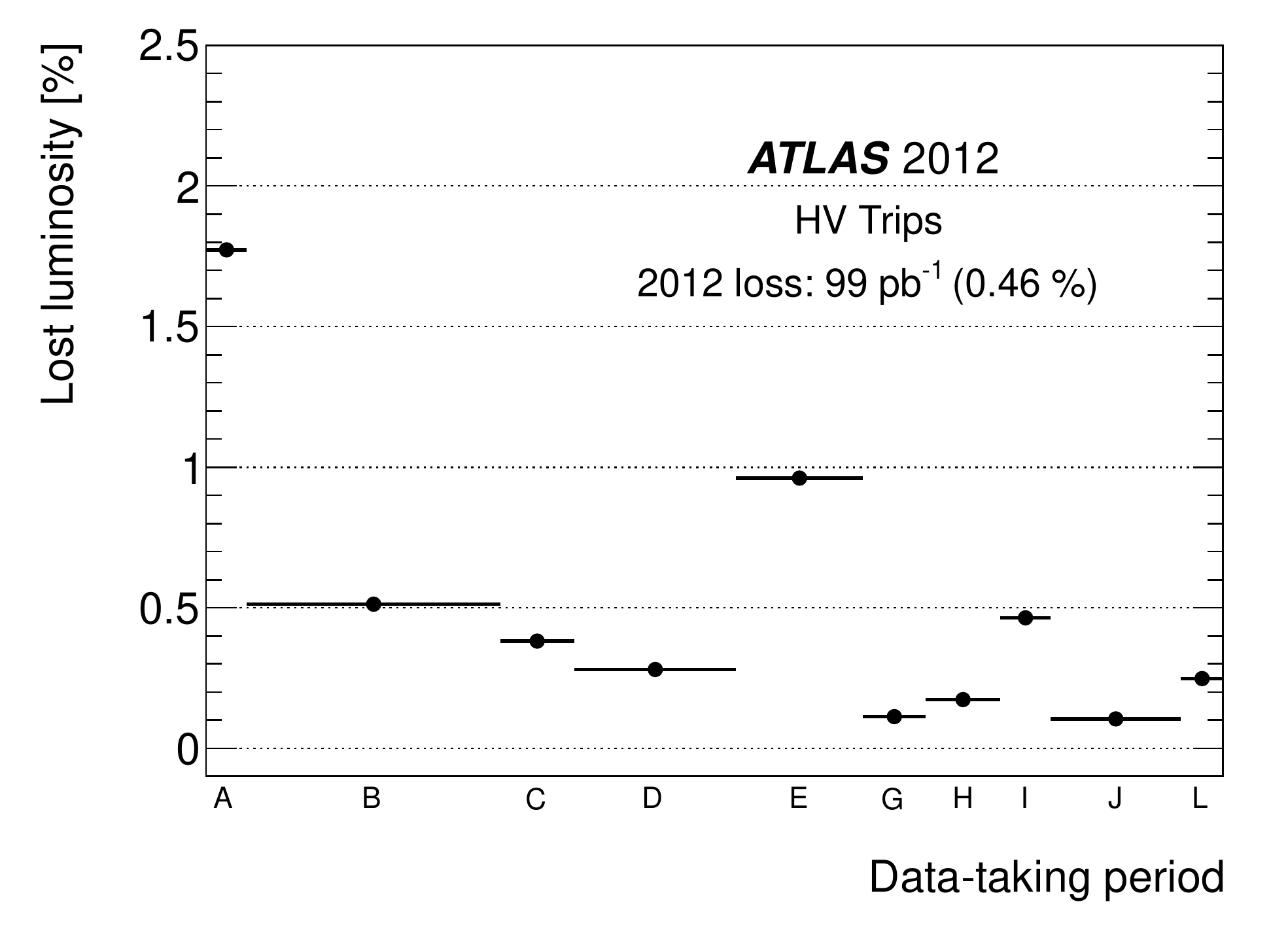}}
    \subfloat[]{\label{fig:2012COVERAGE}\includegraphics[width=0.48\textwidth]{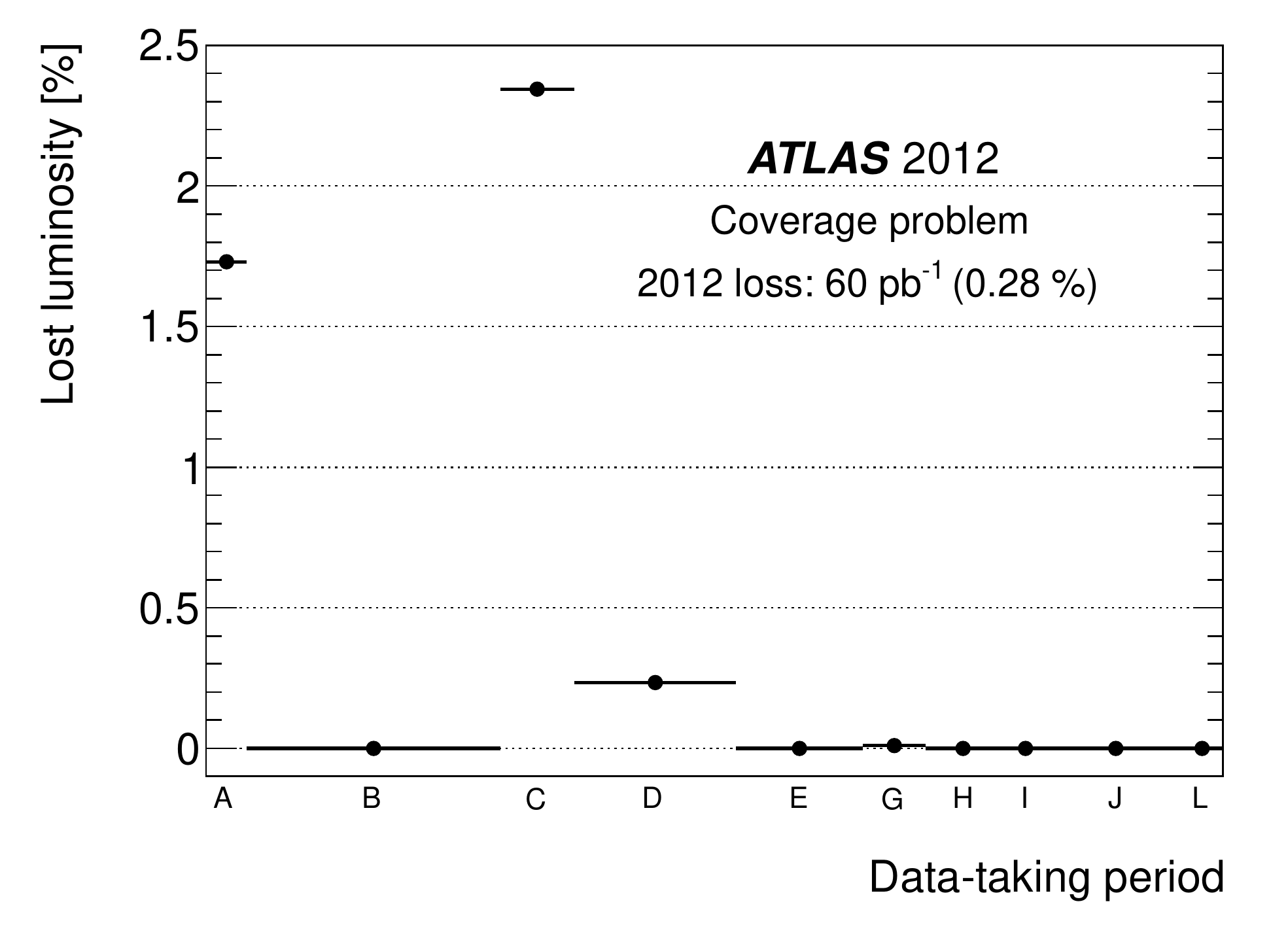}}
   \caption{Lost luminosity due to (a) HV trips and (b) inefficient areas impacting detector coverage as a function of the data-taking period in 2012.}
   \label{fig:2012DCS}
\end{figure}

Figure~\ref{fig:2012DCS}\subref{fig:2012HVTRIP} shows the time evolution of the data rejection level due to HV trips in 2012. In this figure and in all the similar plots of the following sections, the varying bin widths reflect the varying integrated luminosities of the ten 2012 data-taking periods (see section~\ref{sect:intro}).
The remarkable reduction of the losses over the year is mainly due to two effects. 

First, and for reasons not completely understood, the HV trips seemed to occur mainly when the LHC instantaneous luminosity was increasing significantly (typically doubled or tripled) over a few-day period. After a couple of days with stable peak luminosity, the occurrence of trips significantly decreased and then remained very low. When the collisions stopped or if the luminosity was very low for several weeks (machine development, long technical stops, etc.), this transient ``training'' period would recur briefly before a stable HV system was recovered.

Second, the rate of trips was reduced by installing new power supply modules shortly before the start of data taking period B. These new power supplies are able to temporarily switch to a ``current mode'', delivering a user-programmed maximum current resulting in a brief voltage dip instead of a trip~\cite{valerioHV}. Only the most sensitive sectors of the electromagnetic endcap localized at large pseudorapidities (e.g.\ small radius) were equipped with these special power supplies. Additional modules of this type are planned to be installed in 2014 before the LHC restarts.

Figure~\ref{fig:2012DCS}\subref{fig:2012COVERAGE} shows the time evolution of the 2012 data rejection level due to a large inefficient area of detector coverage. The highest inefficiency, observed during period C, comes from special collision runs with the toroidal magnet off, dedicated to the improvement of the relative alignment of the muon spectrometer. During this two-day period, expected to be rejected in any physics analysis, large regions of the HV system were intentionally switched off to investigate the source of noise bursts (see section~\ref{sect:noiseBursts}). The two other sources of data loss in periods A and D are due to two faulty low-voltage power supplies in a front-end readout crate, equivalent to more than 25 missing FEBs or a coverage loss greater than 1\%. These two problems were only transient, lasting less than a couple of hours, the time needed to replace the power supply.

\section{Data integrity and online processing} \label{sect:dataIntegrity}

Each one of the 1524 FEBs amplifies, shapes and digitizes the signals of up to 128 channels~\cite{FEBDesign}. In order to achieve the required dynamic range, the amplification and shaping are performed in parallel with three different gains (of roughly 1, 9.9, and 93). When an event passes the first level of trigger, the signal is digitized. Only the signal with the optimal gain is digitized by a 12-bit analog to digital converter (ADC) at a sampling frequency of 40~MHz. After this treatment, five digitized samples\footnote{In debugging/commissioning mode, up to 32 samples can be readout.} are sent for each cell to the ROD system \cite{RODDesign} via optical links. The ROD boards can either transparently transmit the digitized samples to the data-acquisition system ({\itshape transparent} mode), or compute the energy of the cell and transmit only one number, hence reducing the data size and the offline processing time ({\itshape results} mode). During calibration runs, the ROD can also work in a special mode, where several events are averaged to limit the data size and optimize processing time; however, this is not further considered in this article.

In results mode, the cell energy $E$, directly proportional to the pulse shape amplitude $A$, is computed with a Digital Signal Processing (DSP) chip mounted on the ROD boards, using an optimal filtering technique \cite{clelandOF,larreadiness} and transmitted to the central data-acquisition system. When the energy is above a given threshold $T_{Q\tau}$, the peak time $\tau$ and a quality factor $Q$ are also computed. These quantities can be expressed as:
$$ E = \sum\limits_{i=1}^{5} a_i(s_i - ped) \hspace{1cm} E\times\tau = \sum\limits_{i=1}^{5} b_i(s_i - ped) \hspace{1cm} Q = \sum\limits_{i=1}^{5} (s_i - ped -A(g_i-\tau g'_i))^2 $$ 
where $s_i$ are the five digitized samples, {\itshape ped} is the electronics baseline value, and $g_i$ and $g'_i$ are respectively the normalized ionization pulse shape and its derivative with time. The optimal filtering weights, $a_i$ and $b_i$ are computed per cell and per gain from the predicted ionization pulse shape and the measured noise autocorrelation to minimize the noise and pile-up contributions to the amplitude~A.

The quality factor that reflects how much the pulse shape looks like an argon ionization pulse shape, is lower than 4000 in more than 99\% of argon ionization pulses. Because the quality factor is computed by the DSP chip in a 16-bit word, it is limited to $2^{16}-1 = 65535$; the probability that this saturated value corresponds to a real energy deposit in the calorimeter is estimated negligible.
 
For cell energies above a second energy threshold $T_{\rm{samples}}$ (in absolute value), the five digitized samples are also transmitted to the central data-acquisition system. The two energy thresholds $T_{Q\tau}$ and $T_{\rm{samples}}$ are tuned such that approximately 1--2\% of the cells are involved. This corresponds to an energy threshold of around 50~MeV--10~GeV depending on the layer/partition.

\subsection{Basic data integrity}
\label{DI:singleFEB}

Since the FEB output is the basic detector information building block, careful data integrity monitoring at the earliest stages of the processing chain is mandatory. The input FPGA chip on the ROD board performs basic online checks of the FEB data: most importantly it checks for any error word sent by the different chips on each FEB and checks consistency of data (BCID, event identifier, etc.) defined for each channel which are expected to be uniform but not propagated individually to the data-acquisition system. Beyond these online consistency checks, a software algorithm running both online and offline performs additional checks which require: presence of all data blocks, unchanged data block length from the FEBs to the central data acquisition system, uniform data type and number of digitized samples among the 1524 FEBs. The most serious case of data corruption was observed in 2010 and consisted of a spurious loss of synchronization between the FEB clock and the central clock. The origin of this problem was identified in early 2011 as interference between the two redundant clock links available in each FEB: when only one was supplied with a signal, the inactive link could induce a desynchronization. The problem was fixed by permanently sending a fixed logic level to the inactive clock circuit.

An FEB integrity error indicates a fatal and irrecoverable data corruption. To ensure as uniform a readout coverage as possible within a run, any event containing a corrupted block is discarded. This event rejection is performed offline by applying the time-window veto procedure described in section~\ref{dqLogging}. To limit the offline rejection when a permanent corruption error is observed during data taking, the run must be paused (or stopped and restarted) as promptly as possible to reconfigure the problematic FEBs. However, if the data corruption is limited to less than four FEBs, the ATLAS run coordinator may consider this loss as sustainable and keep the run going to maximize the data-taking efficiency. In this case, the problematic FEBs are masked offline (the data integrity issue translates into a coverage inefficiency), and the data are not rejected but marked with a tolerable defect. This unwanted case happened only twice during 2012.

\begin{figure}[!htb]
    \subfloat[]{\label{fig:Saturation}\includegraphics[width=0.48\textwidth]{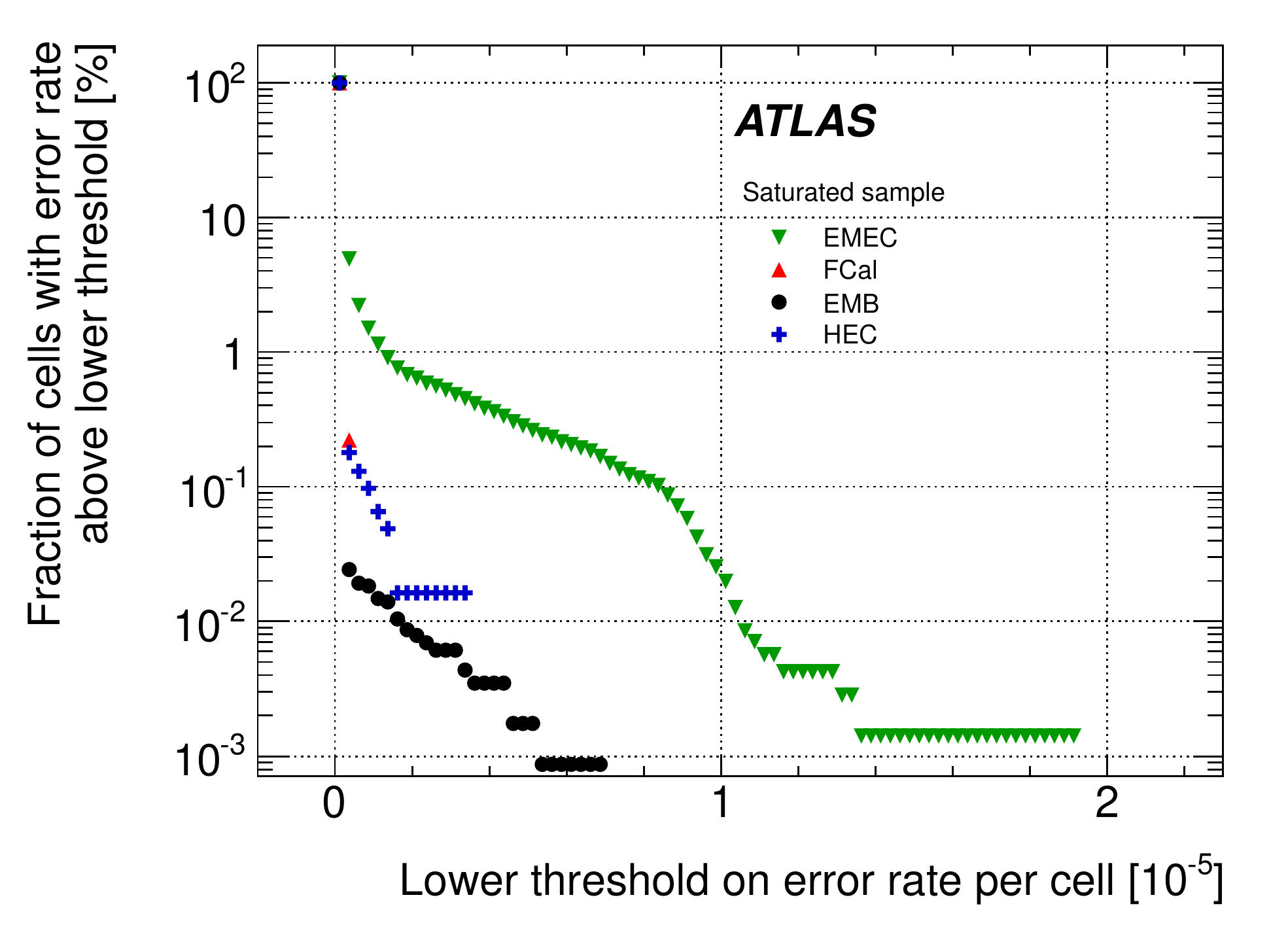}}
    \subfloat[]{\label{fig:NullDigit}\includegraphics[width=0.48\textwidth]{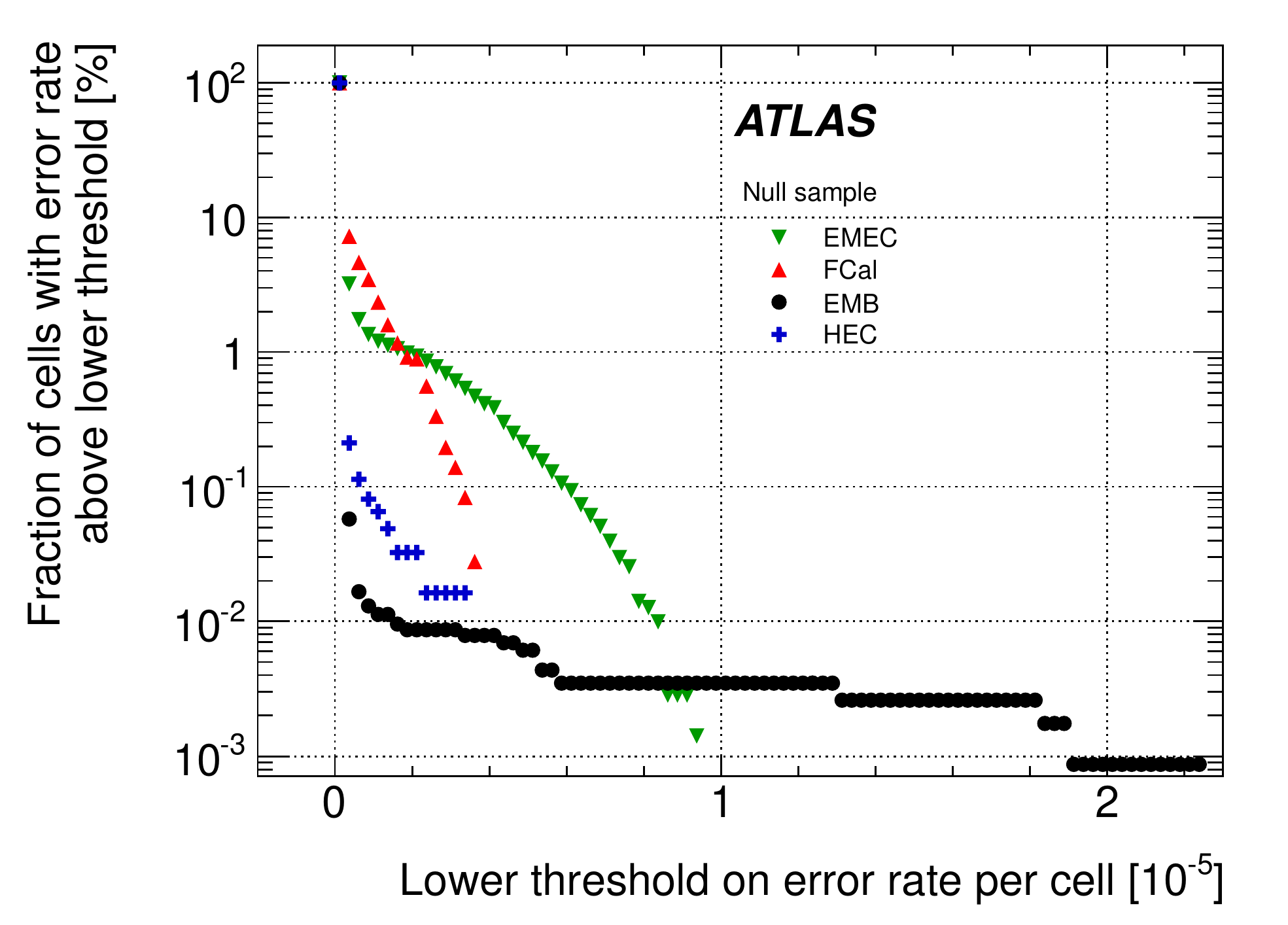}}
   \caption{(a) Percentage of cells that exhibit a saturated sample in a certain proportion of events. (b) Percentage of cells that exhibit a null sample in a certain proportion of events. In both plots, the $x$-axis shows the lower threshold on the error rate per cell. The results are shown for the run~205071.}
   \label{fig:6}
\end{figure}

When the digitized samples are available, the yield of events with a null or saturated sample (i.e. an ADC value equal to 0 or 4095) is monitored. Several problems could induce a large yield of saturated or null samples: a malfunctioning ADC or gain selector, large out-of-time channel/FEB signal, data fragment loss, etc. The proportions of affected events per cell for the run 205071 are presented in figures~\ref{fig:6}\subref{fig:Saturation} and \ref{fig:6}\subref{fig:NullDigit}. In the electromagnetic barrel and the hadronic endcaps, the proportions are close to zero. In the electromagnetic endcaps and forward calorimeter, the yield is slightly higher but still very low: around 0.01\% of EMEC channels exhibit a saturated (null) sample in more than $10^{-5}$ ($0.8 \cdot 10^{-5}$) of events. Moreover, this observation is not due to a defect in the readout chain but simply to the out-of-time pile-up. For these events, the signal peak of the cell is shifted, and the gain selection based on the in-time signal is not appropriate. 
The endcaps are most affected because of a higher particle flux at large pseudorapidity. It is, however, less pronounced in the FCal than in EMEC due to the decision to allow only the medium and low gains in the FCal readout chain specifically for this reason. With a pile-up noise systematically greater than the medium gain electronics noise, this setting does not affect the overall performance. Neither does the very low occurrence of null/saturated samples measured in other partitions (EMB, EMEC and HEC).

\subsection{Online computation accuracy}
\label{DI:DSP}

In results mode, but only for cells where the digitized samples are available, the energy can be recomputed offline with the same optimal filter and compared to the online value to test the DSP computation reliability. Due to the intrinsic hardware limitations of the DSP chip, the precision of the energy computation varies from 1~MeV to 512~MeV, the least significant bit, depending on the energy range \cite{electronicPerf}.

\begin{figure}[!htb]
    \subfloat[]{\label{fig:DSPComputation1}\includegraphics[width=0.48\textwidth]{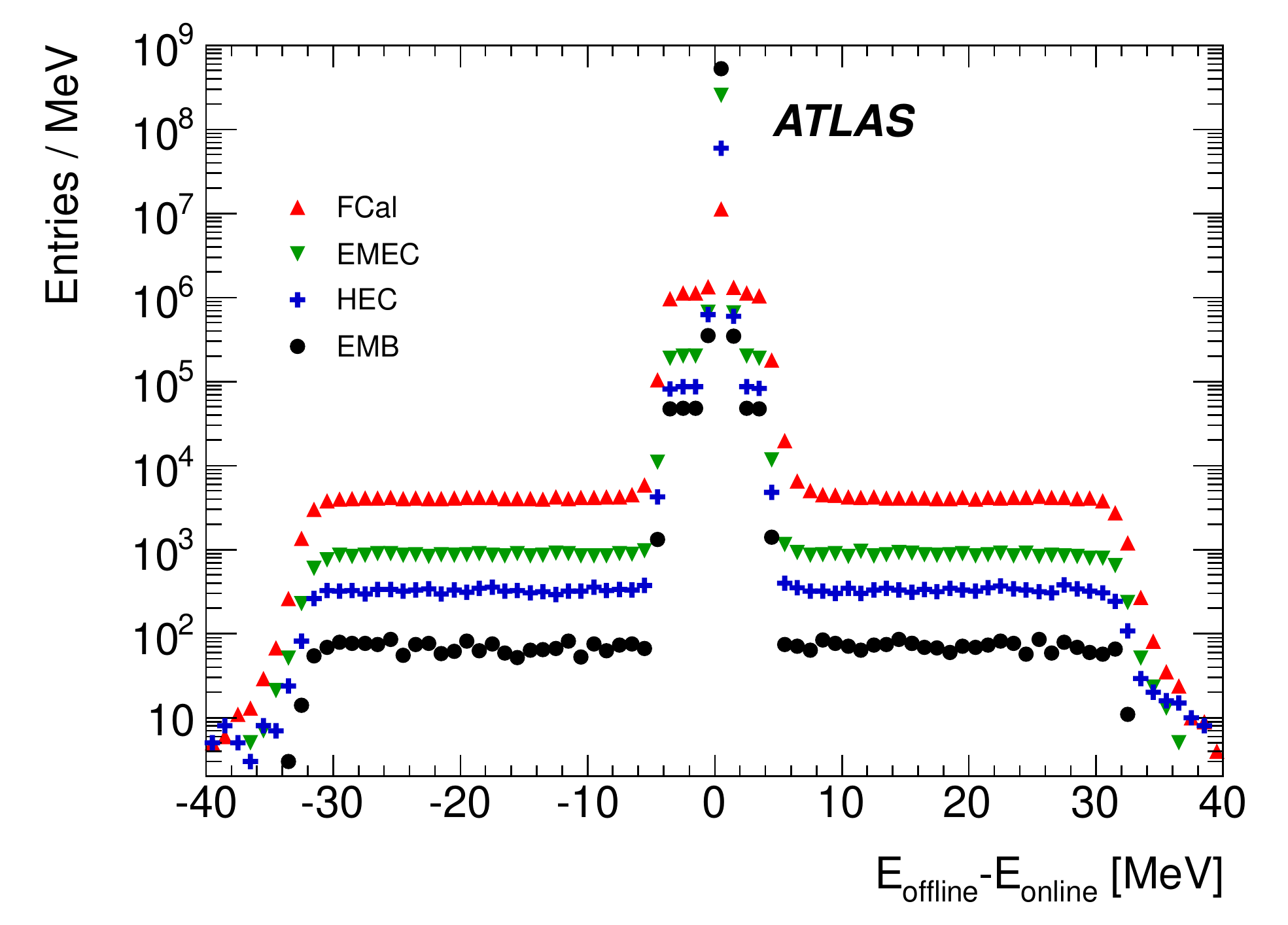}}
    \subfloat[]{\label{fig:DSPComputation2}\includegraphics[width=0.48\textwidth]{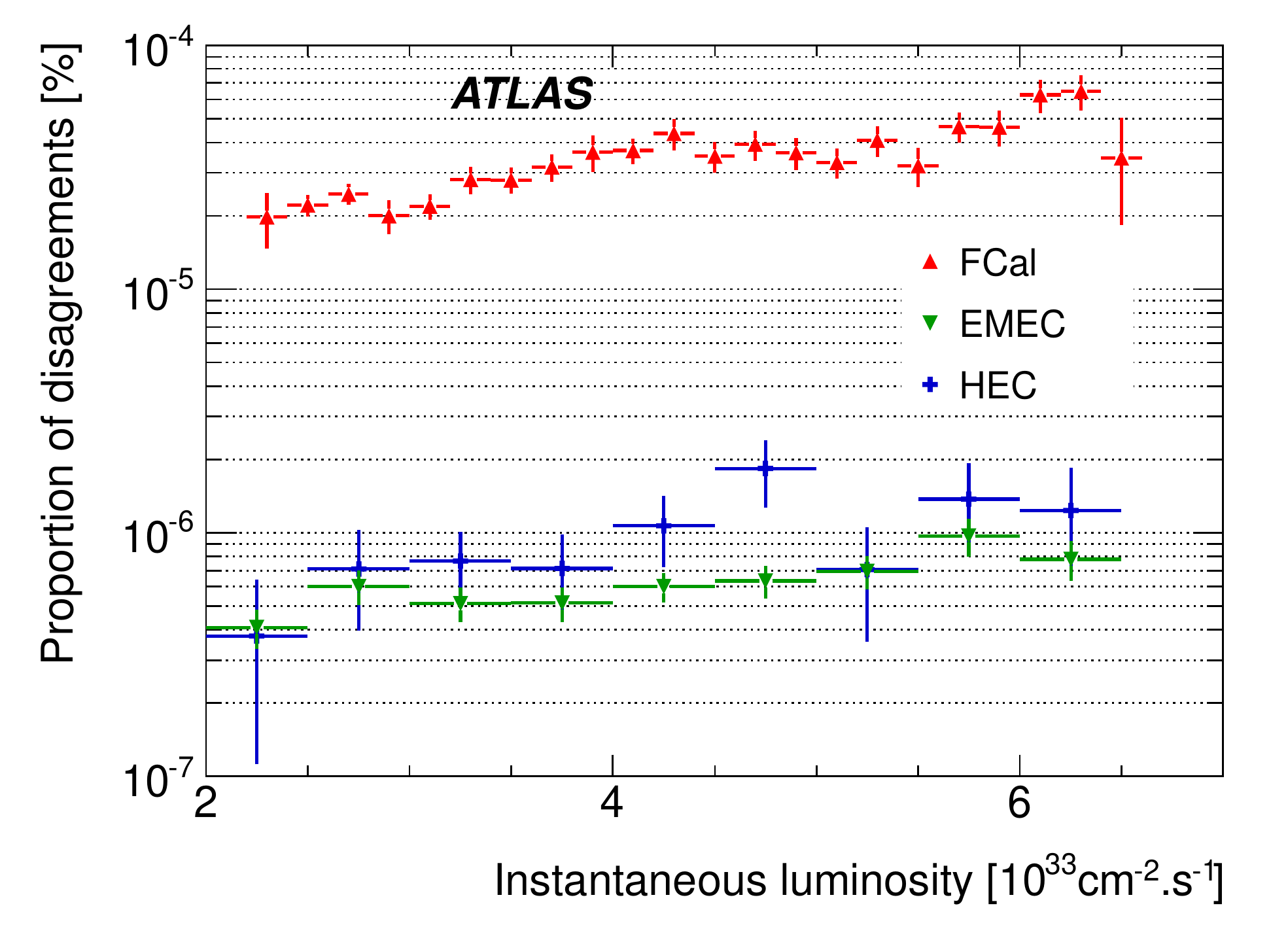}}
   \caption{DSP energy computation accuracy in run 205071. (a) Distribution of the difference between the energy computed offline for the four different partitions and the energy computed online by the DSP. (b) Proportion of computation disagreements (i.e. when the online/offline energy difference lies outside the expected DSP accuracy range) per partition as a function of the instantaneous luminosity. The barrel proportion is not displayed, as only one single disagreement (one channel in one event) is observed.}
   \label{fig:7}
\end{figure}

Figure~\ref{fig:7}\subref{fig:DSPComputation1} shows the distribution of the difference between the online and offline energy computations. A satisfactory agreement between the two calculations is found for the four partitions. Here again, the tails of the distributions are slightly more pronounced in the partitions most affected by out-of-time pile-up (EMEC, FCal). This can be explained by the limited size of the DSP registers (16 bits) that implies specific coefficients rounding rules optimized to deal with in-time signals. This explanation is supported by figure~\ref{fig:7}\subref{fig:DSPComputation2}, which shows an increase in the computation-disagreement yield (normalized by the number of events and the number of channels in each partition) as a function of the instantaneous luminosity.

A similar analysis was also performed to check the correctness of the time and quality factor computed online, and similar accuracies were observed. Since the first LHC collisions, the DSP computation has proved to be fully accurate and never induced any data loss.

\subsection{Missing condition data}
\label{DI:corruptReco}

To limit the effect of out-of-time pile-up, the FEB shaping stage is bipolar (see figure~\ref{fig:1}\subref{fig:PulseShape}), allowing a global compensation between the signal due to the following collisions and the signal due to the previous ones. However, this remains inefficient for the collision events produced in the first (last) bunches of a train: the electronic baselines are then positively (negatively) biased. To correct this bias, the average energy shift is subtracted offline based on the position of the colliding bunches in the train. The pile-up correction makes use of the instantaneous luminosity per bunch provided by the ATLAS luminosity detectors. Due to hardware or software failures, the database information about the instantaneous luminosity may be missing. In that case, the reconstruction of the LAr calorimeter energy is considered non-optimal, and the data are rejected by assigning a dedicated intolerable defect associated with the luminosity detectors. Even if the origin of this feature is not related to the LAr calorimeters, an additional intolerable defect associated with the LAr calorimeter is also assigned to keep track of the non-optimal reconstruction.

\subsection{Associated data rejection in 2012}

\begin{figure}[!htb]
    \subfloat[]{\label{fig:2012DATACORRUPT-TWV}\includegraphics[width=0.48\textwidth]{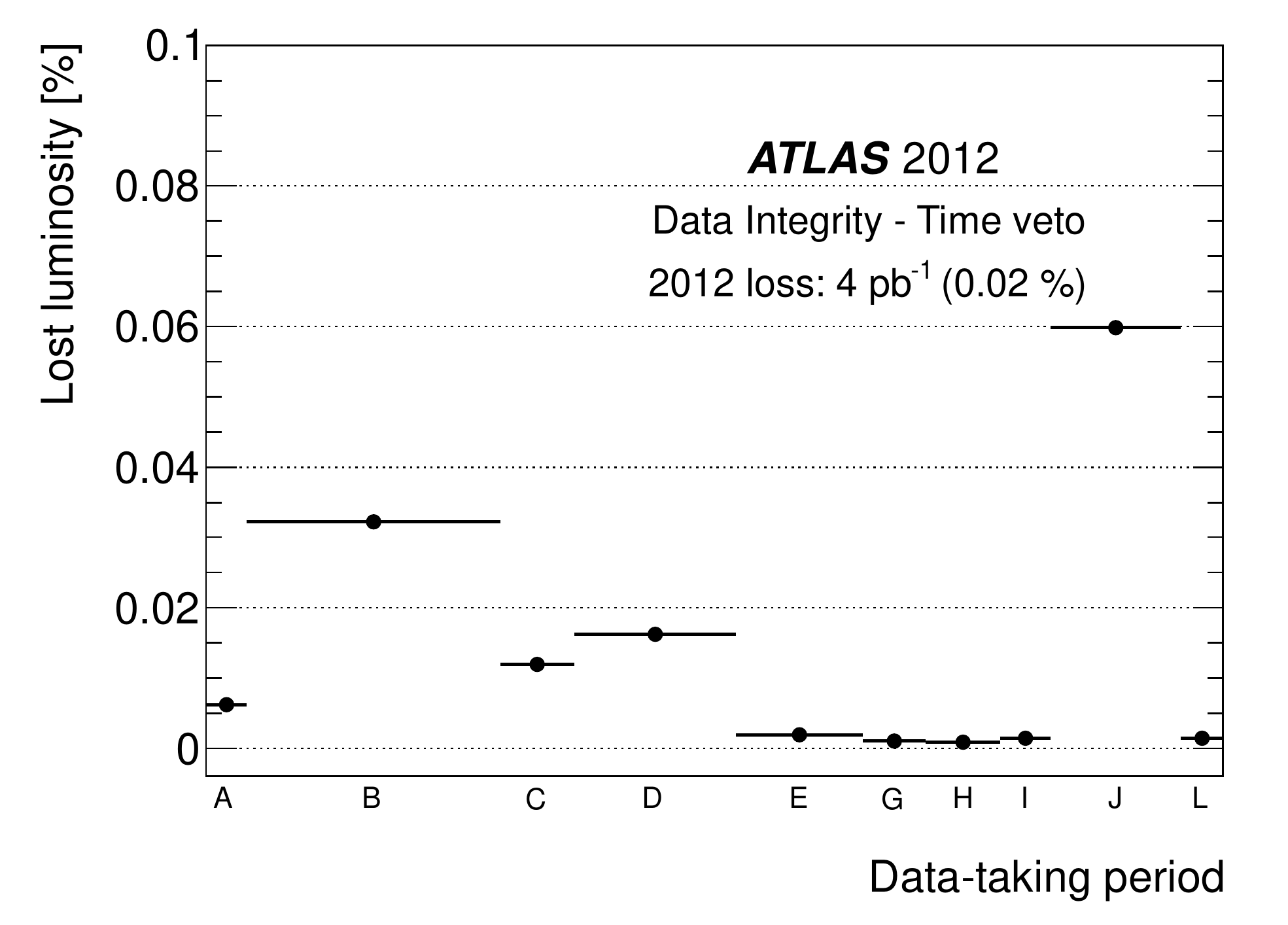}}
    \subfloat[]{\label{fig:2012DATACORRUPT-GRL}\includegraphics[width=0.48\textwidth]{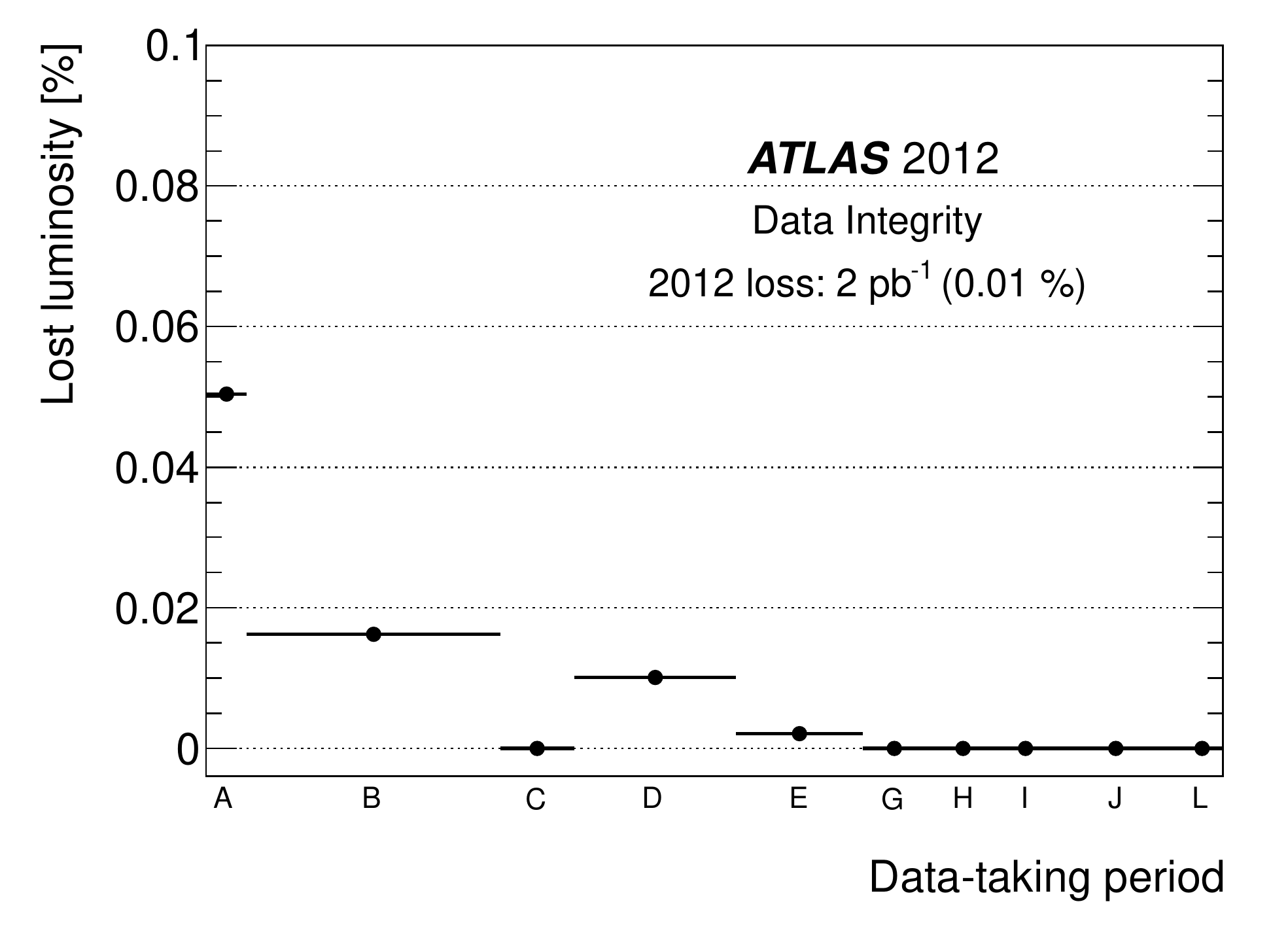}}
   \caption{Lost luminosity due to data corruption as a function of the data-taking period in 2012. (a) Loss due to the time-window veto procedure. (b) Loss due to defect assignment.}
   \label{fig:2012DataCorrupt}
\end{figure}

Figure~\ref{fig:2012DataCorrupt} shows the time evolution of data corruption in 2012 in terms of lost luminosity. The rejection rate is computed from two complementary sources: ~\subref{fig:2012DATACORRUPT-TWV} the time-window veto when the data corruption does not affect the whole {\LB}, and ~\subref{fig:2012DATACORRUPT-GRL} the list of defects corresponding to a totally corrupted {\LB}. In both cases, the rejection rate remains very low throughout the year and below 0.02\% on average. 

\begin{figure}[!htb]
  \center
   \includegraphics[width=0.48\textwidth]{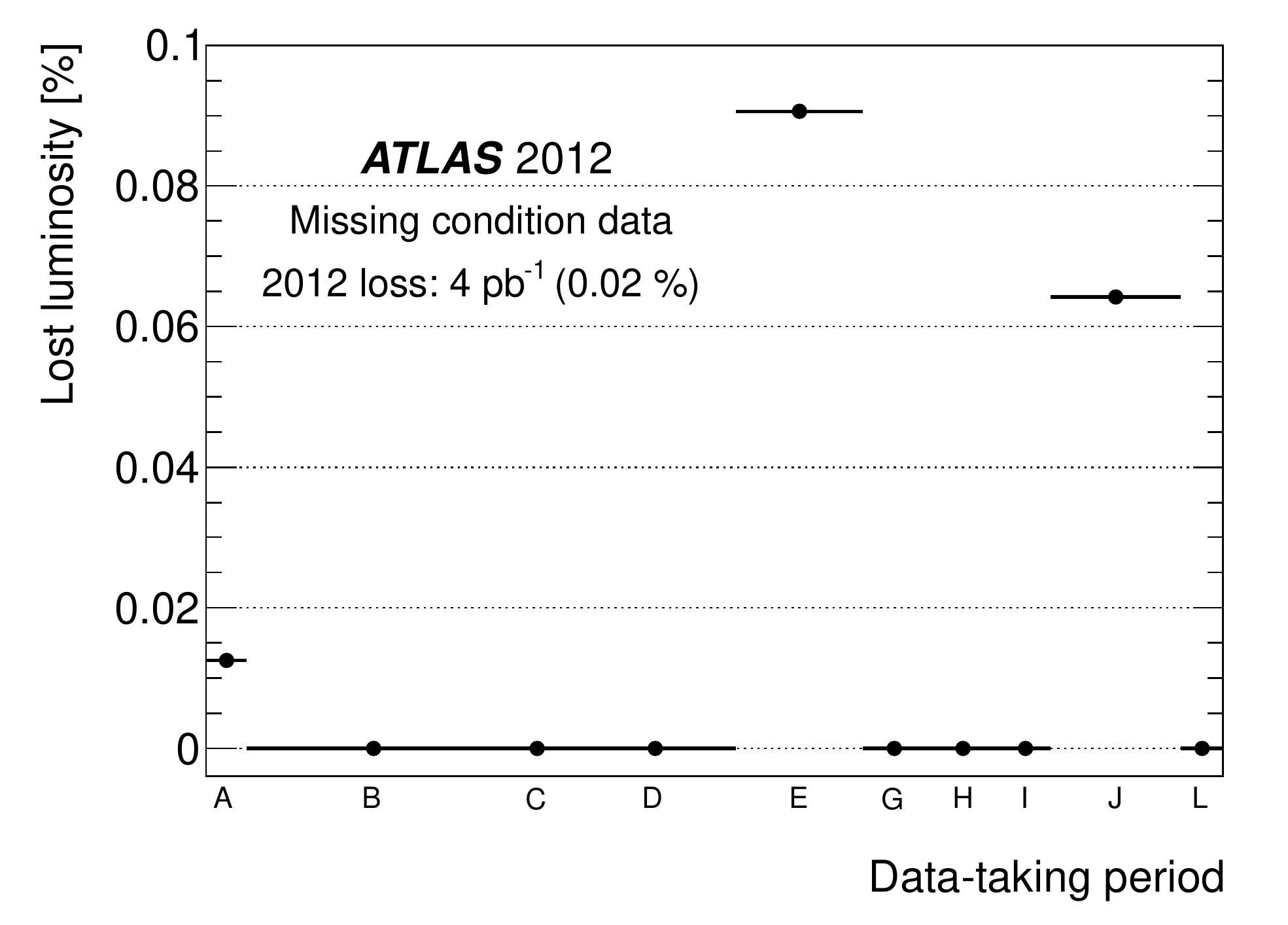}
   \caption{Lost luminosity due to missing conditions as a function of the data-taking period in 2012.}
   \label{fig:2012RecoCorrupt}
\end{figure}

Figure \ref{fig:2012RecoCorrupt} shows the data rejection due to missing conditions data. It remains very low and affects mainly isolated {\LB}s with corrupted instantaneous luminosity per bunch crossing.

\section{Calorimeter synchronization} \label{sect:timing}

A precise measurement of the time of the signal peak, derived from the optimal filter, is a valuable input to searches for exotic particles with a long lifetime or for very massive stable particles. Proper synchronization also contributes to improving the energy resolution. For these reasons, it is important to constantly monitor the calorimeter synchronization, both on a global scale and with finer granularity.

\subsection{Global synchronization} \label{subs:globalTiming}

A mean time is derived for each endcap by considering all cells of FCal (EMEC inner wheel\footnote{The EMEC inner wheel covers $2.5 < |\eta| < 3.2$.}) above 1.2~GeV (250~MeV) and by averaging their signal peak time. At least two energetic cells are requested to limit the impact of noisy cells. When both are available, the average time of the two endcaps is derived to monitor the global synchronization, while the time difference allows a check of the beam spot's longitudinal position and the presence of beam halo. Since the two endcaps are electrically decoupled, the presence of simultaneous signals in both endcaps is very likely to be due to real energy deposits and not due to noise. The high particle flux observed at the considered pseudorapidities allows refined monitoring as a function of time ({\LB}).

\begin{figure}[!htb]
    \subfloat[]{\label{fig:collisionTime1}\includegraphics[width=0.48\textwidth]{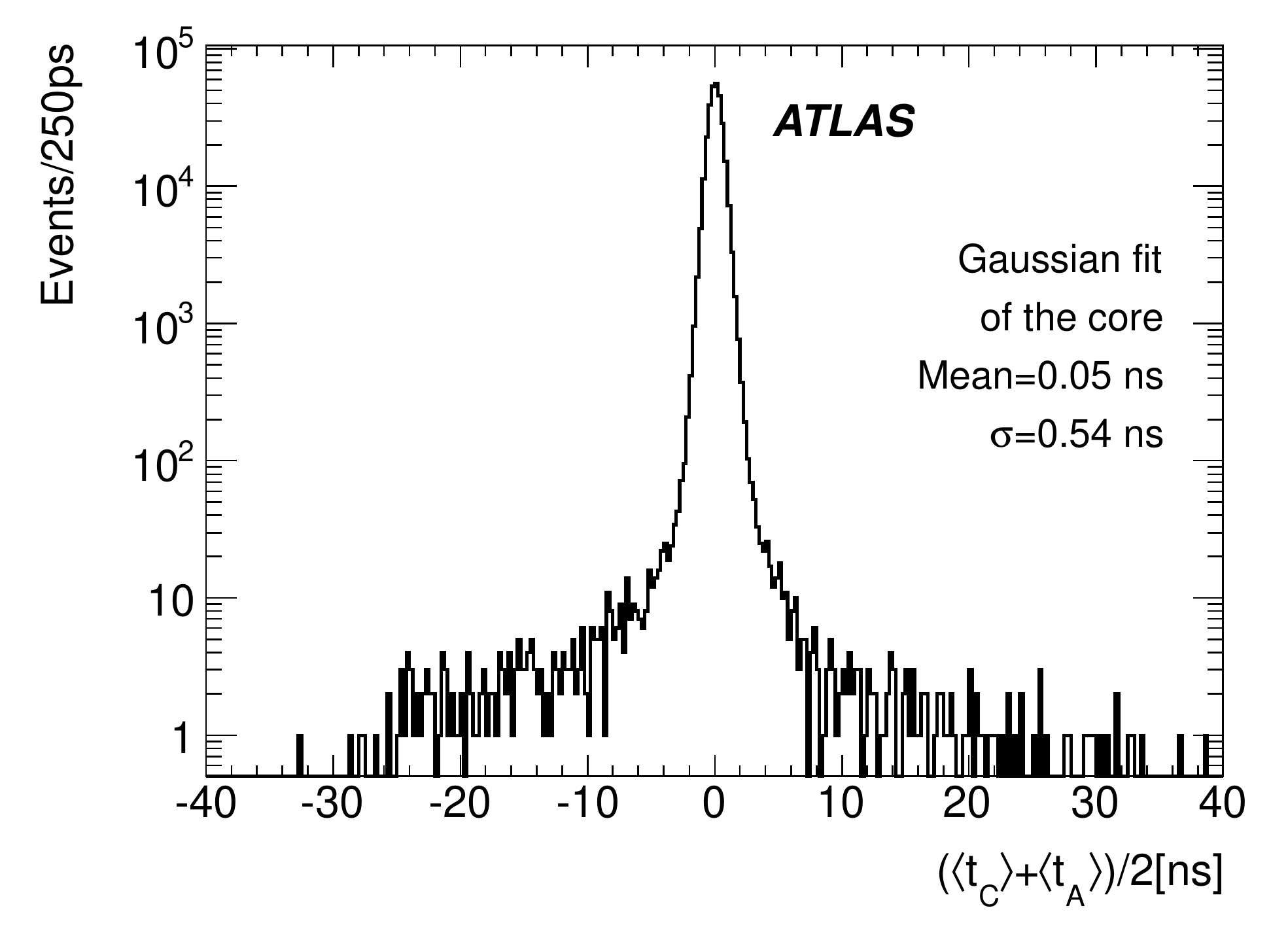}}
    \subfloat[]{\label{fig:collisionTime2}\includegraphics[width=0.48\textwidth]{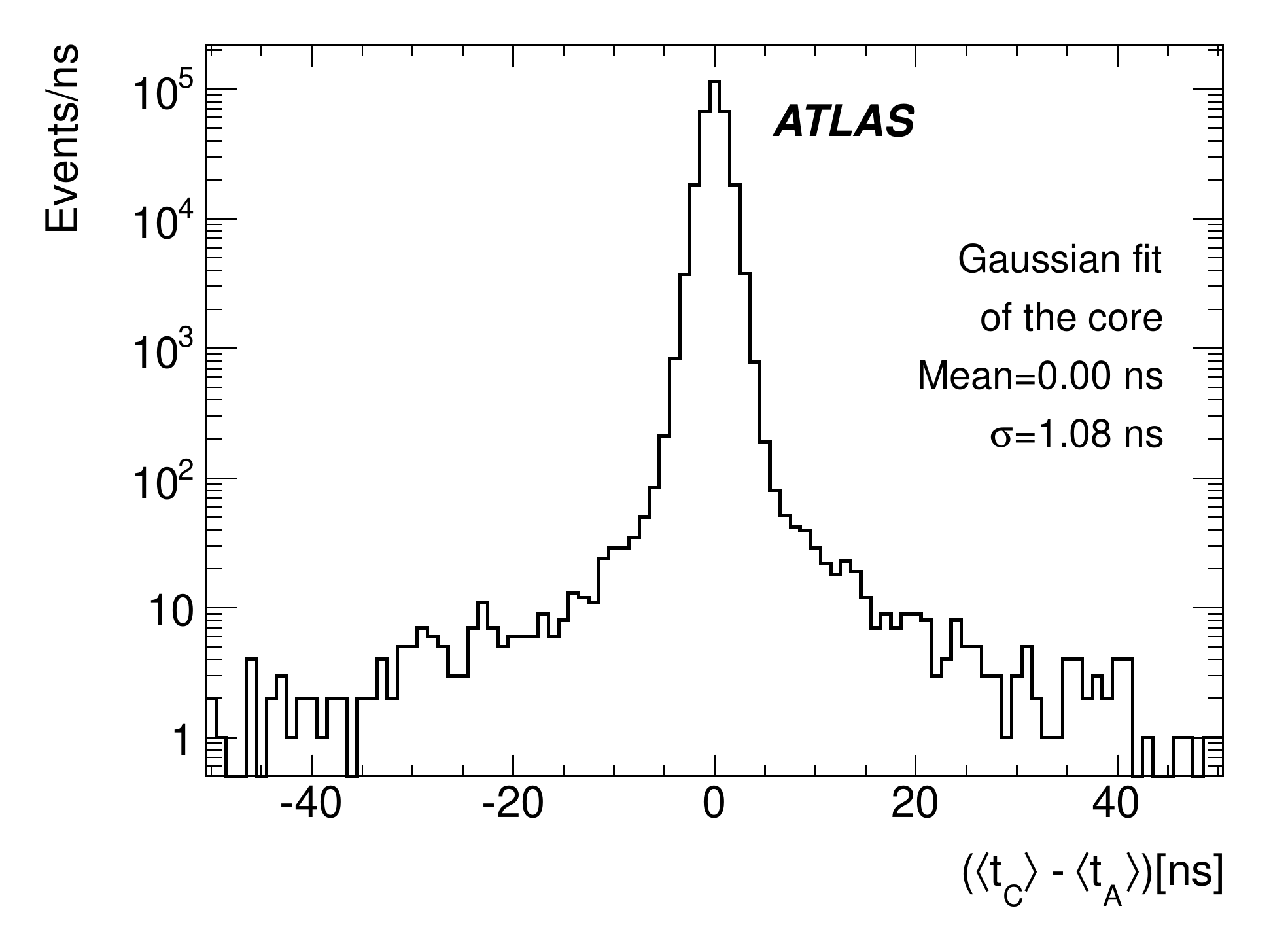}}
   \caption{(a) Average and (b) difference of the two endcap mean times, as defined in the text. The results are shown for the run 205071.}
   \label{fig:timing}
\end{figure}

Figure~\ref{fig:timing}\subref{fig:collisionTime1} shows the average value of the two endcaps' times for the run 205071. The distribution is centred around zero, indicating that the calorimeter (at least the FCal and EMEC inner wheel) is properly synchronized with the LHC clock, as is also shown in section~\ref{FEBtiming}.

Figure~\ref{fig:timing}\subref{fig:collisionTime2} shows the time difference between the two endcaps. The distribution is also centred around zero, indicating that the recorded events are mostly collisions well centred along the beam axis: the particles travel from the centre of the detector, and both endcaps send a signal synchronously. Some secondary peaks may arise due to beam halo, where particles cross the detector along the $z$-axis, from one endcap towards the other. Given the 9~m distance between the endcaps, and assuming that the particles travel at the speed of light, the difference between the signal arrival times from the two endcaps should peak at 30~ns for beam halo. These peaks were observed mainly in 2010; just a tiny bump is observed in the negative tail in figure~\ref{fig:timing}\subref{fig:collisionTime2}. The small continuous tails are due to out-of-time pile-up that may bias the average time of an endcap's signal.

\subsection{Synchronization at front-end board level} \label{FEBtiming}

\begin{figure}[!b]
\begin{center} 
    \subfloat[]{\label{fig:febtime_emb}\includegraphics[scale=0.39]{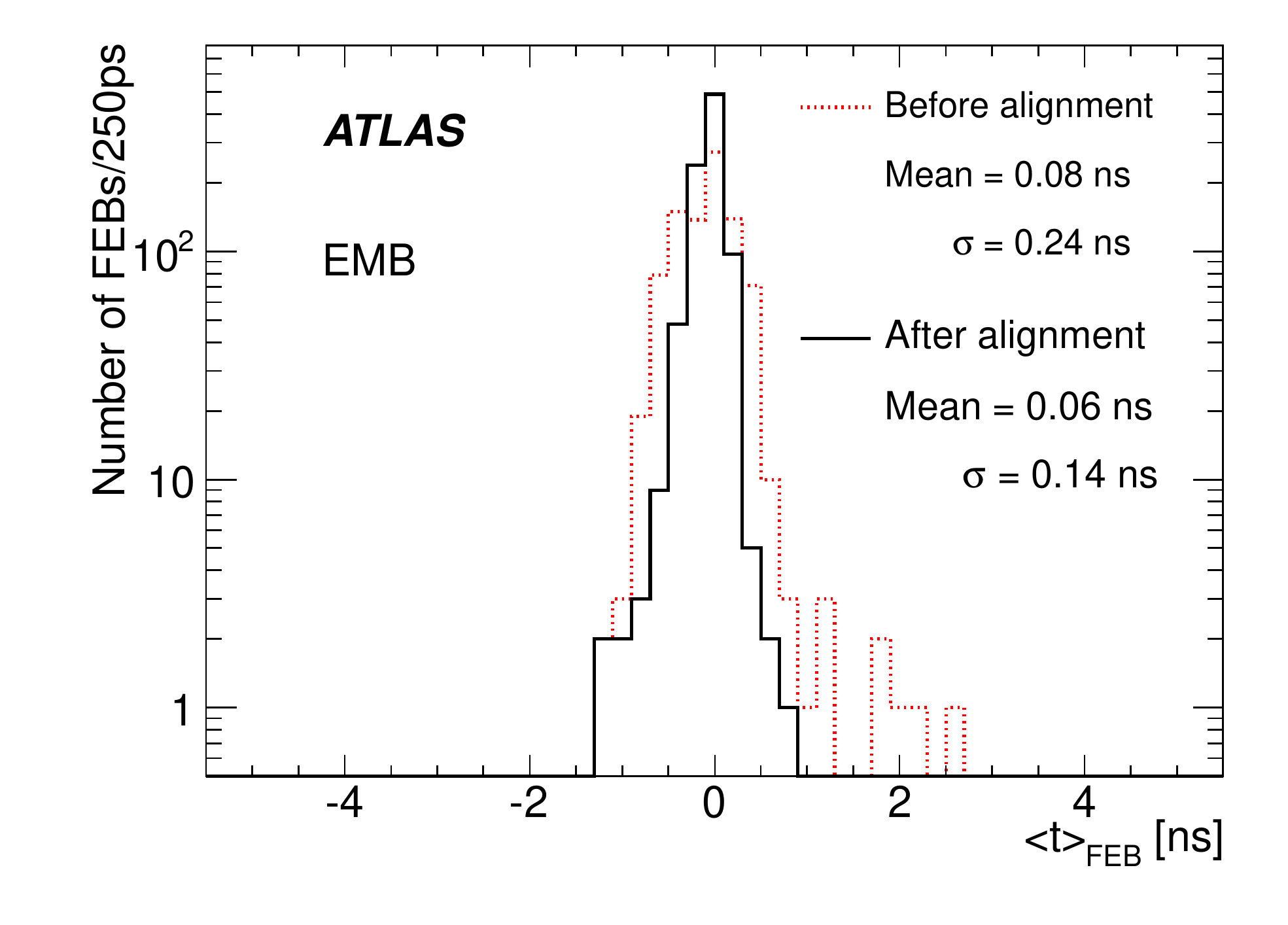}}
    \subfloat[]{\label{fig:febtime_emec}\includegraphics[scale=0.39]{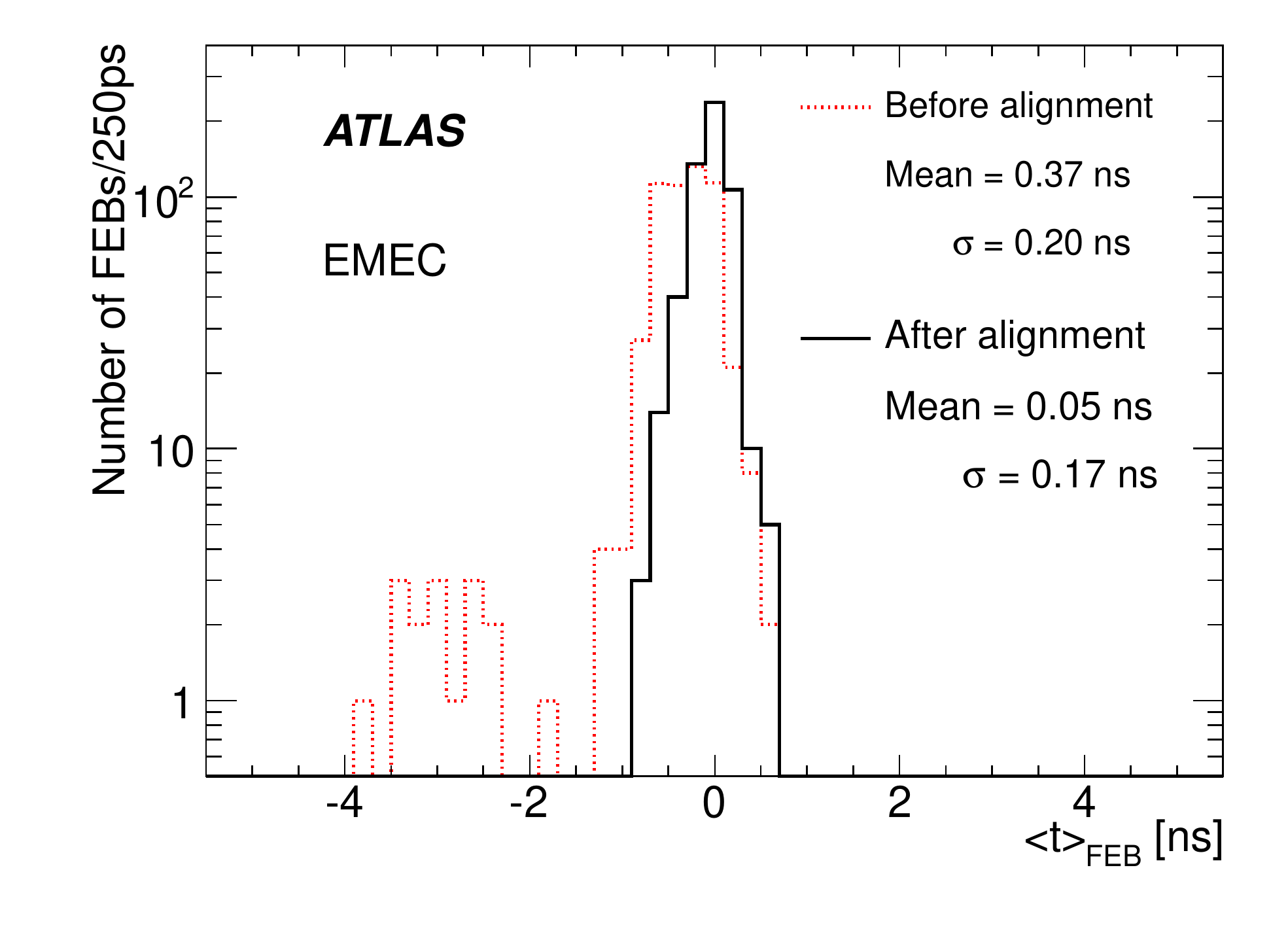}} \\  
    \subfloat[]{\label{fig:febtime-hec}\includegraphics[scale=0.39]{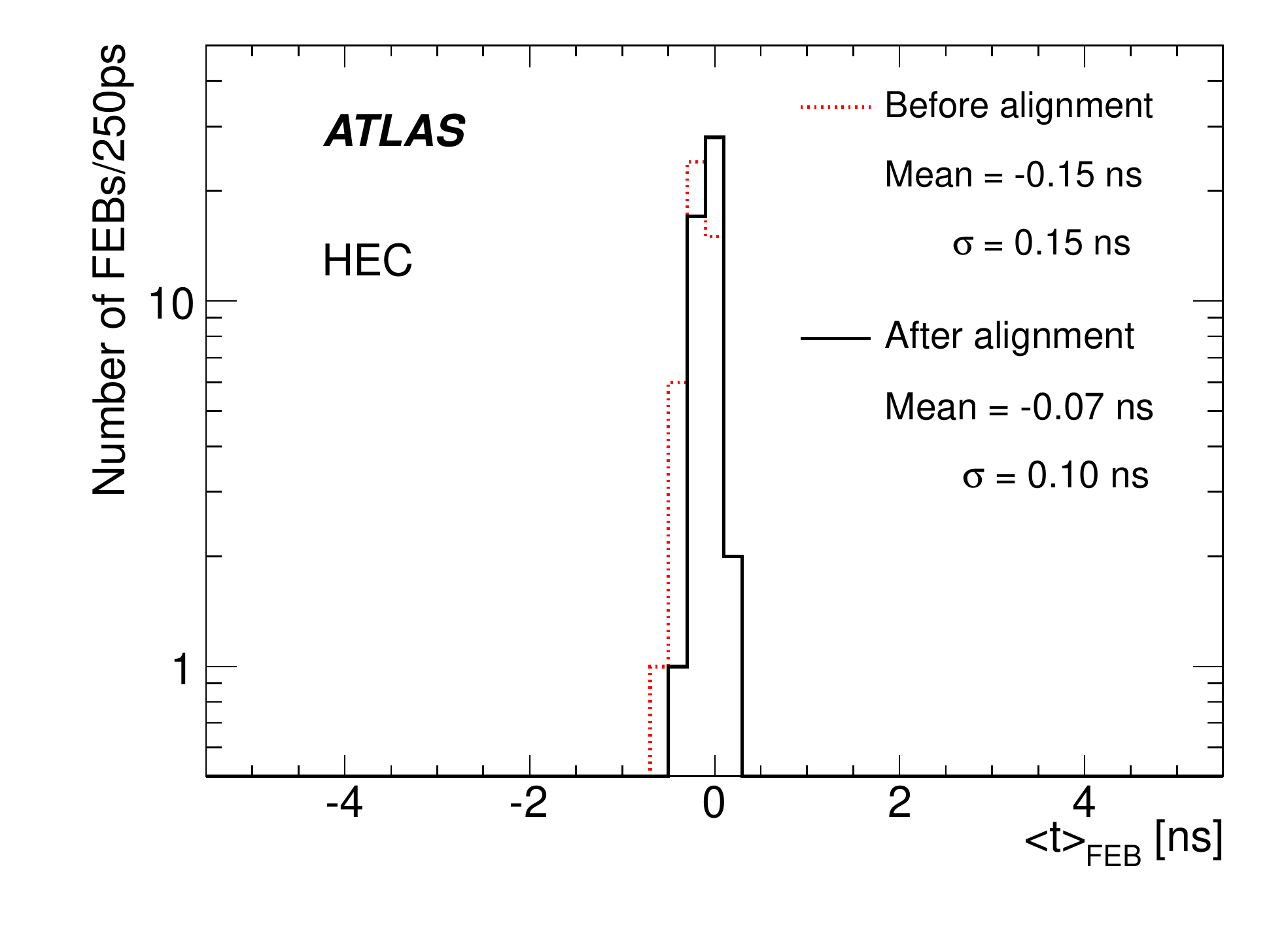}}
    \subfloat[]{\label{fig:febtime_fcal}\includegraphics[scale=0.39]{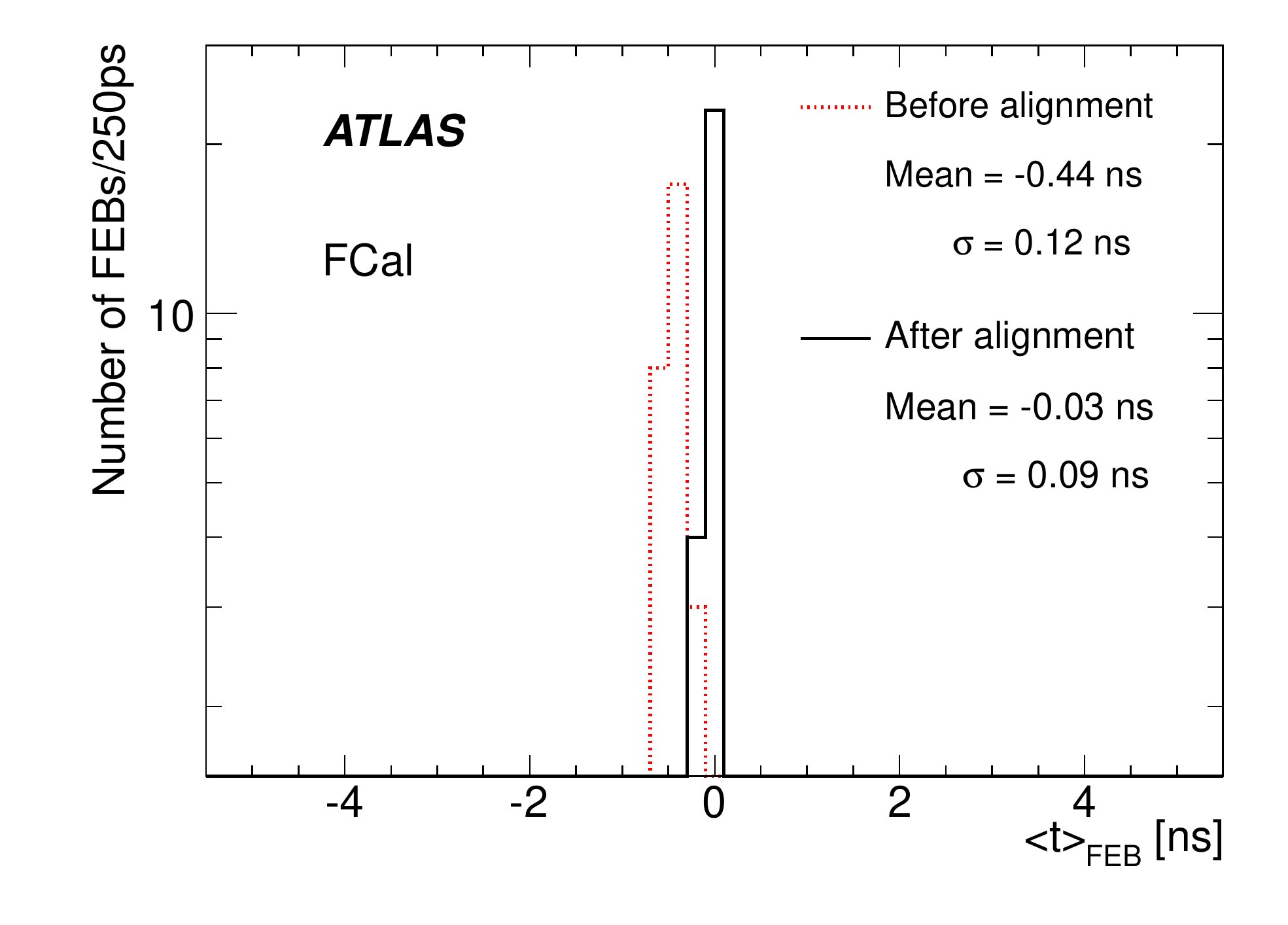}}    
\caption{Distributions of the average FEB times in the four subdetectors: (a) EMB, (b) EMEC, (c) HEC, and (d) FCal. Distributions before and after spring 2012 timing alignment are superimposed, as described in the legend.}
\label{fig:febtime}
\end{center}
\end{figure}

The procedure detailed in section~\ref{subs:globalTiming} is mainly meant to monitor online the global synchronization of the LAr calorimeter and its evolution throughout the {\LBs} of a run. A refined analysis is also performed offline to monitor the time synchronization of each individual FEB and optimize the phase of the clock delivered to each FEB (adjustable in steps of 104~ps via hardware settings \cite{FEBDesign}). With loose trigger thresholds, the LArCells stream allows collecting enough signals to monitor the individual FEB synchronization in every single run with at least 100~pb$^{-1}$. After rejecting the events affected by a noise burst (see section \ref{sect:noiseBursts}) and masking all the channels flagged as problematic (see section~\ref{sect7_noisyCells}), all cells above a certain energy threshold are selected. The energy thresholds vary between 1~GeV and 3.5~GeV (10~GeV in FCal) depending on the layer/partition and were optimized to lie well above the electronics noise without reducing the sample size too much. An energy-weighted distribution of the time of all cells of each FEB is built. The average time of each FEB is then derived from a two-step iterative procedure using a Gaussian fit of the distribution. In the rare cases of too few events or non-convergence of the fitting procedure, the median value of the distribution is used instead.

The average times of the 1524 FEBs were very accurately measured with the first 1.6~$\rm{fb}^{-1}$ of data accumulated in 2012 (period A and first runs of period B). The results are presented in figure~\ref{fig:febtime}: dispersions up to 240~ps were observed with some outliers. At this time, the clock delivery to each FEB was tuned individually, making use of the 104~ps adjustment facility provided by the timing system. The improvement associated with this alignment procedure is superimposed in figure~\ref{fig:febtime}. The dispersions, originally in the range 120--240~ps, were significantly reduced in each subdetector, and no outlier in the FEB average time distribution was observed above 1.5~ns.
\begin{figure}[!b]
\begin{center}
   \includegraphics[scale=0.48]{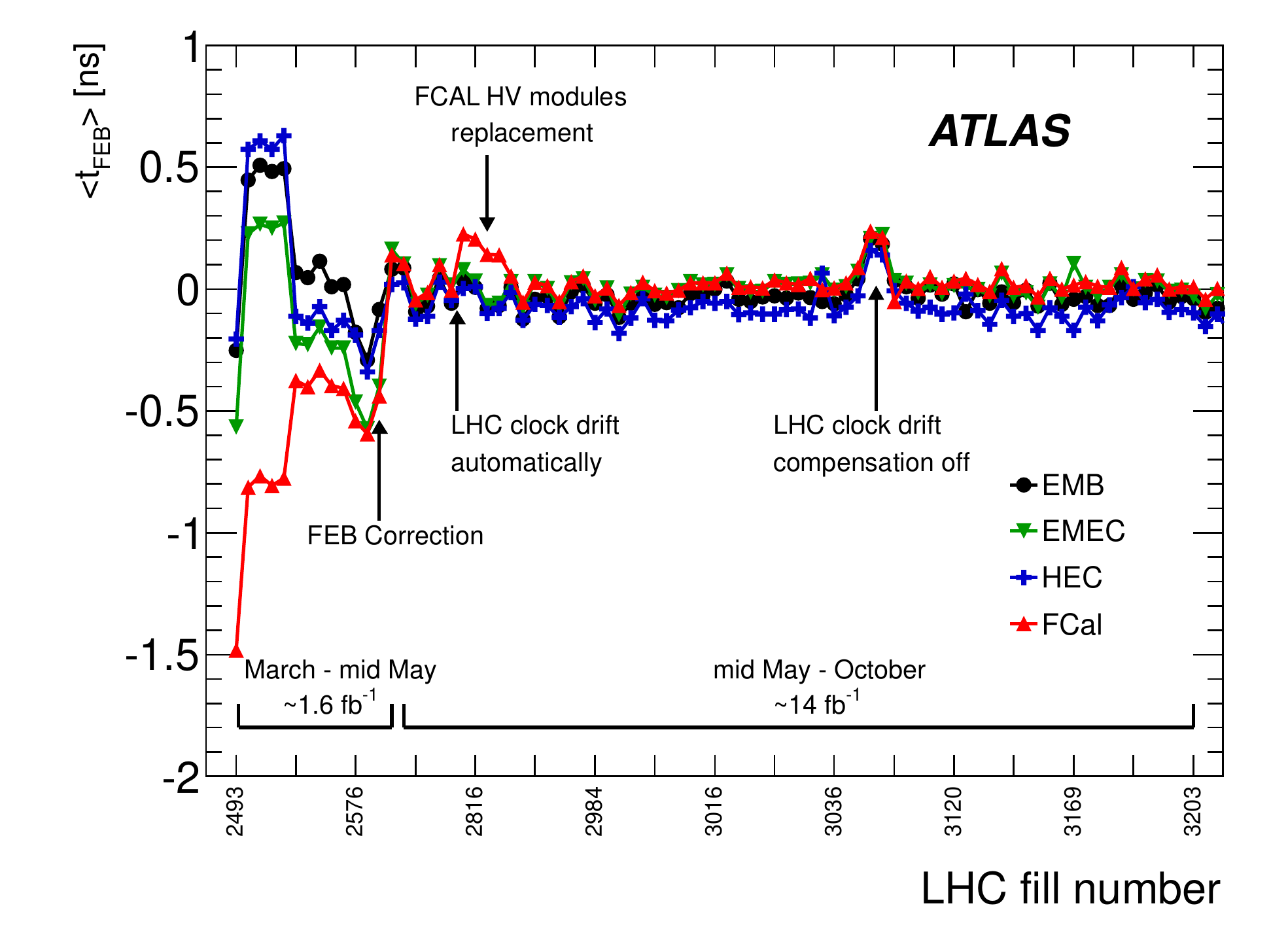} 
\caption{Average FEB time per subdetector as a function of the LHC fill number, during 2012 data taking. To improve readability, the period from October through December 2012, showing similar stability, is omitted.}
\label{fig:febevol}
\end{center}
\end{figure}

With the large data sample accumulated during 2012, it was possible to routinely monitor the FEB synchronization during the year. An automated processing framework was set up on the CAF computing farm~\cite{cernCAF} to provide fast feedback to the {\LADIeS}. The evolution throughout 2012 of the average FEB time per subdetector is shown in figure~\ref{fig:febevol}. The effect of the first 2012 timing alignment previously mentioned is clearly visible at the beginning of the year. Shortly after this alignment, a system that automatically adjusts the ATLAS central clock to align with the LHC clock was commissioned. Originally tuned by hand, this adjustement compensates for the length variation of the optical fibres delivering the LHC clock to the ATLAS experiment due to temperature changes. With the level of synchronization achieved after the FEB synchronization, this automatic procedure became crucial. An illustration of this importance is given by the 200~ps bump observed in summer, when the automated compensation procedure was accidentally switched off (around LHC fill number 3050). Finally, another feature observed during summer 2012 was a $\sim$300~ps time shift in the FCal FEBs around LHC fill number 2816. The origin of this problem was identified as the installation of two faulty HV modules that delivered a voltage lower than expected, hence impacting the electron drift time. As soon as the cause was identified, the faulty modules were replaced to recover the optimal synchronization. Beside this synchronization problem, these faulty modules also impacted the energy response. However, an offline correction was applied to recover an appropriate calibration. Except for these two minor incidents, which had negligible impact on data quality, figure~\ref{fig:febevol} shows impressive global stability within 100~ps during the 2012 data taking. 

A more refined synchronization at the cell level was implemented during a data reprocessing campaign. This should allow further improvement of the calorimeter timing accuracy that was measured in 2011 to around 190~ps for electrons and photons~\cite{timing190ps}.

\section{Treatment of large-scale coherent noise} \label{sect:noiseBursts}

When the instantaneous luminosity reaches $10^{32}~\rm{cm^{-2}s^{-1}}$ and above, the LAr calorimeter is affected by large bursts of coherent noise, mainly located in the endcaps. Since the occurrence rate increases with instantaneous luminosity, a specific treatment had to be developed in summer 2011 to limit the data loss.

\subsection{Description of the pathology}

Between its installation inside the cavern in 2005 and the first collisions in 2009, the LAr calorimeter was extensively commissioned, and many detailed performance studies were pursued, with a special emphasis on the Gaussian coherent noise of the front-end boards. This Gaussian coherent noise was measured to be at a level lower than 10\% of the total electronics noise per channel \cite{electronicPerf}. 

On a larger detector scale, the coherent noise can be estimated by considering the variable {\ythrees} for each partition, defined as the fraction of channels with a signal greater than three times the Gaussian electronics noise.\footnote{The electronics noise is measured in calibration runs, using simple clock-generated trigger.} Assuming a perfect, uncorrelated Gaussian noise behaviour in the entire calorimeter, the {\ythrees} variable is expected to peak around 0.13\%. In the early days of commissioning, the {\ythrees} variable exhibited sizeable tails above 1\% in randomly triggered events, characteristic of large coherent noise. Its source was identified as a major weakness of the high-voltage filter box supplying the presampler, which was fixed in 2007. After the fix, minor tails were still observed in the {\ythrees} variable distribution, but only within calorimeter self-triggered events (i.e. events triggered by a large signal in the LAr calorimeter). 

Further studies were carried out before closing the detector in 2009, which led to the conclusion that the remaining coherent noise was likely to be introduced again inside the detector via the HV system: when all the HV power supplies were turned off, no noise was observed. Although some areas of the detector were obviously more affected than others, switching off only the specific HV lines powering the noisiest regions did not cure the problem. This indicated that the noise was most likely radiated by unshielded HV cables inside the cryostat, rather than directly injected. Imperfections or peculiarities of the cable routing inside the cryostat may explain why some regions are more affected than others, but given the limited range of the problem and the difficult access to the hardware components, no further action was taken at that time.

\begin{figure}[!htb]
  \center
  \includegraphics[width=0.55\textwidth]{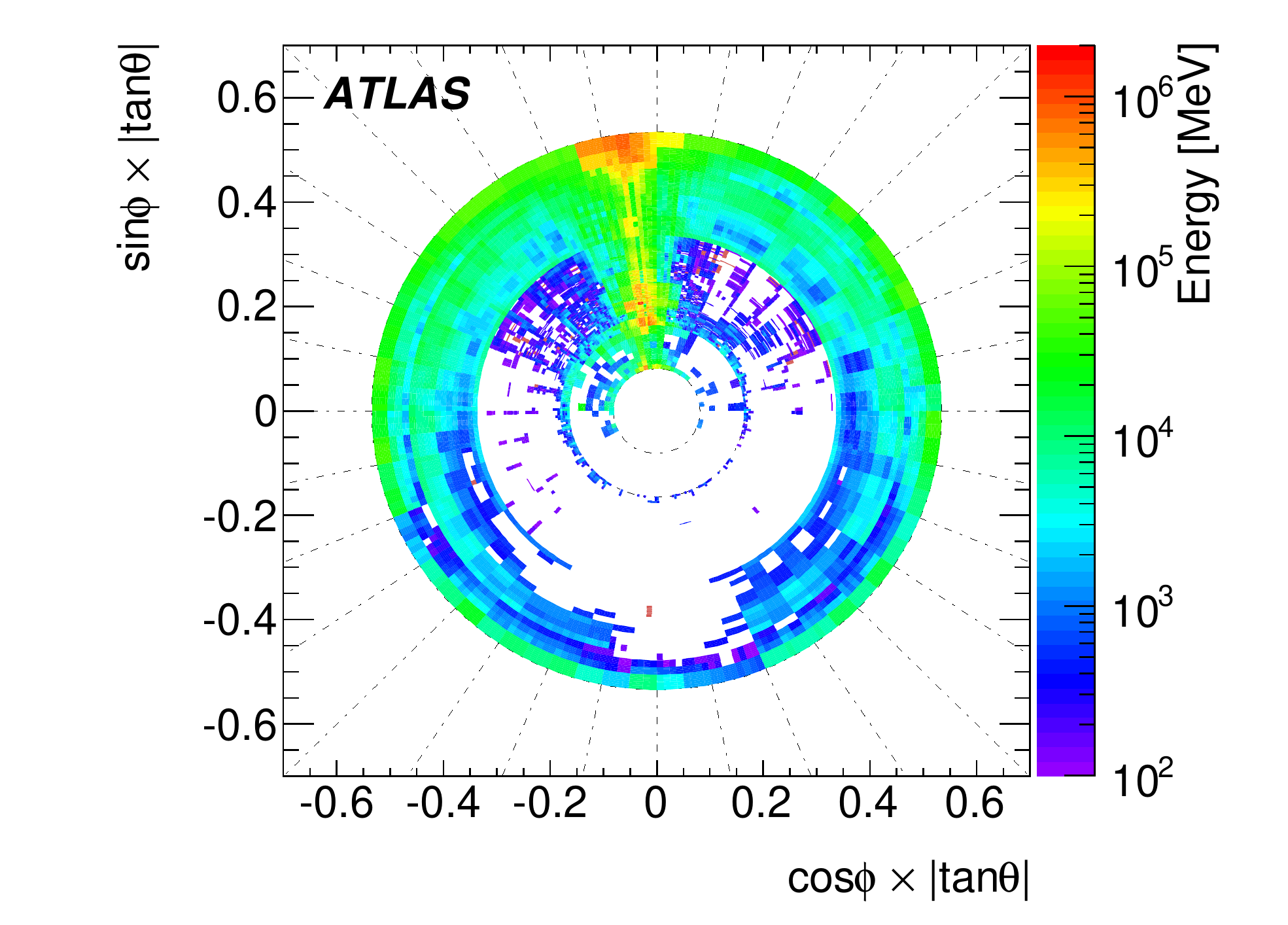}
  \caption{Example of a typical coherent noise event observed in the EMECA partition in run 205071 -- {\streamCosmic}. The energies of the 3--4 layers were summed along a fixed value of pseudorapidity and partially along the azimuthal angle.}
  \label{fig:noiseBurstEvent}
\end{figure}

During autumn 2010, the instantaneous luminosity reached $10^{32}~\rm{cm^{-2}s^{-1}}$. At this time, pathological events with a very large signal (equivalent to several TeV) affecting a whole partition were observed in the empty bunches ({\streamCosmic}), when the LHC was in collision mode. The electromagnetic endcaps were especially affected. In the worst cases, some noise could be also observed in the hadronic endcap and the forward calorimeter at the same time as in the electromagnetic endcap. Figure \ref{fig:noiseBurstEvent} shows a typical event in the transverse plane of the electromagnetic endcap (A side) recorded at an instantaneous luminosity of $6\times 10^{33}~\rm{cm^{-2}s^{-1}}$: the total energy peaks around 2~TeV, and the {\ythrees} variable reaches 25\%. Although the topologies and occurrence rates differ slightly, both endcaps are affected. They are treated in the same way and merged into the same distributions in all the following studies. The barrel distributions are also merged.

\begin{figure}[!htb]
  \subfloat[]{\label{fig:y3sigma}\includegraphics[width=0.48\textwidth]{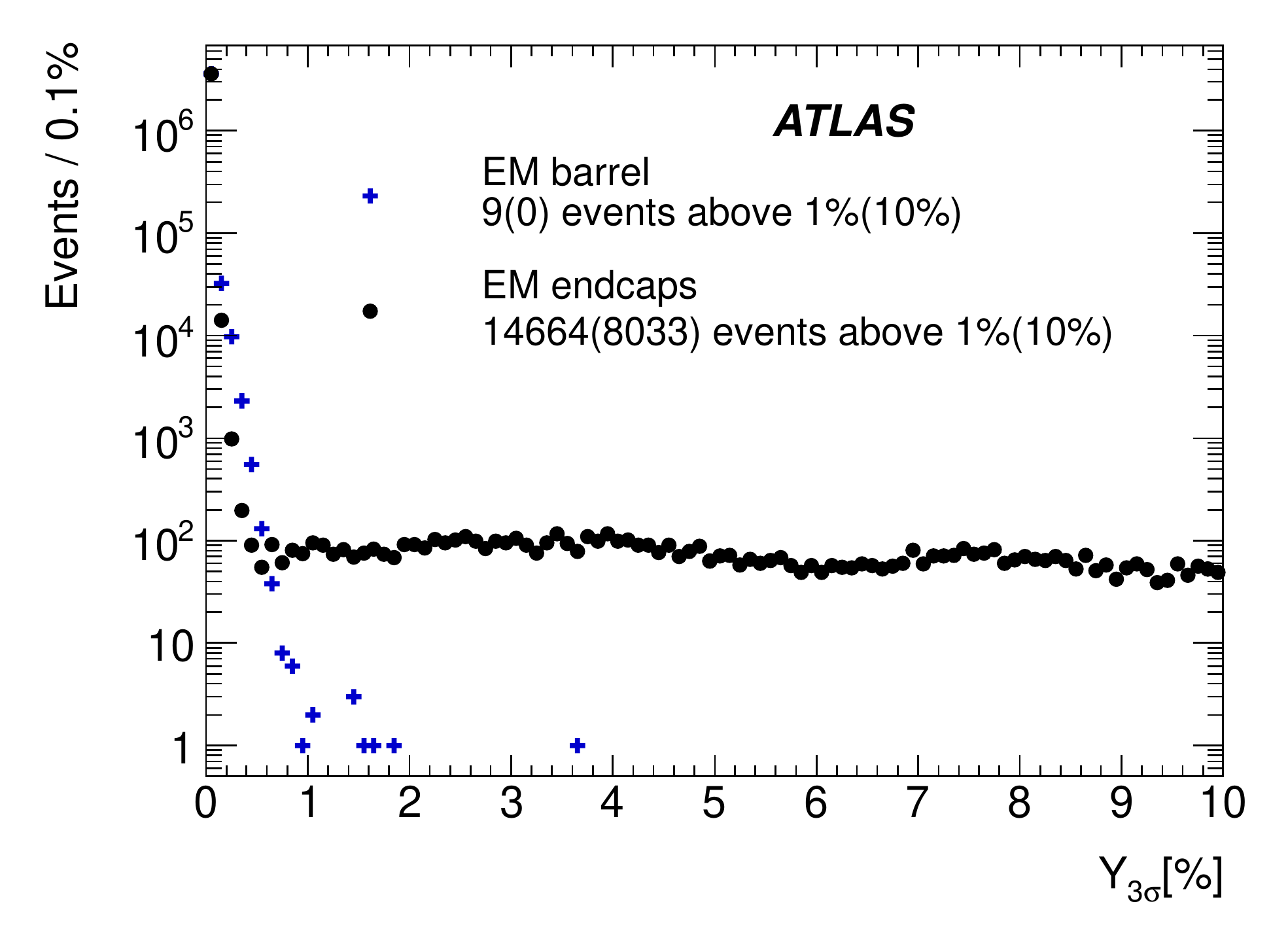}}
  \subfloat[]{\label{fig:y3sigma-Std}\includegraphics[width=0.48\textwidth]{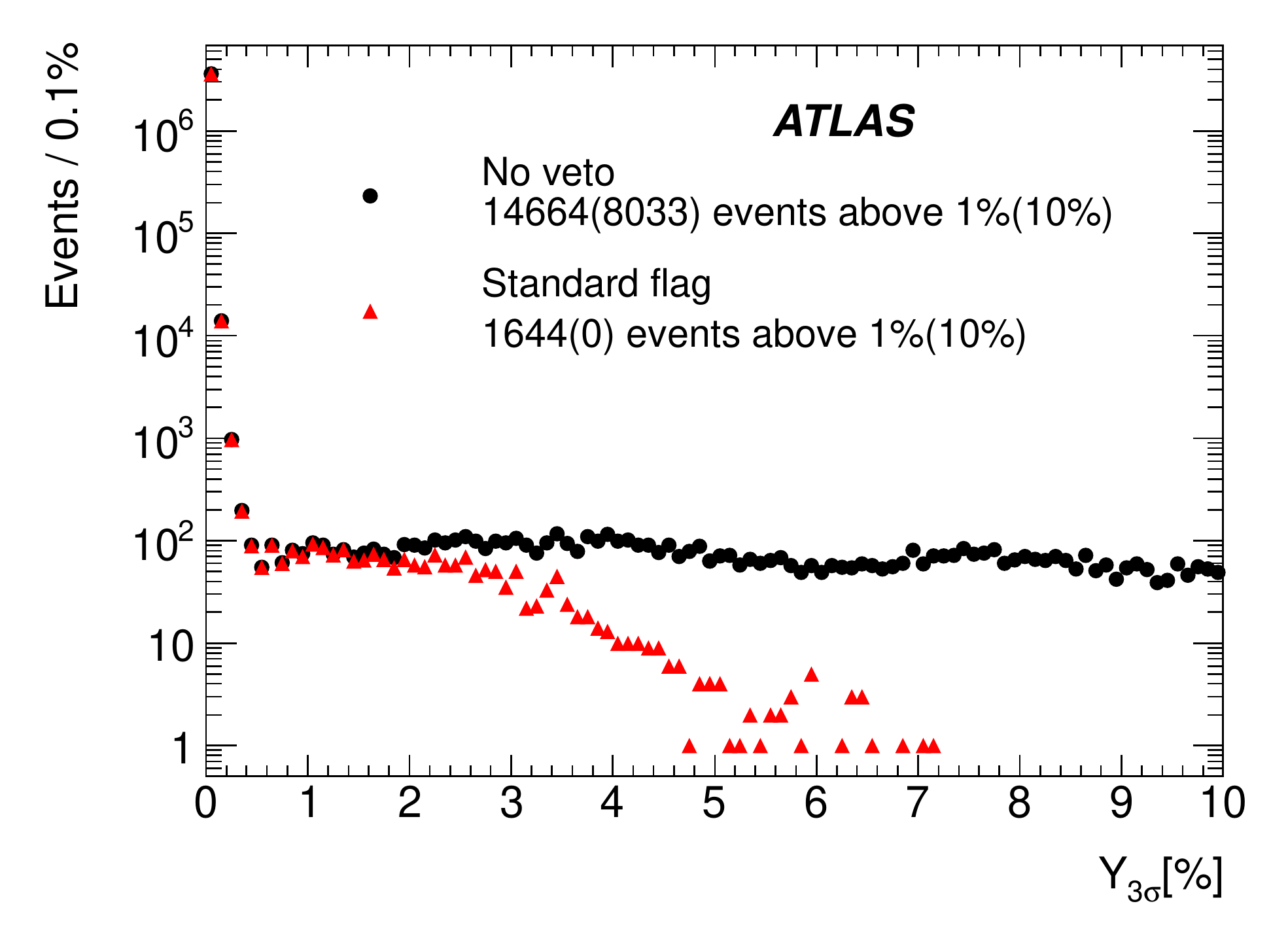}}
  \caption{(a) {\ythrees} distributions for the four electromagnetic partitions (positive and negative pseudorapidities merged in a single distribution). (b) Effect of the {\itshape Standard} flag veto on the electromagnetic endcap {\ythrees} distribution. The event sample was acquired in the {\streamCosmic} during 135 hours corresponding to an integrated luminosity of 1.7~fb$^{-1}$.}
  \label{fig:y3sigma_both}
\end{figure}

Figure~\ref{fig:y3sigma_both}\subref{fig:y3sigma} shows the {\ythrees} distribution, computed for the barrel and endcap partitions over a period of roughly 135 hours of data taking; during this period, 1.7~$\rm{fb^{-1}}$ of data were accumulated, with an instantaneous luminosity greater than $3\times10^{33}~\rm{cm^{-2}s^{-1}}$. The distribution appears as expected in the barrel, with a sharp peak around 0.13\% and negligible tails. But in the endcaps, the distribution exhibits very large tails, typical of coherent noise, with a very large fraction (up to 70\% -- not visible on this figure) of channels fluctuating coherently within a partition. 

The noise burst topology shown in figure~\ref{fig:noiseBurstEvent} is very similar to the one observed during the commissioning phase, but its amplitude is significantly larger. The very similar topologies at different times excluded the possibility that this pathology could be due to beam background or parasitic collisions. The HV lines were again suspected, the increased rates and amplitudes being explained by the larger drawn currents. This hypothesis was favoured because the endcaps are the most involved and their behaviour is almost Gaussian outside the LHC collision mode.

\subsection{Use of the quality factor for noise identification} \label{subs:noiseBurstFlag}

In collision streams, the {\ythrees} variable is positively biased by the presence of energy deposits in the calorimeter due to collisions (typically peaking around 1--2\% at high luminosity) and cannot be used to identify coherent noise. It is therefore crucial to define alternative ways to study this coherent noise in the presence of collisions. New Boolean variables, hereafter named {\itshape flags}, had to be introduced.
\begin{itemize}[topsep=3pt,itemsep=2pt,parsep=3pt]
\item The {\itshape \Std} flag requires strictly more than five FEBs containing more than 30 channels each with a quality factor greater than 4000.
\item The {\itshape \Sat} flag requires more than 20 channels with an energy greater than 1~GeV and a saturated quality factor (i.e. equal to 65535). 
\end{itemize}

The flag definitions are based on the observation of poor quality factors in the noisy events, indicative of abnormal pulse shapes and very unlikely to be due to argon ionization. The {\Std} flag is sensitive to phenomena largely spread over a partition. The {\Sat} flag, with a much higher constraint on the quality factor, is triggered in very atypical phenomena but possibly confined to a very reduced area. With this criterion, limited in terms of geometrical extent, the {\Sat} flag is less reliable than the {\Std} flag. However, it is useful for particular cases, described in the following. Figure~\ref{fig:y3sigma_both}\subref{fig:y3sigma-Std} illustrates the {\Std} flag efficiency in reducing the tails of the {\ythrees} endcaps distribution. When vetoing on this flag, only 11\% of events with {\ythrees} above 1\% remain and no event remains with {\ythrees} above 10\%.

\subsection{Time duration of the pathology}

To measure the time extent of the coherent noise, events with {\ythrees} greater than 1\% and separated by less than one second are clustered, assuming that they belong to the same burst of noise. By this method, the time extent (defined as the difference between the first and last clustered events), was measured to be around a few hundreds nanoseconds. However, this method is limited, since it relies on empty bunches: the {\emptyBG} is composed of a group of BCIDs of length 600--1000~ns between two trains of populated bunches of approximately 3.6--7.2~$\rm{\mu s}$ (see section~\ref{offlinePolicy}). This method is therefore potentially biased by the {\emptyBG}'s timelength being comparable to the measured time extent.

\begin{figure}[!htb]
    \subfloat[]{\label{fig:duration1}\includegraphics[width=0.48\textwidth]{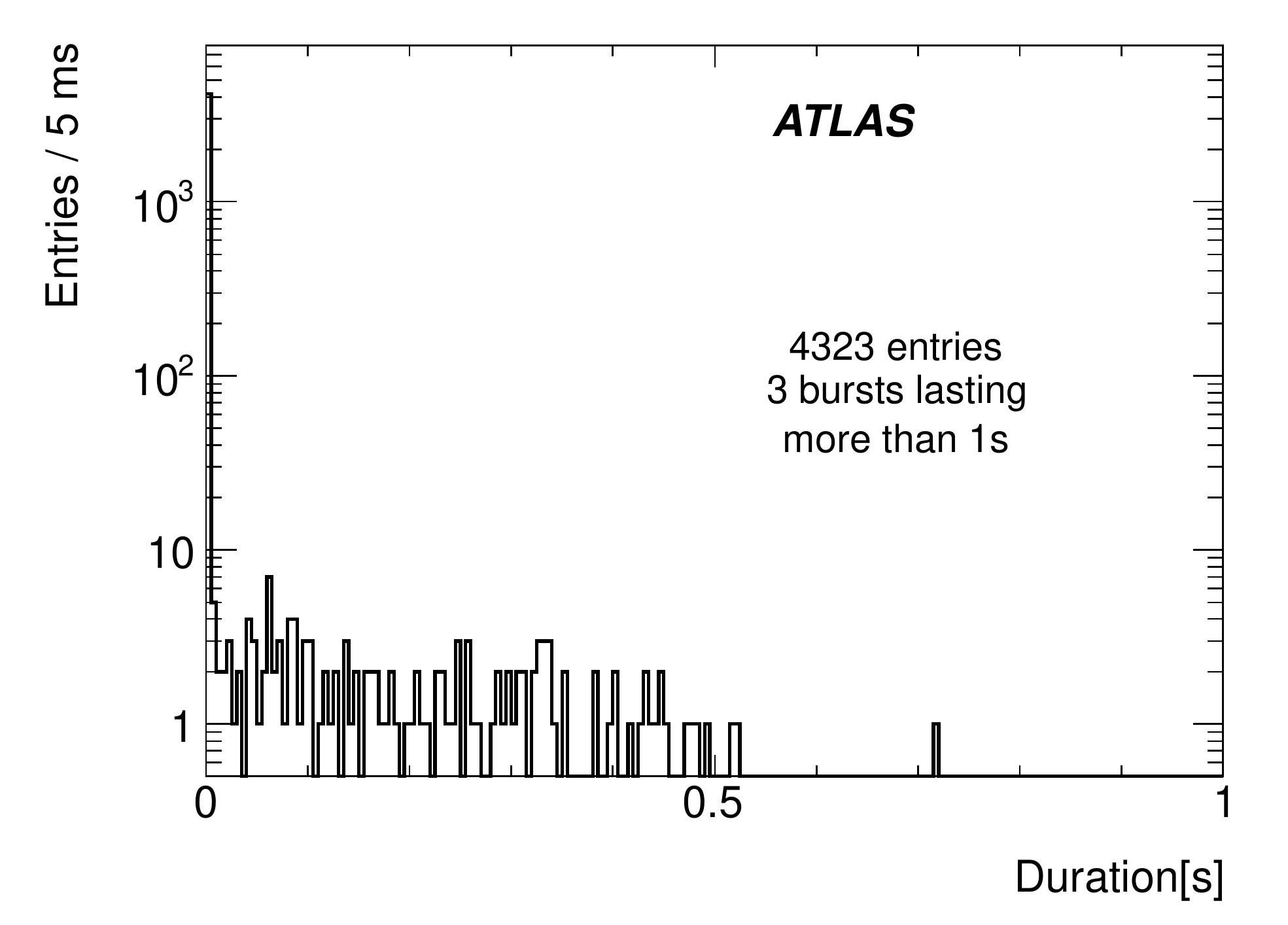}}
    \subfloat[]{\label{fig:duration2}\includegraphics[width=0.48\textwidth]{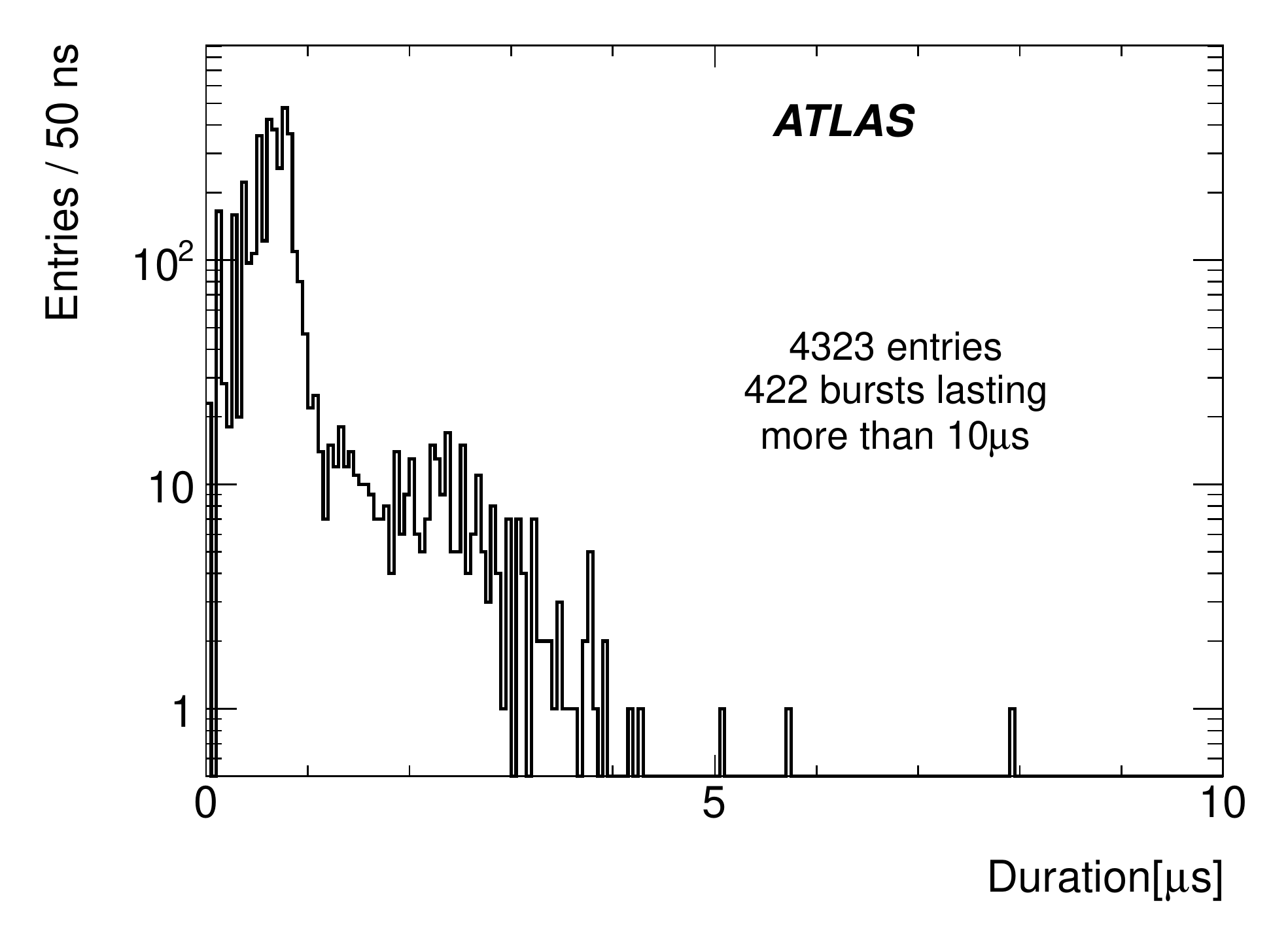}}
   \caption{Distributions of the noise burst duration. Figure (b) zooms in on the shortest times of figure~(a).}
   \label{fig:duration}
\end{figure}

To overcome this limitation, the same event clustering method can be applied by replacing the criterion for the {\ythrees} variable by a criterion for the {\Std} flag. To be conservative, events flagged by the {\Sat} method are also clustered with events flagged by the {\Std} method if they are separated by less than one second. Since the {\Sat} method was found to be less reliable, requesting the event to be close to an event flagged by the {\Std} flag limits the risk of considering fake noisy events. With this clustering definition independent of the {\ythrees} variable, it is possible to consider both the CosmicCalo and {\streamExp}s, and hence empty and filled bunches. The result is shown in figure \ref{fig:duration}. Virtually all pathologies are found to be shorter than 0.5~s (see figure~\ref{fig:duration}\subref{fig:duration1}), and more than 90\% of them are shorter than 5~$\rm{\mu}$s (see figure~\ref{fig:duration}\subref{fig:duration2}). Due to the short duration of the phenomenon, the pathologies are referred to as {\itshape noise bursts}. The very limited duration of the bursts, much shorter than the {\LB} length, also suggested the development of a dedicated offline treatment with a time-window veto procedure to limit the amount of data rejected.

\subsection{Time-window veto procedure}

The scanning of a sample of noise bursts showed that most of them consist of a peak of {\itshape hard} events surrounded (before and after) by peripheral {\itshape soft} events: the hard events are characterized by a large {\ythrees} and are properly identified by the {\Std} flag, whereas the soft ones are characterized by a {\ythrees} variable around 2-3\% (if recorded in empty bunches) and are not identified by the {\Std} flag. It was therefore proposed to apply a time-window veto procedure around the well-identified hard events to remove the soft ones. 

Technically, the noise burst cleaning is achieved by storing a status word in the event header as explained in section~\ref{dqLogging}. This requires the extraction of the noise burst's peak timestamp with the express processing of the CosmicCalo and {\streamExp}s: a clustering procedure is performed on the same events as detailed in the previous section. The timestamps of the first and last flagged events are used to define a unique time interval. To veto the peripheral events of the noise burst, the time interval is extended by $\pm\delta t/2$, where $\delta t$ is a parameter to be optimized. The computed time window is then stored in a dedicated conditions database during the calibration loop, and read back for the bulk reconstruction to fill the status word of all events falling inside the time-window veto.

Finally, it is important to emphasize that a noise burst candidate with a single flagged event in the peak is not vetoed. This is done deliberately, to avoid discarding unusual events where the decays of exotic particles deposit energy in the calorimeter at much later time than the bunch crossing (a delayed signal is very likely to have a poor quality factor). By requesting at least two events flagged within a short time, the risk of throwing away unexpected new physics events is considered negligible.

\begin{figure}[!htb]
  \center
  \includegraphics[width=0.48\textwidth]{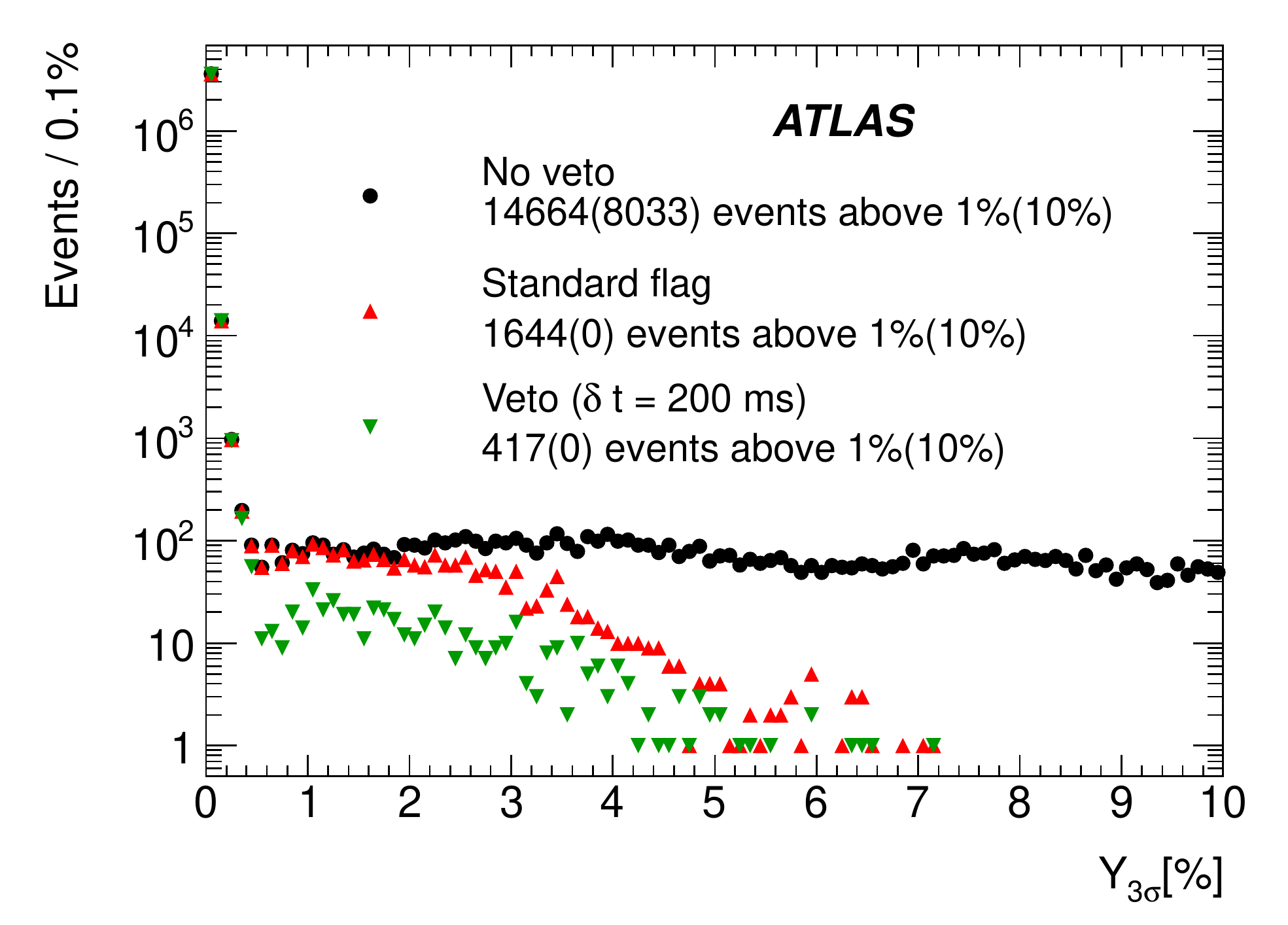}
  \caption{Effect of the time-window veto on the electromagnetic endcap {\ythrees} distribution for a value of $\delta t$ = 200~ms. Same dataset as in figure 14.}
  \label{fig:y3sigmaCleaned}
\end{figure}

The improvement to the {\ythrees} distribution resulting from applying the time-window veto is shown in figure~\ref{fig:y3sigmaCleaned}. The quantitative performance of the procedure is also summarized in table~\ref{tab:cleanEff}. In the two most sensitive partitions (the two electromagnetic endcaps), the time-window veto procedure reduces by a factor of four the number of events with a {\ythrees} greater than 1\% remaining after having applied the {\Std} flagging method or by a factor of 35 when comparing with the uncleaned data sample.
Several values of $\delta t$ were tried, between 100~ms and 2~s, leading to the same efficiency as the one quoted in the table for a value of 200~ms. Compared to the measured time extent of the noise bursts, these numbers are very conservative, as confirmed by the stable efficiency. There is probably some room left for tuning this parameter, but given the very low associated data loss (see section~\ref{sect:nbAssociated}), a conservative value of 200~ms (1~s) was applied in the 2012 (2011) data processing.

\begin{table}[!hbt]\label{tab:nbsum}
\caption{Number of events with a {\ythrees} greater than 1\% after applying the simple {\Std} flagging and the time-window veto procedure. The efficiencies $\epsilon$  of each cleaning procedure are given in parentheses. Same dataset as in figure 14.}
\begin{center}
\begin{tabular}{|p{2.1cm}|p{2.1cm}|p{4.2cm}|p{5cm}|}
  \hline
Partitions  & No cleaning procedure & After applying the {\Std} flag method & After applying the time-window veto procedure ($\delta t$ = 200~ms)  \\ 
  \hline
EM barrel  & 9 & 3 ($\epsilon$ = 66.6\%)  & 1 ($\epsilon$ = 88.9\%) \\
EM endcaps & 14664 & 1644 ($\epsilon$ = 88.8\%)  & 417 ($\epsilon$ = 97.2\%) \\
  \hline
\end{tabular}
\end{center}
\label{tab:cleanEff}
\end{table} 

The number of affected events per hour (\ythrees$>$1\%) was originally around 108. With the time-window veto method, it decreased to only three events per hour. 

\subsection{Luminosity dependence}\label{subs:noiseBurstLumi}
\begin{figure}[!htb]
    \subfloat[]{\label{fig:occurenceVsLumi}\includegraphics[width=0.48\textwidth]{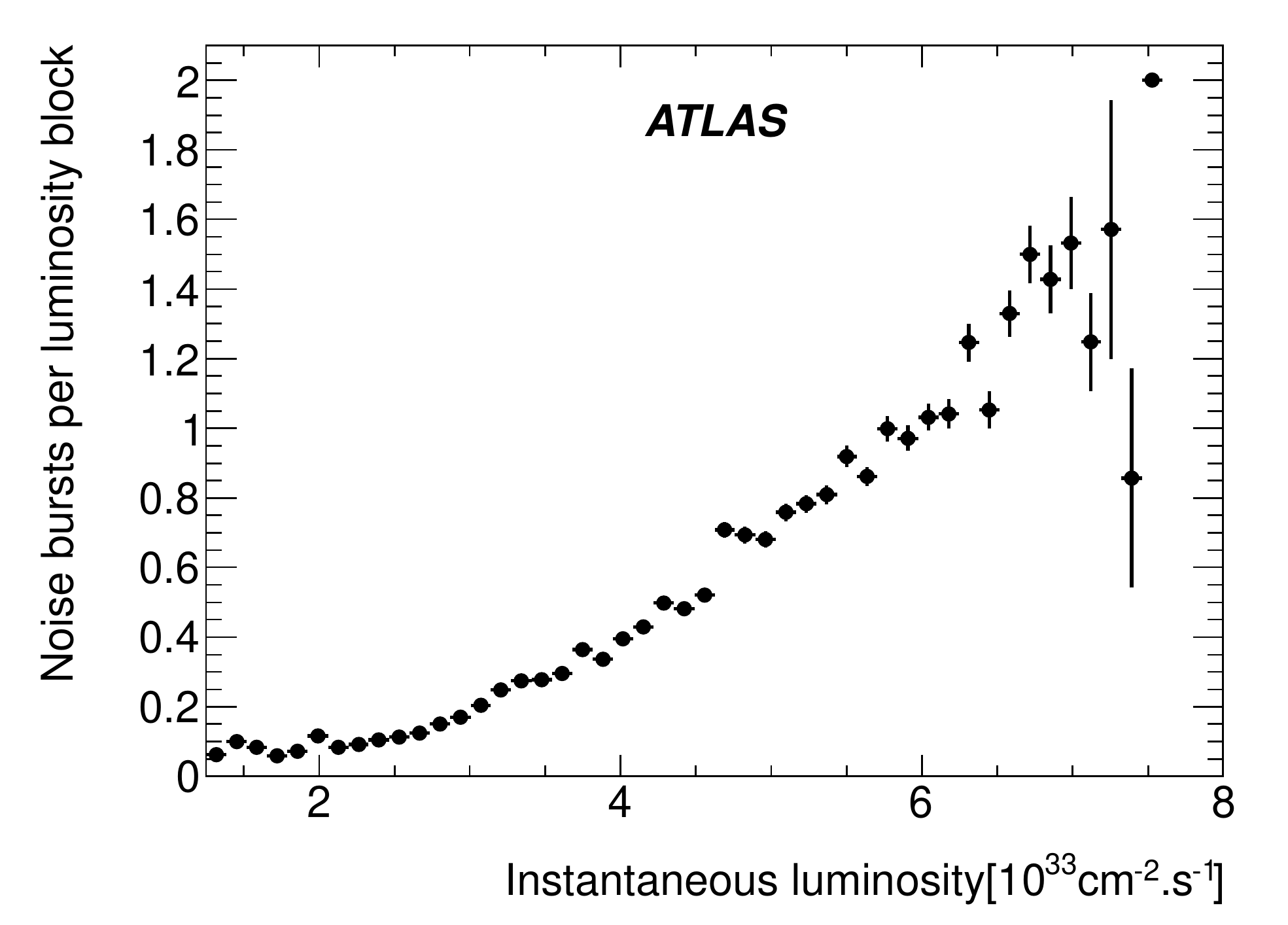}} 
    \subfloat[]{\label{fig:durationVsLumi}\includegraphics[width=0.48\textwidth]{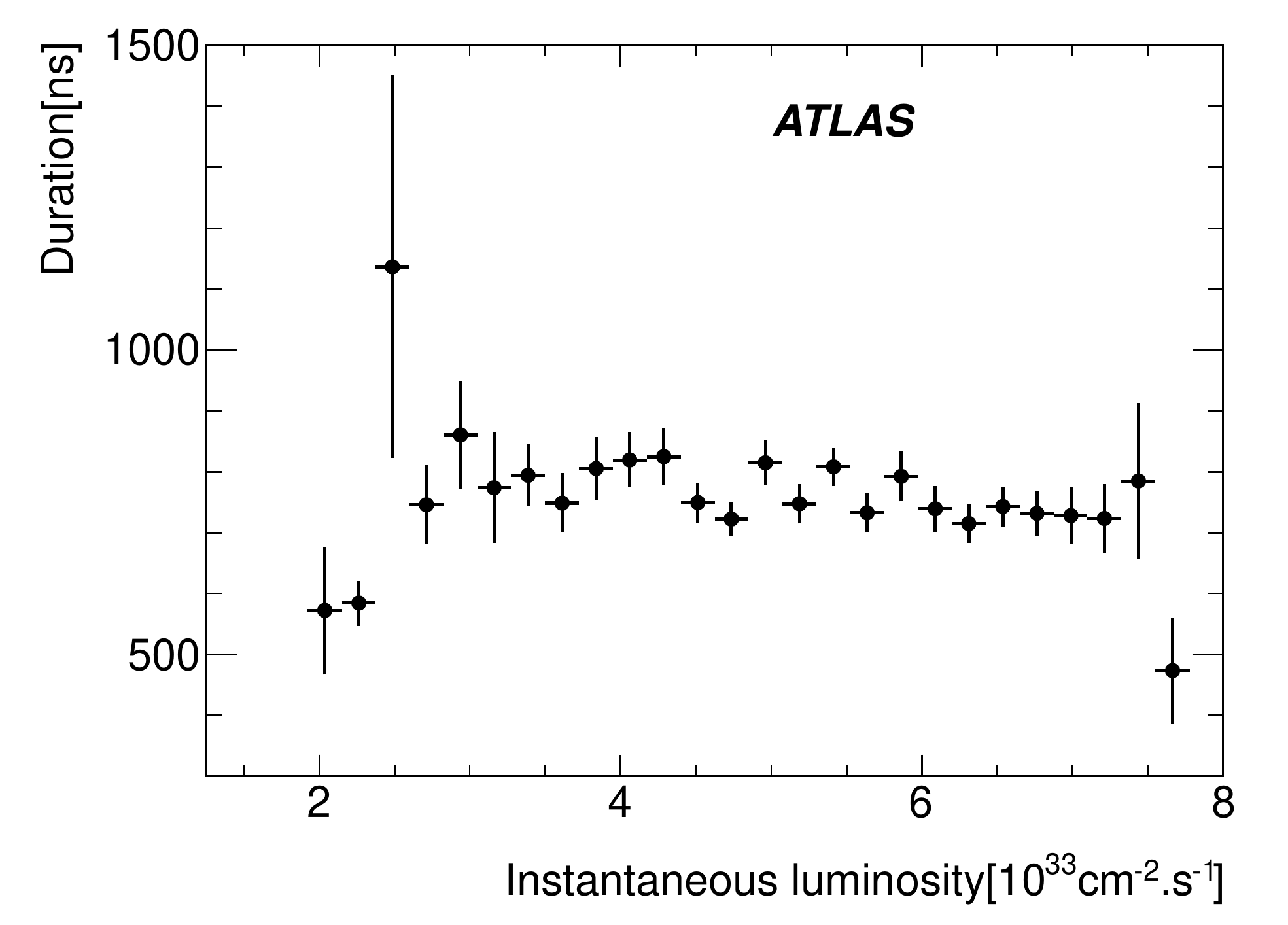}}
   \caption{(a) Noise burst occurrence frequency per {\LB} and (b) luminosity dependence of the mean duration of the noise bursts as a function of instantaneous luminosity.}
   \label{fig:lumiDep}
\end{figure}
In 2012, around 15\% (40\%) of the {\LBs} contained a noise burst in the CosmicCalo (JetTauEtmiss) streams. Figure~\ref{fig:lumiDep}\subref{fig:occurenceVsLumi} illustrates the number of noise bursts per {\LB} as a function of the instantaneous luminosity (in any stream). A steady dependence is observed. Parabolic extrapolations from this plot indicate that each {\LB} will contain around five noise bursts at the peak luminosity expected after 2015 ($1$--$2\times 10^{34}~\rm{cm^{-2}s^{-1}}$). However, even if the rate evolves as a function of the instantaneous luminosity, the noise bursts' mean duration remains stable, as shown in~figure~\ref{fig:lumiDep}\subref{fig:durationVsLumi}. As the current choice of the $\delta t$ parameter is very conservative with respect to the noise burst time extent, its reduction can be envisaged to fully compensate for the future increased occurrence yield.

\subsection{Associated data rejection in 2012}
\label{sect:nbAssociated}

\begin{figure}[!b]
    \subfloat[]{\label{fig:2012SevNoiseBurst_veto}\includegraphics[width=0.48\textwidth]{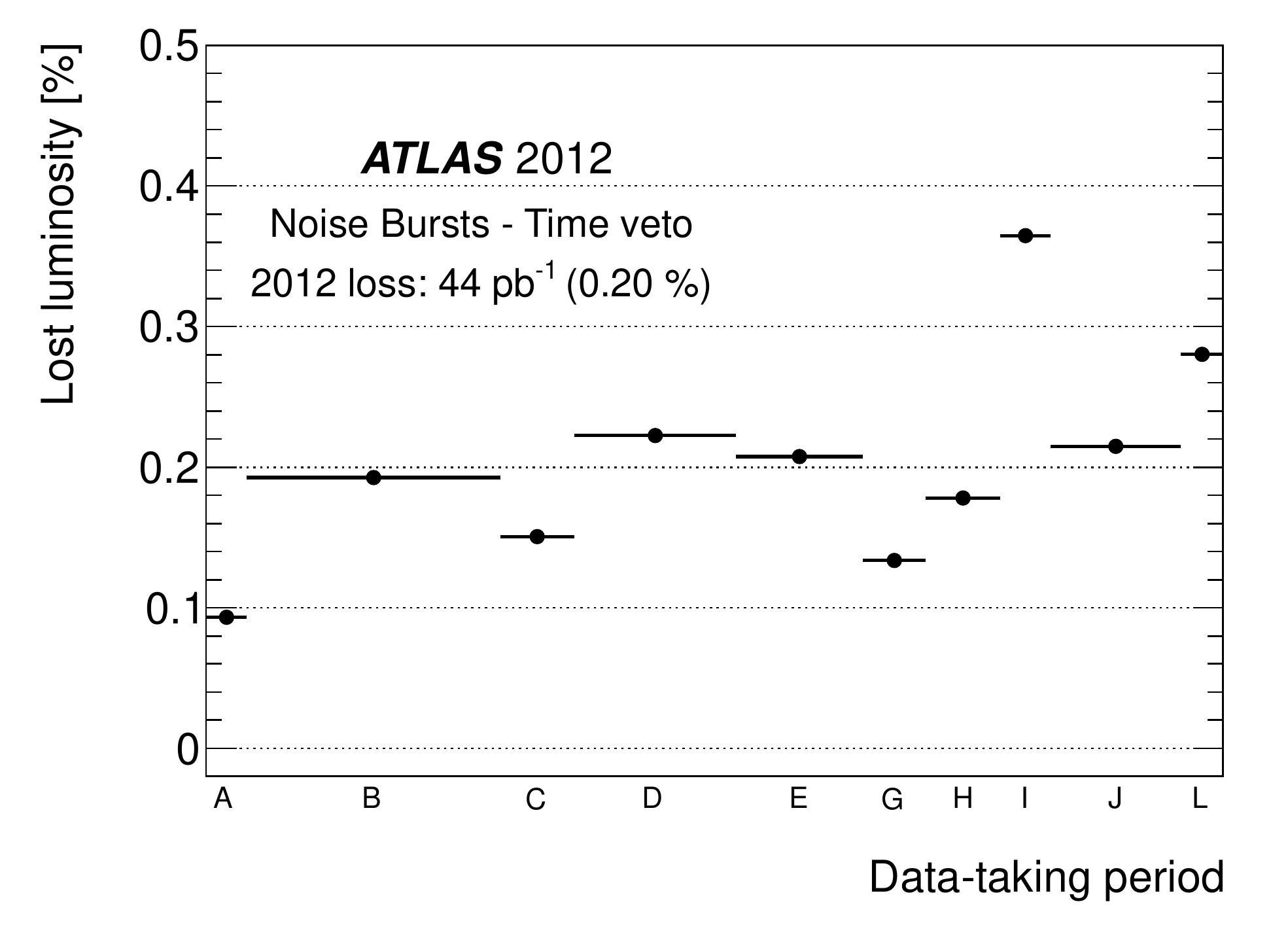}}
    \subfloat[]{\label{fig:2012SEVNOISEBURST}\includegraphics[width=0.48\textwidth]{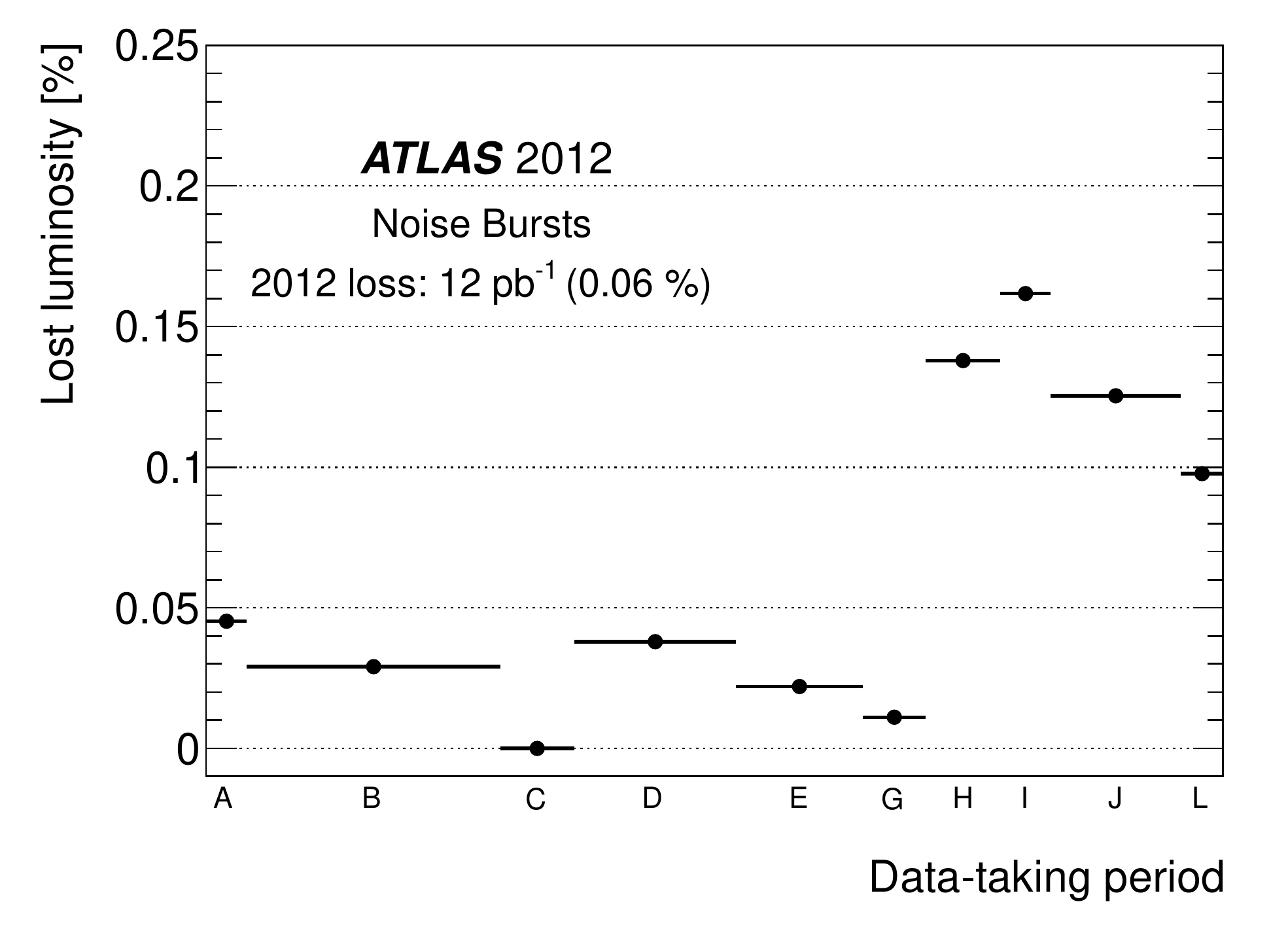}}
   \caption{Lost luminosity due to noise bursts as a function of the data-taking period in 2012. (a) Loss due to the time-window veto procedure. (b) Loss due to defect assignment.}
   \label{fig:2012NoiseBurst}
\end{figure}

The data loss associated with the time-window veto procedure  as a function of the data-taking period is presented in figure~\ref{fig:2012NoiseBurst}\subref{fig:2012SevNoiseBurst_veto}. It amounts to 0.2\%. The observed variation is explained by the differences in the instantaneous luminosity profiles impacting the noise burst rates, as explained in the previous section.

The efficiency of the time-window veto procedure is cross-checked in the {\streamJet}, by searching for remaining events flagged as noise bursts by the {\Std} method outside the defined time veto periods. Their treatment depends on whether such events are isolated in time or close to another one.
\begin{itemize}[topsep=3pt,itemsep=2pt,parsep=3pt]
\item If such an event is isolated in time, no action is taken, as it might be due to a delayed decay of exotic particles. In 2012, only 192 such events remain in the dataset considered for physics analysis. Furthermore, a complementary cleaning at the jet level is also available offline, to make sure that any remaining noise bursts do not bias physics analysis \cite{BeamBackground2011}.
\item If two or more events close in time remain, they are very likely to belong to a single noise burst not observed in the express processing streams. The only solution is to reject them by assigning an intolerable defect to the whole {\LB}. This induces a much larger data loss, not recoverable until the next full data reprocessing where the time-window veto procedure can be applied again using updated database information.
\end{itemize}

Consequently, the efficiency of the time-window veto procedure heavily relies on the ability to select the noise burst peak events in the Express and {\streamCosmic}s in order to compute the veto interval periods before the start of the bulk processing. To achieve this, four dedicated trigger chains were designed to ensure efficient streaming. The trigger chains are seeded at the first-level trigger step from standard jet or $E_{\rm{T}}^{\rm{miss}}$ triggers, and make use of quality factor ($Q$) information to design a pseudo-Standard-flag algorithm given as input to the higher trigger levels (second level and high level trigger).

Figure~\ref{fig:2012NoiseBurst}\subref{fig:2012SEVNOISEBURST} summarises the 2012 data rejection due to noise bursts that were not identified in the express processing, hence not available for the definition of a time veto window.
The overall inefficiency is found to be very low but for different reasons depending on the data-taking period. The low inefficiency observed in the data-taking periods A--G is due to the reprocessing campaign of autumn 2012 : the time windows for the veto were refined based on the original bulk processing output. The periods H--L did not benefit from this second update, and the low level of data rejection comes only from the noise burst identification in the calibration streams, showing a satisfactory trigger efficiency.

\section{Treatment of per-channel noise} \label{sect7_noisyCells}

The regular calibration procedure \cite{electronicPerf} is the main input to identify problematic calorimeter channels. However, a specific source of non-Gaussian noise was found to occur only in the presence of LHC collisions. A reliable procedure to extract these channels had to be designed to treat them within the calibration loop.

\subsection{Regular calibration procedure}\label{subs:calib}

The extraction of the electronic calibration constants requires three types of calibration runs: pedes-tal, ramp and delay. Pedestal runs allow the measurement of the baseline level and noise properties of the readout electronics, ramp runs allow the measurement of the readout gain, and delay runs allow the measurement of the pulse shape as a function of time. These special calibration runs are acquired between LHC fills, in absence of collisions, requiring only simple clock-generated triggers. Pedestal and ramp runs are taken several times a week, while the high stability of the calibration constants observed during the calorimeter commissioning~\cite{electronicPerf} indicates that delay runs are needed only once a week. These calibration runs are also the primary input to identify and classify problematic channels in a dedicated database. The different pathologies imply different offline treatments. Three main treatments that are applied are listed below.
\begin{itemize}[topsep=3pt,itemsep=2pt,parsep=3pt]
\item When a cell is not operational (deteriorated signal routing in cryostat, dead readout channel, large noise, etc.), it is {\itshape unconditionally masked} offline. Its energy is then estimated from the average energy of the eight neighbouring channels in the same calorimeter layer. In this case, the peak time and quality factor are not available.
\item A cell may be operational, but affected by large noise with very different characteristics compared to a real physics signal. The cell quality factor can be used to disentangle the signal due to a real energy deposit from the noise on an event-by-event basis. When the quality factor is lower than a fixed value (4000), the cell is considered as operational and no treatment is applied; when the quality factor is large, the cell energy is estimated from the eight neighbours of the same layer, as for an unconditionally masked cell. In this case, the cell is said to be {\itshape conditionally masked}.
\item When a cell cannot be calibrated due to a faulty calibration line, its electronic calibration constants are estimated from those of similar cells in the same layer and at the same azimuthal position. In this case, the cell is {\itshape patched}. 
\end{itemize}
At the beginning of 2012, less than 0.9\% of the calorimeter channels were patched due to a faulty calibration line. The impact of this patching being almost negligible,\footnote{The inaccuracy on the calibration was estimated about 3\%.} it is not discussed further in the following. Table \ref{tab:badChannels2012} summarises the proportion of cells unconditionally or conditionally masked at the beginning of 2012 that remained masked during the whole year. More than 99.9\% of the channels were fully functional. The 119 pathological channels being widespread across all the calorimeter regions, no large inefficient area emerges, hence the impact on the performance is considered negligible. These pathological channels remain masked (conditionally or unconditionally) during the whole data-taking period, but in addition, some other channels exhibited transient pathologies to be treated on a per-run basis, as is explained in the following.

\begin{table}[bhp]
\caption{Total number of channels unconditionally or conditionally masked at the beginning of 2012 in different partitions.}
\begin{center}
\begin{tabular}{|p{5.2cm}|p{2.4cm}|p{1.75cm}|p{1.7cm}|p{2.3cm}|}
  \hline
  & Electromagnetic & Hadronic & Forward & Global \\ 
  & calorimeter     & endcap   & calorimeter & \\
  \hline
Total number of channels & 173312 & 5632 & 3524 & 182468 \\
  \hline
Channels unconditionally masked & 76 (0.04\%) &  22 (0.39\%) & 8 (0.23\%) &  106 (0.06\%) \\
Channels conditionally masked &  8 (5$\times10^{-3}$\%) & 5 (0.09\%) & 0  & 13 (7$\times10^{-3}$\%) \\
  \hline
\end{tabular}
\end{center}
\label{tab:badChannels2012}
\end{table} 

\subsection{Monitoring of Gaussian noise during collision runs}\label{subs:cleaning}

Individual channel behaviour is also constantly monitored during collision runs. This monitoring largely relies on data streams with empty bunches (CosmicCalo and {\streamLCE}s), where no energy deposit is expected in the LAr calorimeter. The collision streams (Express, EGamma and {\streamJet}s) are mainly used for data quality assessment in the reconstruction of higher-level objects (such as electron/photon, J/$\psi$ candidates, etc.) beyond the scope of this article. However, these streams are especially useful in confirming non-operational or misbehaving channels spotted in calibration runs.

The Gaussian noise and electronics baseline, accurately characterized during the calibration runs, are cross-checked by looking at three distributions:
\begin{itemize}[topsep=3pt,itemsep=2pt,parsep=3pt]
\item mean energy and noise per cell;
\item fraction of cells with energy above 3$\sigma$, where $\sigma$ is the measured electronics noise.
\end{itemize}

If pathologies are observed in these distributions, the team responsible for the calibration is informed and they either inquire further and/or trigger urgently a new calibration procedure. No immediate systematic action is required by the {\LADIeS}. The 2012 experience showed that the Gaussian part of the electronics noise was very stable in the presence of collisions. But beside this reassuring statement, a sizeable non-Gaussian behaviour seriously complicated the data quality procedure.

\subsection{Monitoring of non-Gaussian noise during collision runs}

\begin{figure}[!htb]
    \subfloat[]{\label{fig:noiseBurstCleaning_1}\includegraphics[width=0.48\textwidth]{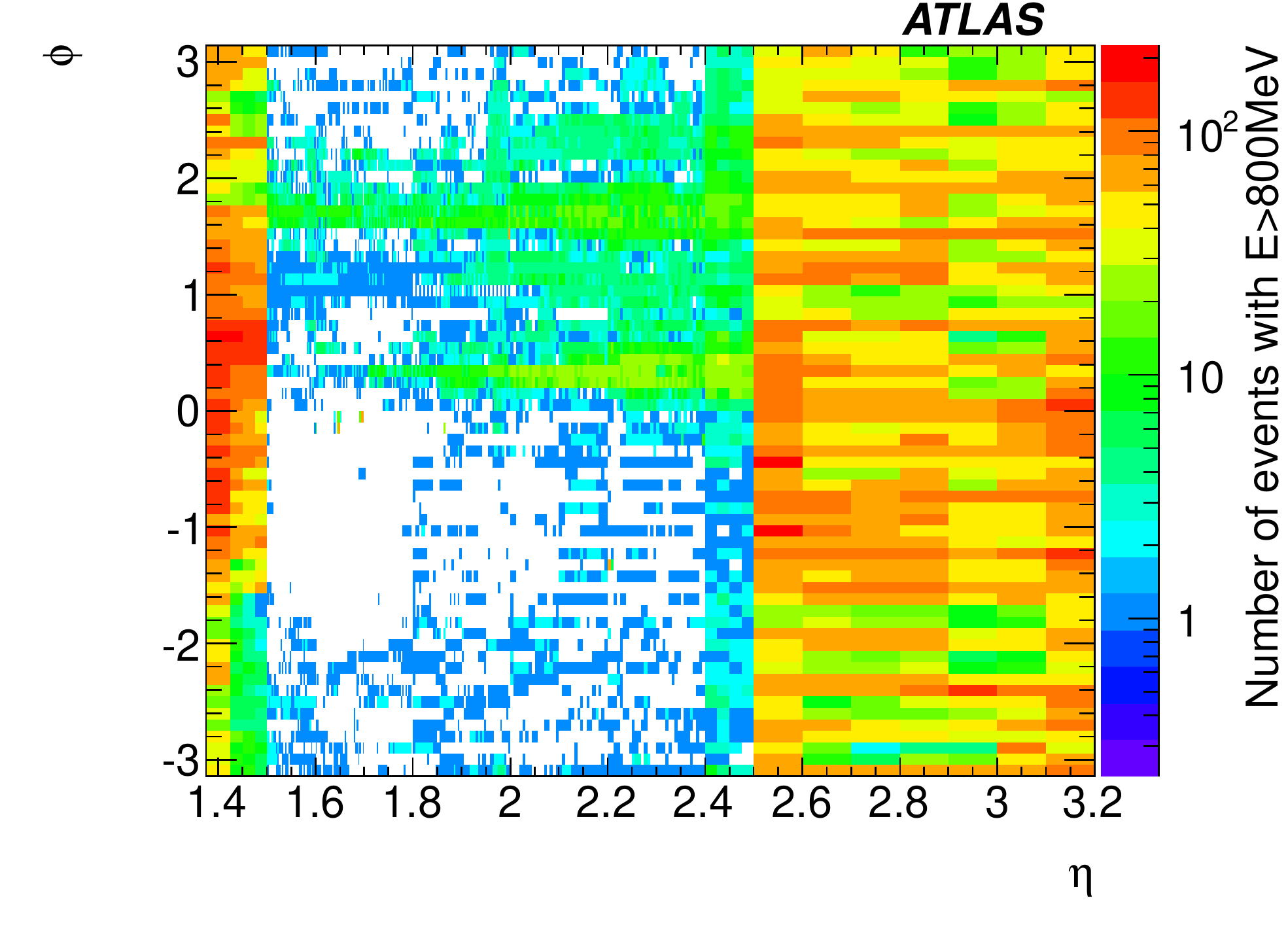}}
    \subfloat[]{\label{fig:noiseBurstCleaning_2}\includegraphics[width=0.48\textwidth]{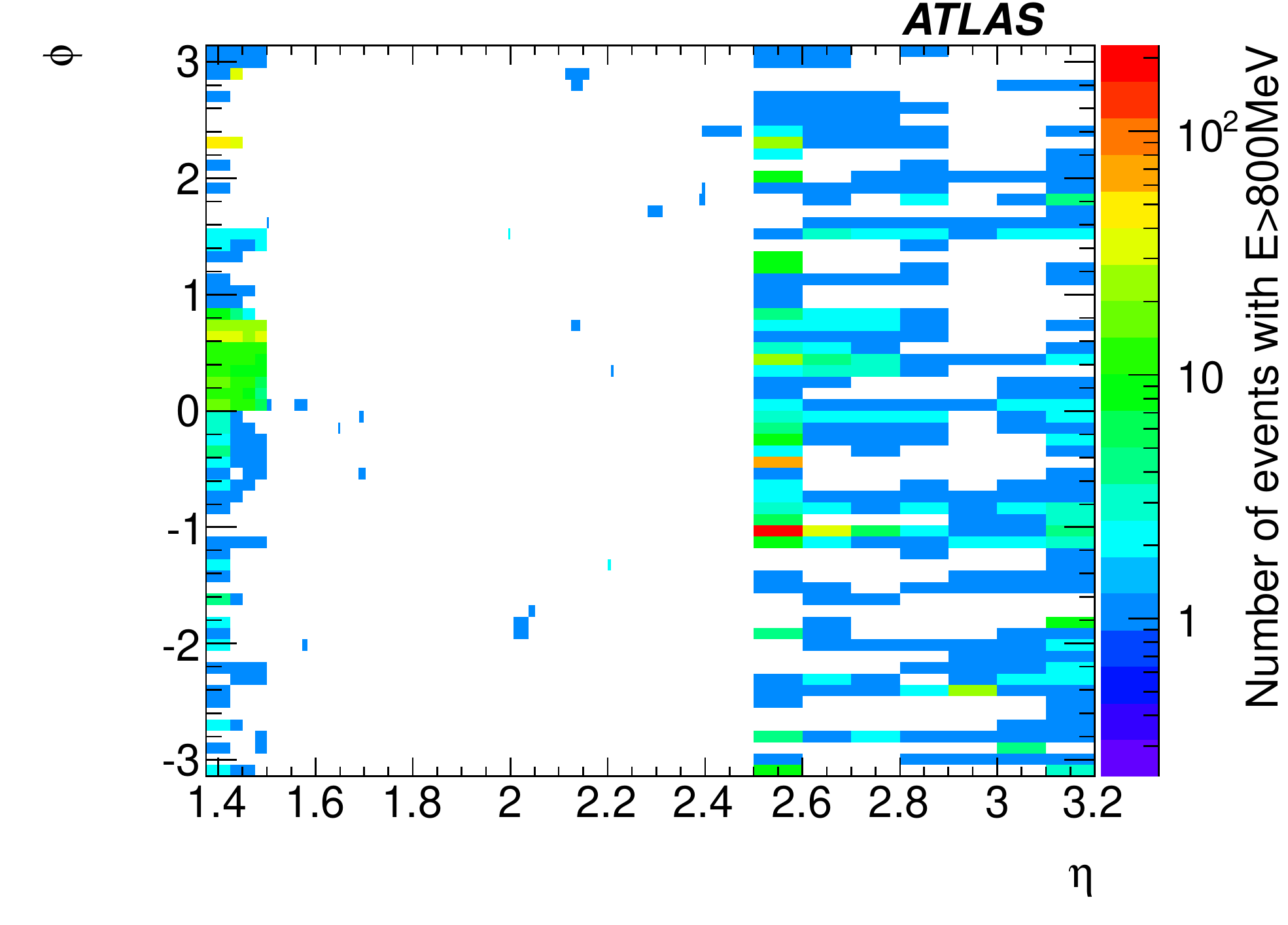}}
   \caption{($\eta$,$\phi$) distributions of the number of events per cell with cell energy greater than 800 MeV in the first layer of EMECA (a) before and (b) after removal of {\LB}s affected by a noise burst. The results are shown for the run 205071 (only {\streamCosmic}).}
   \label{fig:noiseBurstCleaning}
\end{figure}

The non-Gaussian behaviours were identified in the {\streamCosmic}, where no large energy deposit is expected, from distributions showing the number of events with an energy far exceeding the expected electronics noise (typically 20--30$\sigma$). At the express processing level, these distributions cannot be directly used, as they are polluted by noise bursts (the time-window veto cleaning procedure described in section~\ref{sect:noiseBursts} is applied only at the bulk processing stage). Such pollution can be seen in figure~\ref{fig:noiseBurstCleaning}\subref{fig:noiseBurstCleaning_1}: the large signal observed in the azimuthal ring at $\eta=1.4$ is typical of noise bursts and can be also recognized in figure~\ref{fig:noiseBurstEvent} (outer ring of the endcap).

To remove this pollution, the primary (temporary) Tier 0 monitoring outputs per luminosity block are merged, excluding the luminosity blocks affected by noise bursts. This procedure reduces the monitoring dataset by 15\%, as explained in section~\ref{subs:noiseBurstLumi}, but is crucial to avoid masking channels that would look perfectly normal after the time-window veto is applied. An example of this custom merging procedure is shown in figure~\ref{fig:noiseBurstCleaning}. 

\begin{figure}[!htb]
    \subfloat[]{\label{fig:cscCleaning_1}\includegraphics[width=0.48\textwidth]{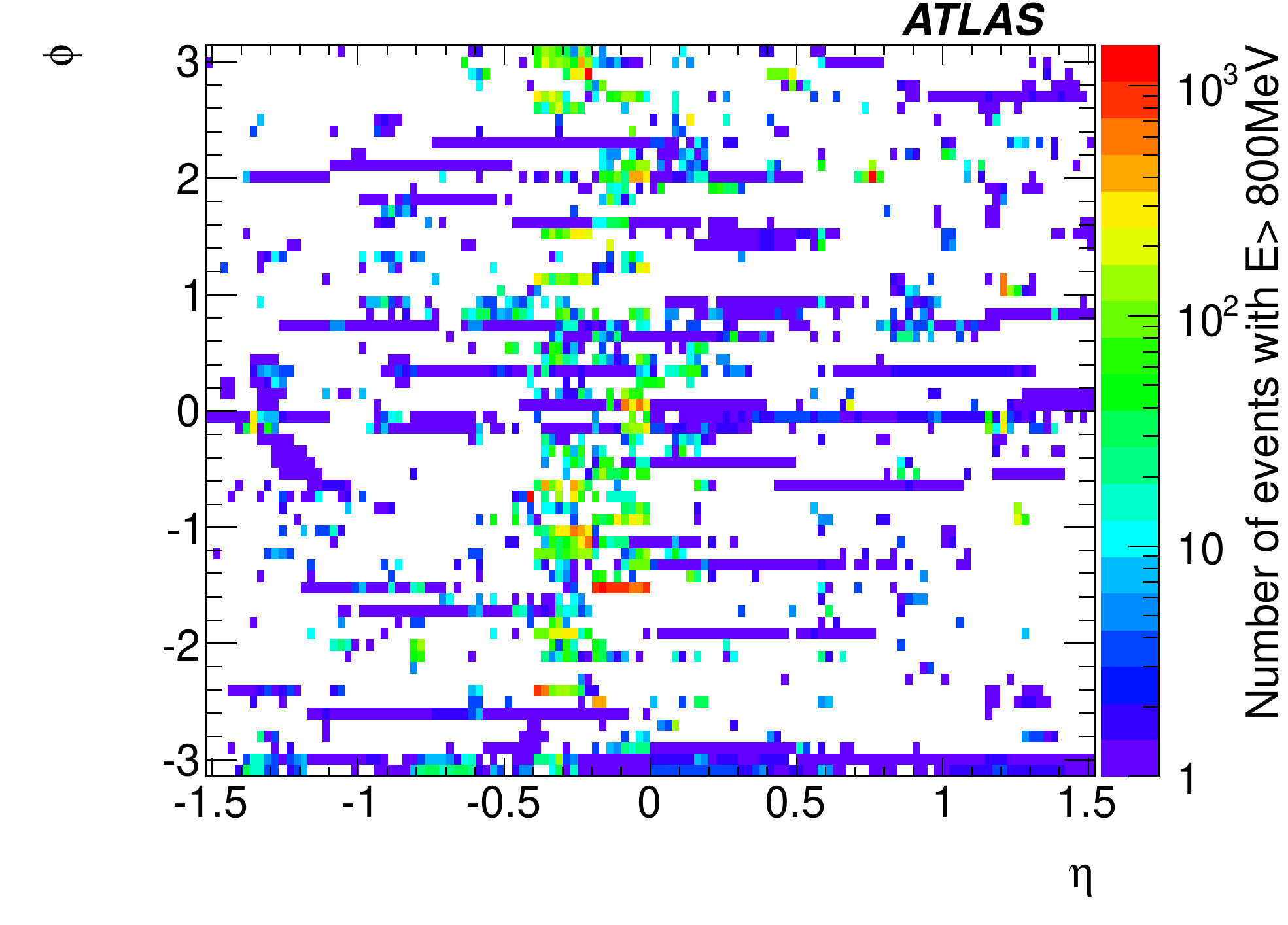}}
    \subfloat[]{\label{fig:cscCleaning_2}\includegraphics[width=0.48\textwidth]{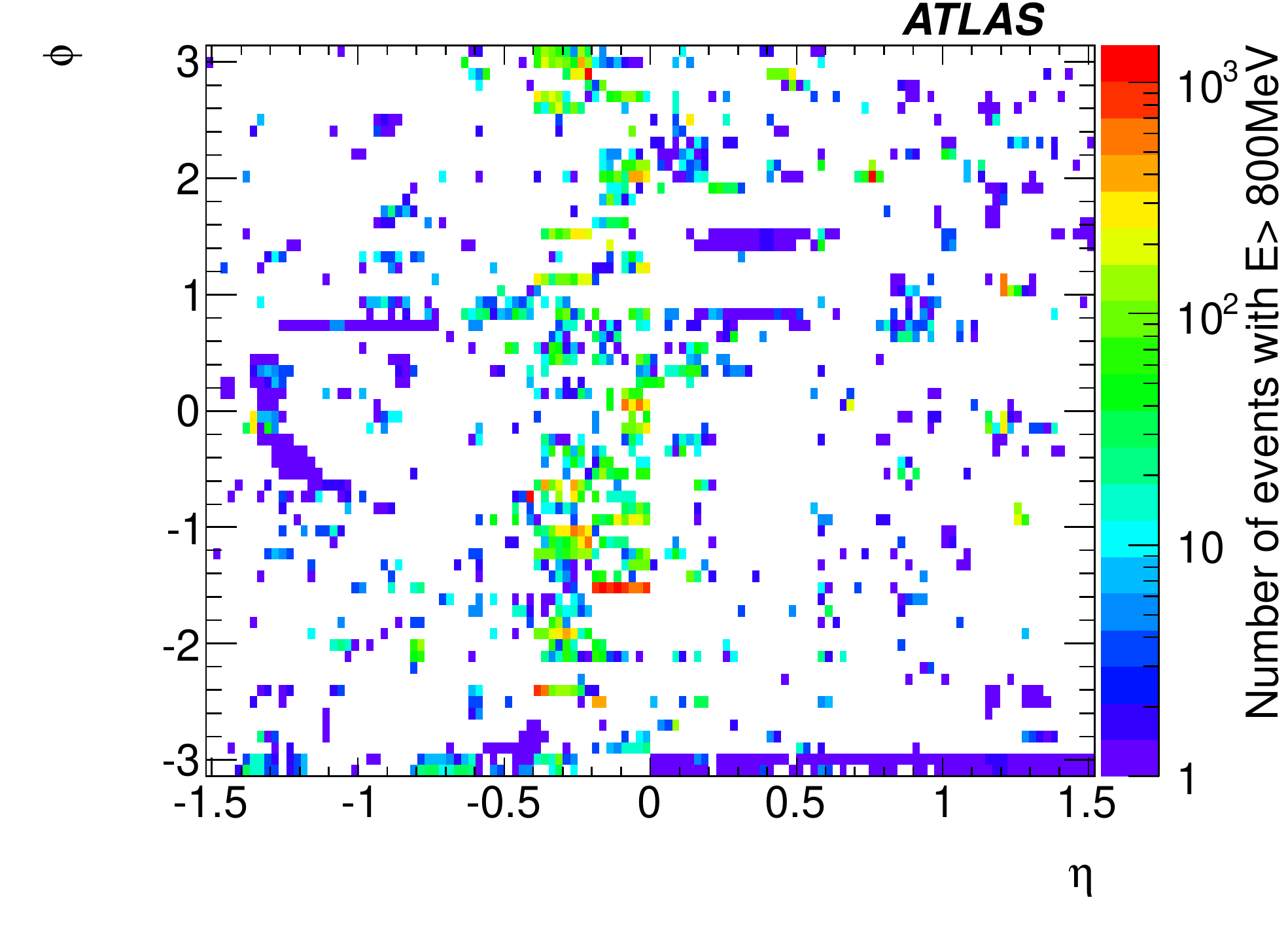}}
   \caption{($\eta$,$\phi$) distributions of number of events per cell with cell energy greater than 800 MeV in the barrel presampler (a) before and (b) after the beam background removal with the CSC tagging method. The results are shown for the run 205071 (only {\streamCosmic}).}
   \label{fig:cscCleaning}
\end{figure}

Beside the noise burst pollution, the {\streamCosmic} distributions were also found to be polluted by the LHC beam-induced background. This background -- halo or beam-gas events -- mainly originates far away from the interaction point (at more than 150~m\cite{BeamBackground2011}) and the trajectories are therefore almost parallel to the beam line. An example of such pollution is given in figure~\ref{fig:cscCleaning}\subref{fig:cscCleaning_1}, where energy deposits above 800~MeV are observed in several contiguous cells at the same azimutal position. As the radial coverage of the LAr calorimeters is very similar to the Cathode Strip Chamber (CSC) coverage of the muon spectrometer~\cite{muonTDR}, it is possible to use the coincidence of signals registered in the CSC detectors to identify this background. The improvement due to this tagging method can be visualized by comparing figures~\ref{fig:cscCleaning}\subref{fig:cscCleaning_1} and \ref{fig:cscCleaning}\subref{fig:cscCleaning_2}. In the remainder of this section, the CSC tagging method is applied to all the monitoring distributions. Finally, given the trigger conditions (thresholds and prescales) and the typical energy deposit of the cosmic-ray muons~\cite{larreadiness}, these distributions are not biased by the cosmic rays reaching the LAr calorimeter.

Despite the obvious improvement observed in figure~\ref{fig:cscCleaning}\subref{fig:cscCleaning_2} after vetoing the CSC tagged events in figure~\ref{fig:cscCleaning}\subref{fig:cscCleaning_1}, a large accumulation of noisy cells remains, especially in the pseudorapidity region $-0.3<\eta<0$. This residual noise is mainly visible in the presampler, and is interpreted as a non-Gaussian noise source, originating from inside the cryostat. Further studies were carried on to characterize this noise.
\begin{itemize}[topsep=3pt,itemsep=2pt,parsep=3pt]
\setlength{\itemsep}{0pt}
\item This noise is not visible in clock-generated triggered events.
\item It is not constant over time, only appearing for a few to several minutes before disappearing.
\item The measured signal can reach up to 100~GeV in a single cell.
\item This noise does not always affect the same cells from one run to another.
\item Some regions are more affected than others, like the $-0.3<\eta<0$ region quoted above with no obvious correlation between the affected regions and any calorimeter components or integration conditions (electrode batches or vendors, assembly conditions, etc.).
\item Lowering the HV settings in specific sectors reduces the noise amplitude in these sectors.
\item No coherent behaviour is observed between the affected channels.
\end{itemize}
This phenomenon is very different from the noise bursts considered in section~\ref{sect:noiseBursts}: the typical time scale is much longer and no coherent fluctuation is observed. The long time scale makes treatment with the time-window veto procedure impractical, as it would reject too much data. It was therefore decided to correct this noise by masking the affected channels. Given the non-permanent nature of this noise (usually named {\itshape sporadic}) and the large variations from one run to another, the list of affected channels has to be extracted per run and uploaded to the corresponding database during the calibration loop. As already explained in section~\ref{subs:calib}, the masking choice -- conditional or unconditional -- depend on the noise shape, i.e. depend on the ability to distinguish between noise and real physics signal with the cell quality factor.

\begin{figure}[!htb]
   \center
   \includegraphics[width=0.48\textwidth]{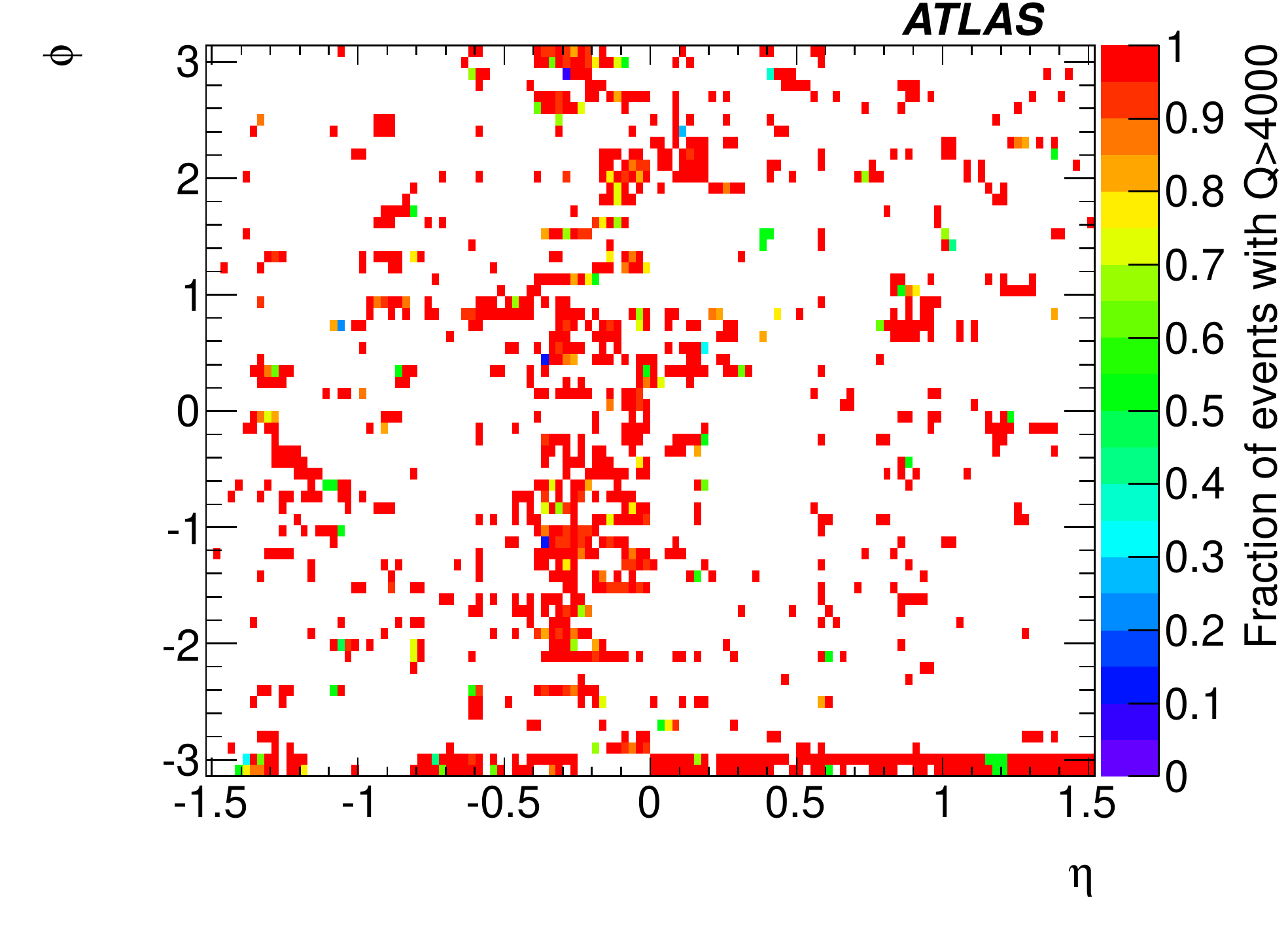}
   \caption{Fraction of events in which cells with cell energy greater than 800 MeV also have quality factor greater than 4000 in the barrel presampler.  The results are shown for run 205071 (only {\streamCosmic}).}
   \label{fig:qFactor}
\end{figure}

Figure~\ref{fig:qFactor} shows the fraction of high-energy events with a cell quality factor greater than 4000, i.e. the fraction of events where masking cells conditionally would be efficient. This distribution, convolved with the distribution shown in figure~\ref{fig:cscCleaning}\subref{fig:cscCleaning_2}, provides the number of high-energy events per channel surviving a conditional masking. A conservative upper threshold of 80 events per cell per run was arbitrarily chosen to decide whether or not a channel should be conditionally masked. If more than 80 noisy events survive for a given channel, an unconditional masking has to be applied, more severely impacting the calorimeter performance.

\begin{figure}[!htb]
  \center
  \subfloat[]{\label{fig:upd4effect_1}\includegraphics[width=0.48\textwidth]{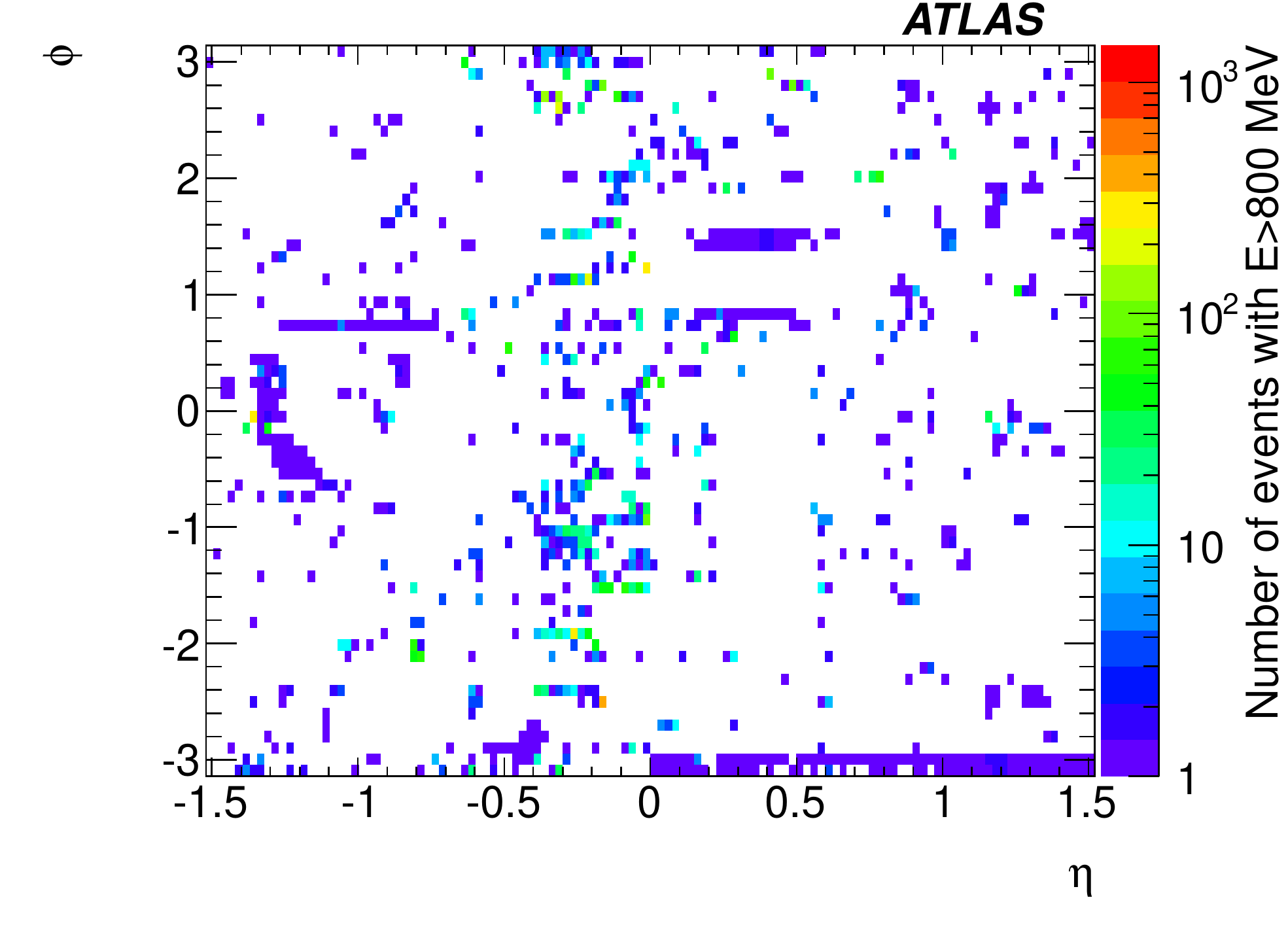}}
  \subfloat[]{\label{fig:upd4effect_2}\includegraphics[width=0.48\textwidth]{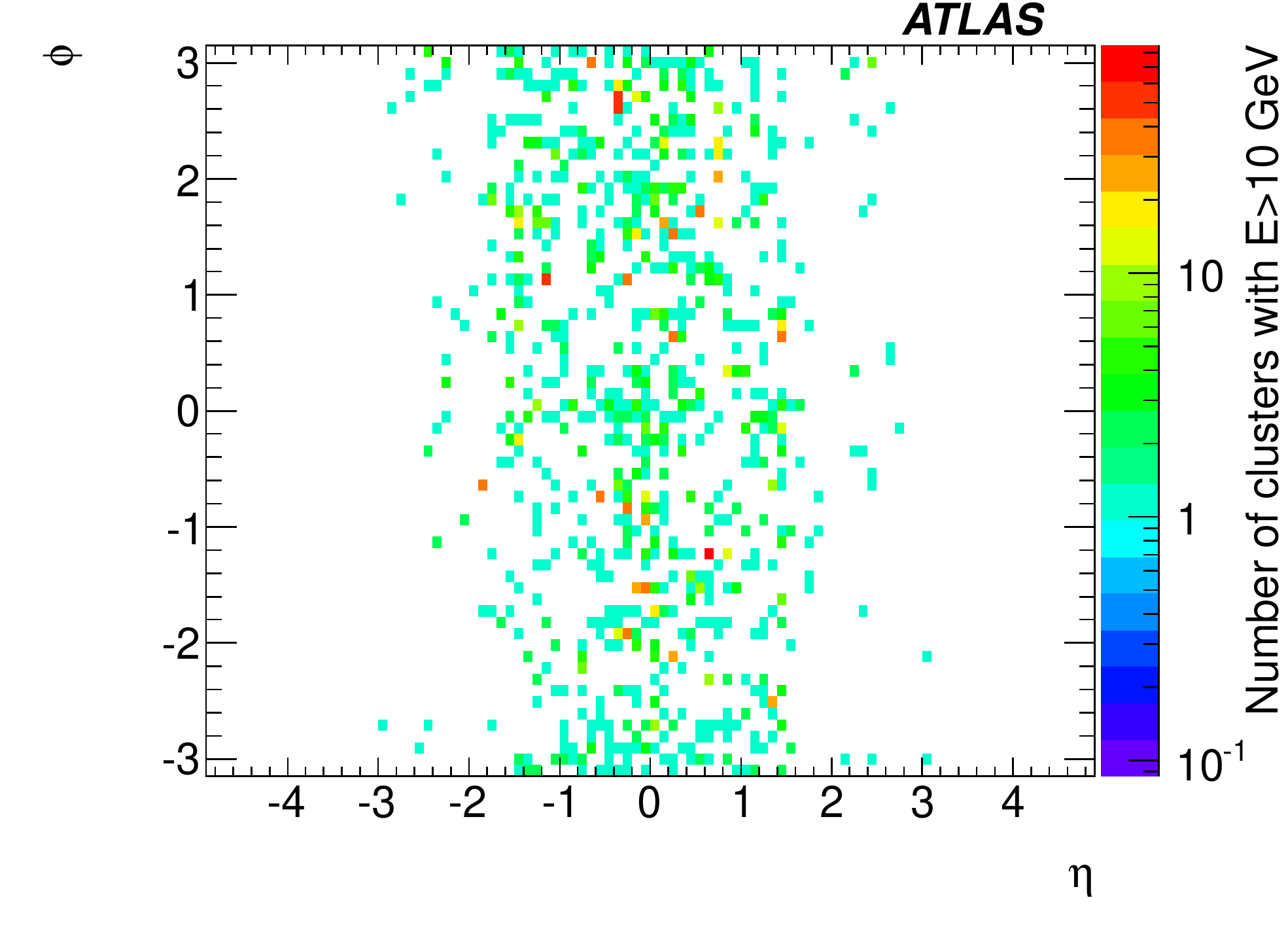}}
  \caption{($\eta$,$\phi$) distributions produced after database updates with the same dataset as previously (run 205071 -- {\streamCosmic}).
(a) Number of events per cell with an energy greater than 800 MeV in the barrel presampler. 
(b) Number of clusters with a transverse energy greater than 10 GeV.}
  \label{fig:upd4effect}
\end{figure}

The masking efficiency is double-checked on the same data streams after the bulk processing, where the database updates are included in the reconstruction. Figure~\ref{fig:upd4effect}\subref{fig:upd4effect_1}, to be compared with the original map in figure~\ref{fig:cscCleaning}\subref{fig:cscCleaning_2}, illustrates the effect of masking the noisy cells. A large reduction of events with an energy above 800~MeV is observed. A final cross-check is performed by looking at the $(\eta,\phi)$ map of the clusters, the primary objects used in the electron/photon reconstruction, with a transverse energy greater than 10~GeV. This particular threshold was chosen as it corresponds to the minimal energy cut applied in most of the ATLAS analyses. The very limited number of clusters visible in figure~\ref{fig:upd4effect}\subref{fig:upd4effect_2} validates the satisfactory efficiency of the channel-masking procedure. 

However, the masking procedure may sometimes fail. This happens in the very unfortunate cases where a cell is noisy only in {\LB}s affected by a noise burst. The noisy {\LB}s are excluded from the express processing output due to the custom merging procedure detailed in section~\ref{subs:cleaning}, and the noisy cells are missed by the {\LADIeS} during the calibration loop. During the bulk processing, the noise bursts are removed from the {\LB}s with the time-window veto. The missing {\LB}s are thus automatically re-included in the Tier 0 monitoring output of the bulk processing, and the sporadic noise emerges. Since it is too late to correct the data after the bulk processing, the offending {\LB} has to be discarded by assigning an intolerable defect. Still, the database is updated to include the additional noisy channels so that the masking can be applied during any future data reprocessing to recover the lost luminosity.

Given the large number of affected channels and their fluctuating nature, the whole procedure for the cell identification, masking proposal optimization, cluster matching, etc.\ is automatically performed within the dedicated LAr calorimeter data quality web infrastructure described in section~\ref{offlinePolicy}.

\subsection{Proportion of masked cells}
\begin{figure}[!htb]
  \center
  \includegraphics[width=0.48\textwidth]{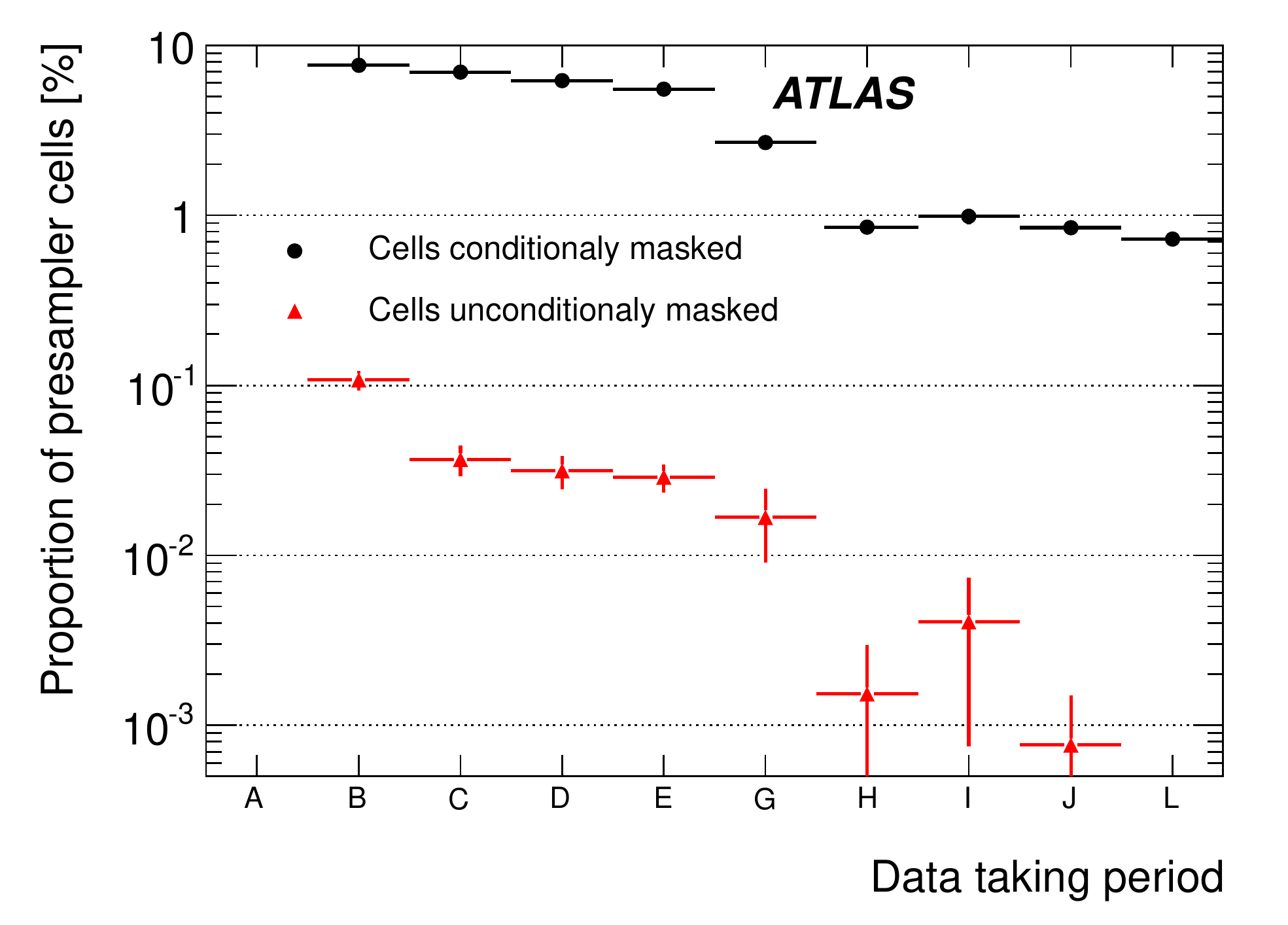}
  \caption{Proportion of presampler cells conditionally or unconditionally masked as a function of time, during the 2012 data-taking periods.}
  \label{fig:upd4Stat-0}
\end{figure}

Figure~\ref{fig:upd4Stat-0} shows the proportion of masked presampler channels as a function of the data-taking period in 2012; as a small dependence on integrated luminosity is observed for short runs, only the 95 runs with an integrated luminosity greater than 100~$\rm{pb}^{-1}$ recorded were considered. The proportion of unconditionally masked presampler cells remained below 0.2\% for the whole LHC running period, while the proportion of conditionally masked presampler channels was greater than 7\% during the first weeks of data taking. During the periods B--E, the HV settings of the most problematic lines were reduced from the original 1.6 kV to limit the sporadic noise, allowing reduction of the proportion of cells conditionally masked. Then, in September 2012 (middle of period G), it was decided to reduce globally the HV settings to 1.2~kV. This reduction  gave a proportion of cells conditionally masked below 1\%. The gain in electron and photon energy resolution due to the presampler is preserved despite a 10\% increase in electronics noise.

\begin{figure}[!htb]
    \subfloat[]{\label{fig:upd4Stat_3}\includegraphics[width=0.48\textwidth]{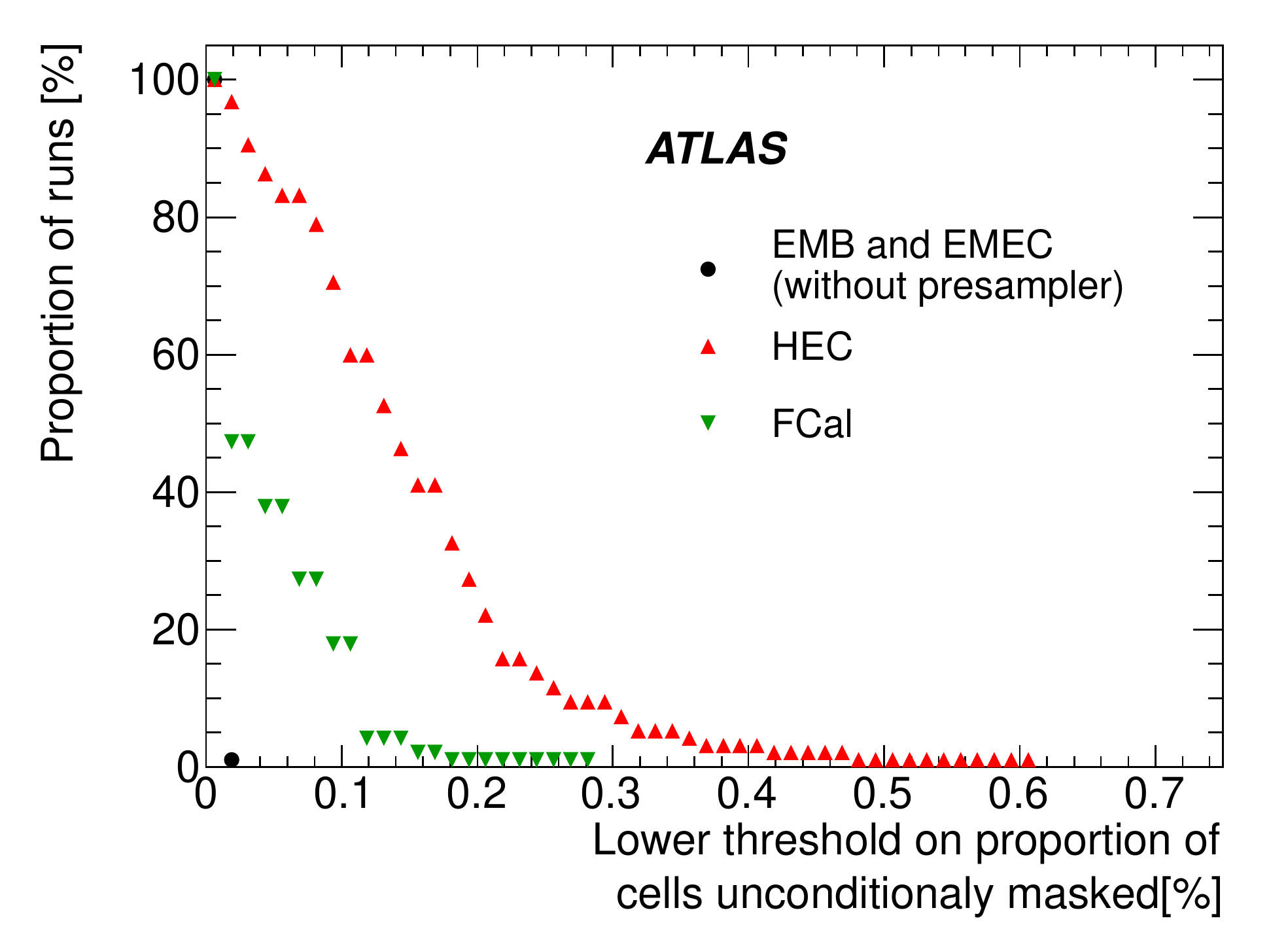}}
    \subfloat[]{\label{fig:upd4Stat_2}\includegraphics[width=0.48\textwidth]{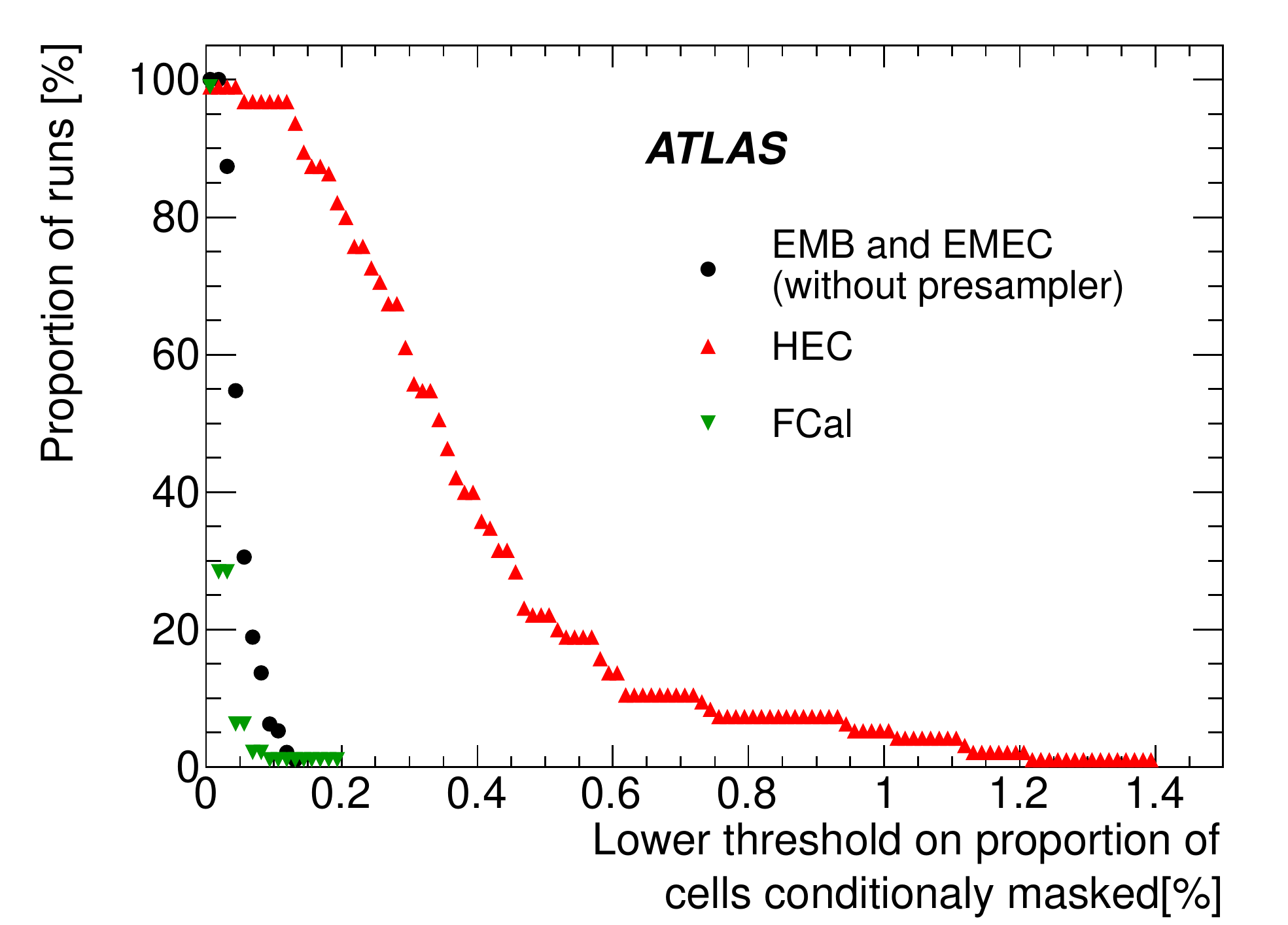}}
  \caption{(a) Proportion of runs for which a proportion of unconditionally masked cells is above a given threshold. (b) Proportion of runs for which a proportion of conditionally masked cells is above a given threshold.}
   \label{fig:upd4Stat}
\end{figure}

Figure~\ref{fig:upd4Stat} shows the proportion of channels masked in the same high-luminosity runs for all partitions except the presampler. The proportion of unconditionally masked channels remains very low in all the partitions: it is negligible in the electromagnetic calorimeter, and lower than 0.4\% (0.2\%) in the HEC (FCal) in 95\% of the runs. The proportion of conditionally masked channels is slightly larger, but the impact on performance is also negligible since only the subset of events with a high quality factor is effectively masked. 

\subsection{Associated data rejection in 2012}

\begin{figure}[htb]
  \center
  \includegraphics[width=0.48\textwidth]{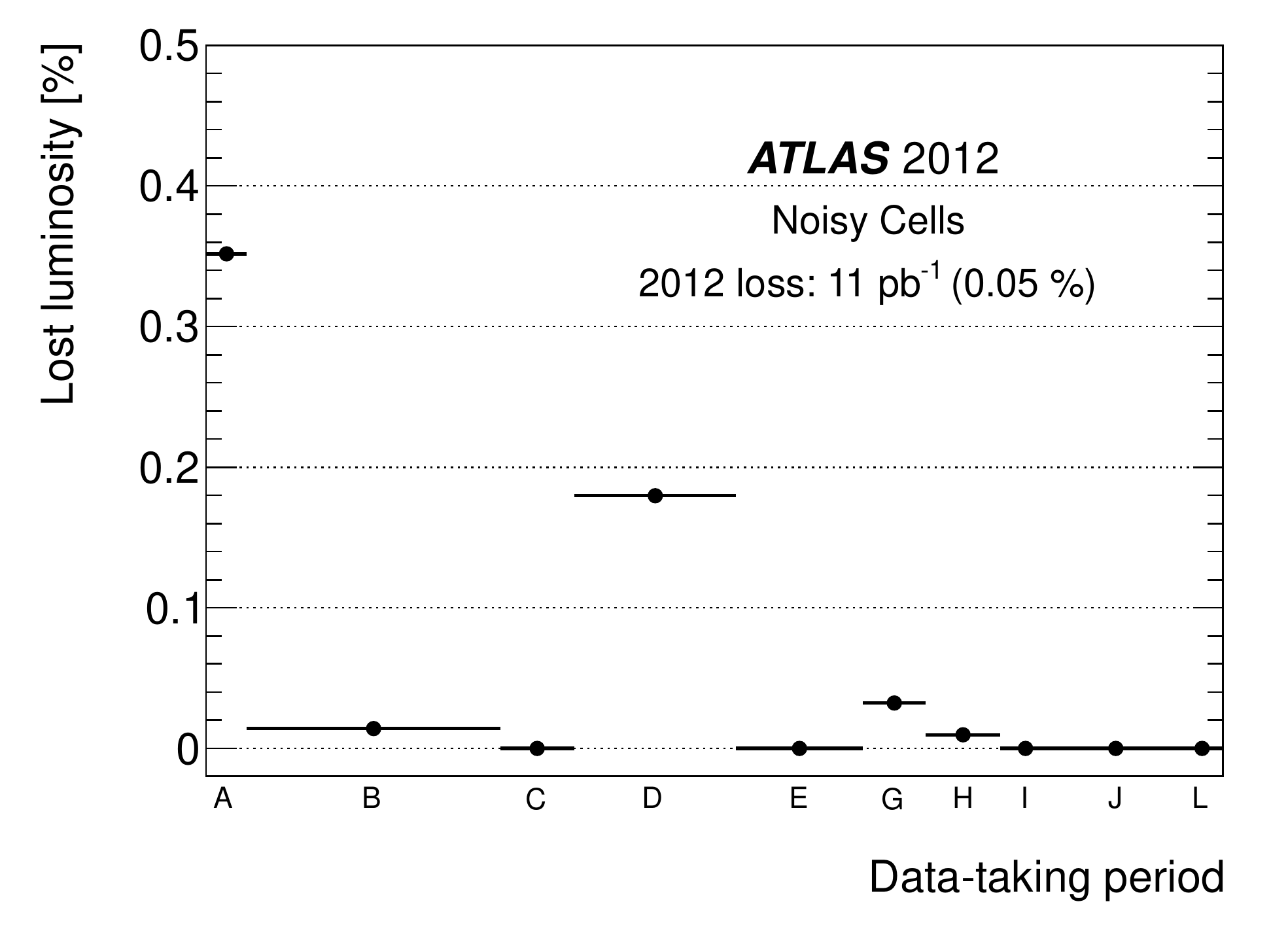}
   \caption{Lost luminosity due to noisy channels as a function of the data-taking period in 2012.}
  \label{fig:2012NoisyChannel}
\end{figure}

The data loss associated with a non-optimal treatment of the noisy channels (within the calibration loop) is shown in figure~\ref{fig:2012NoisyChannel}. This loss remains very low over the whole year, and it even goes to zero for the last 2012 data-taking periods, indicating that the latest version of the diagnostic algorithms was properly tuned and able to catch all the problematic channels within the calibration loop.

\section{Achieved performance and outlook} \label{sect:concl}

\subsection{Performance in proton--proton collision run (2011--2012)}

Table \ref{tab:GRLLoss} summarises the data rejection by defect assignment in the 2011 and 2012 datasets, corresponding respectively to integrated recorded luminosities of $5.2~\rm{fb^{-1}}$ and $21.3~\rm{fb^{-1}}$~\cite{lumi2011}. The 2011 performance is systematically worse than the 2012 performance described in detail in this article, and several reasons can be listed to explain this observation.
\begin{itemize}[topsep=3pt,itemsep=2pt,parsep=3pt]
\item The 2011 larger rejection due to HV trips is explained by luminosity conditions that were less stable in 2011 than in 2012 and by the replacement of several HV power supply modules with more sophisticated ones in 2012.
\item The 2012 reduction of missed noise bursts is related to the implementation of a dedicated trigger chain in early 2012, which added more coherent noise events in the calibration streams and hence allowed a more efficient time-window veto procedure.
\item The reduced data loss observed in 2012 for the other defects is due to the improved software robustness and automation in both the daily calorimeter operation and the data quality assessment.
\end{itemize}
Table \ref{tab:vetoLoss} summarises the data rejection due to the time-window veto in 2011 and 2012. The rejection levels are very similar despite the much higher instantaneous luminosities recorded in 2012, which induced an enhanced yield of noise bursts. The higher noise burst rate in 2012 is counterbalanced by the choice of a narrower time window extension compared to 2011 ($\delta t = 200$~ms in 2012 vs $\delta t = 1$~s in 2011). 
\begin{table}[!hbp]
\caption{Summary of data rejection by defect assignment for the 2011 and 2012 proton--proton collision datasets.}
\begin{center}
\begin{tabular}{|p{0.9cm}|p{1.25cm}||p{1.7cm}|p{2.0cm}|p{1.6cm}|p{1.4cm}|p{1.25cm}|p{1.6cm}|} 
 \hline
Year  & Total & Data corruption & Missing condition data & HV trips & Coverage & Noise bursts & Noisy\newline channels \\
  \hline
2011   & 3.20\% & 0.04\% & 0.11\% & 0.96\% & 0.70\% & 1.24\% & 0.15\% \\
2012   & 0.88\% & 0.01\% & 0.02\% & 0.46\% & 0.28\% & 0.06\% & 0.05\% \\
  \hline
\end{tabular}
\end{center}
\label{tab:GRLLoss}
\end{table} 
\begin{table}[!hbp]
\caption{Summary of data rejection by the time-window veto procedure for the 2011 and 2012 proton--proton collision datasets.}
\begin{center}
\begin{tabular}{|p{1.15cm}|p{1.15cm}||p{2.1cm}|p{2.1cm}|}
  \hline
Year & Total & Data corruption & Noise bursts \\
  \hline
2011 & 0.28\% & 0.09\% & 0.20\% \\ 
2012 & 0.22\% & 0.02\% & 0.20\% \\
  \hline
\end{tabular}
\end{center}
\label{tab:vetoLoss}
\end{table}

\begin{figure}[!htb]
    \subfloat[2011 dataset]{\label{fig:GRL2011}\includegraphics[width=0.48\textwidth]{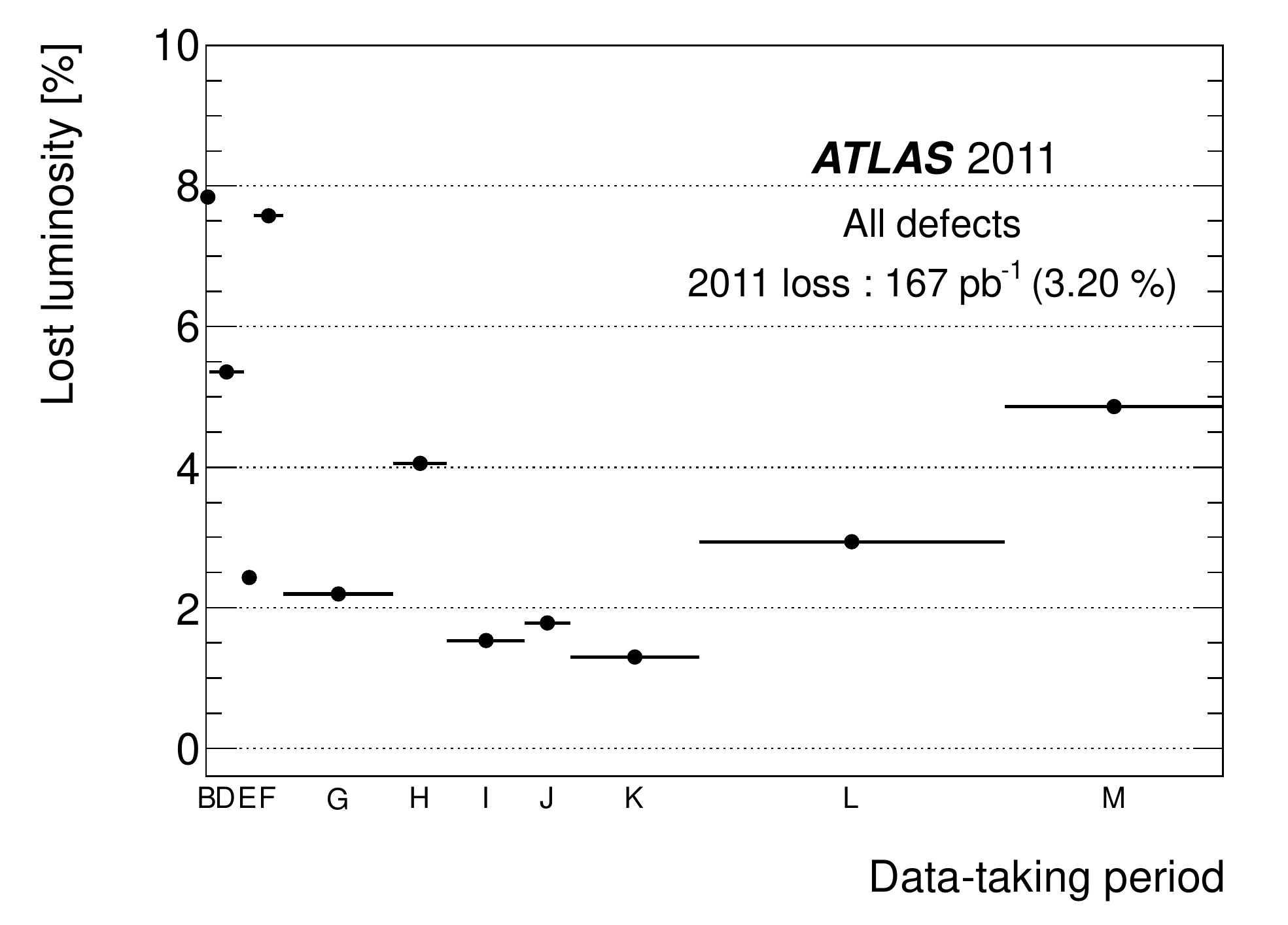}}
    \subfloat[2012 dataset]{\label{fig:GRL2012}\includegraphics[width=0.48\textwidth]{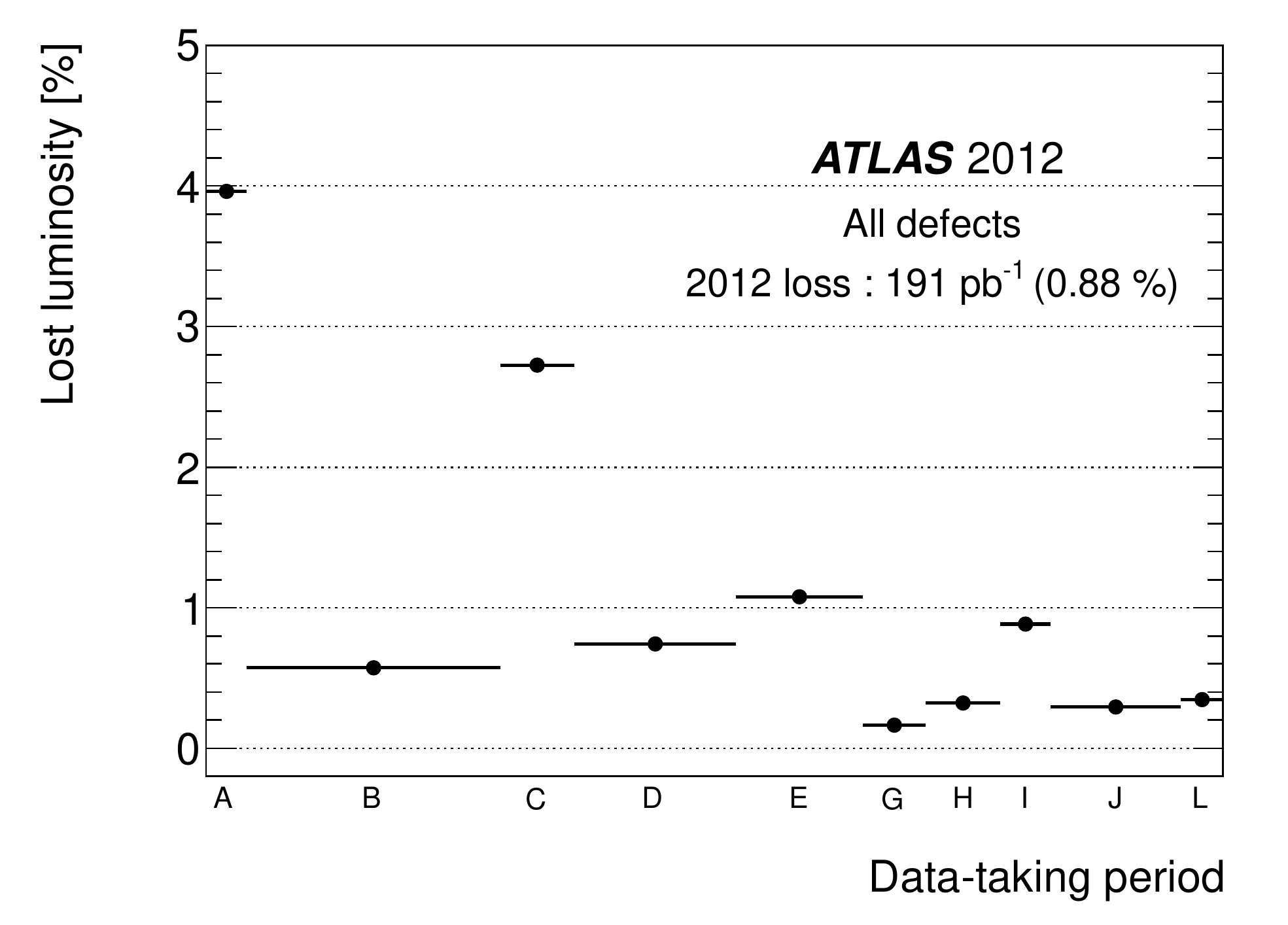}}
   \caption{Lost luminosity associated to a defect assignment as a function of the data-taking period in 2012.}
   \label{fig:GRLLoss_period}
\end{figure}

\begin{figure}[!htb]
    \subfloat[2011 dataset]{\label{fig:veto2011}\includegraphics[width=0.48\textwidth]{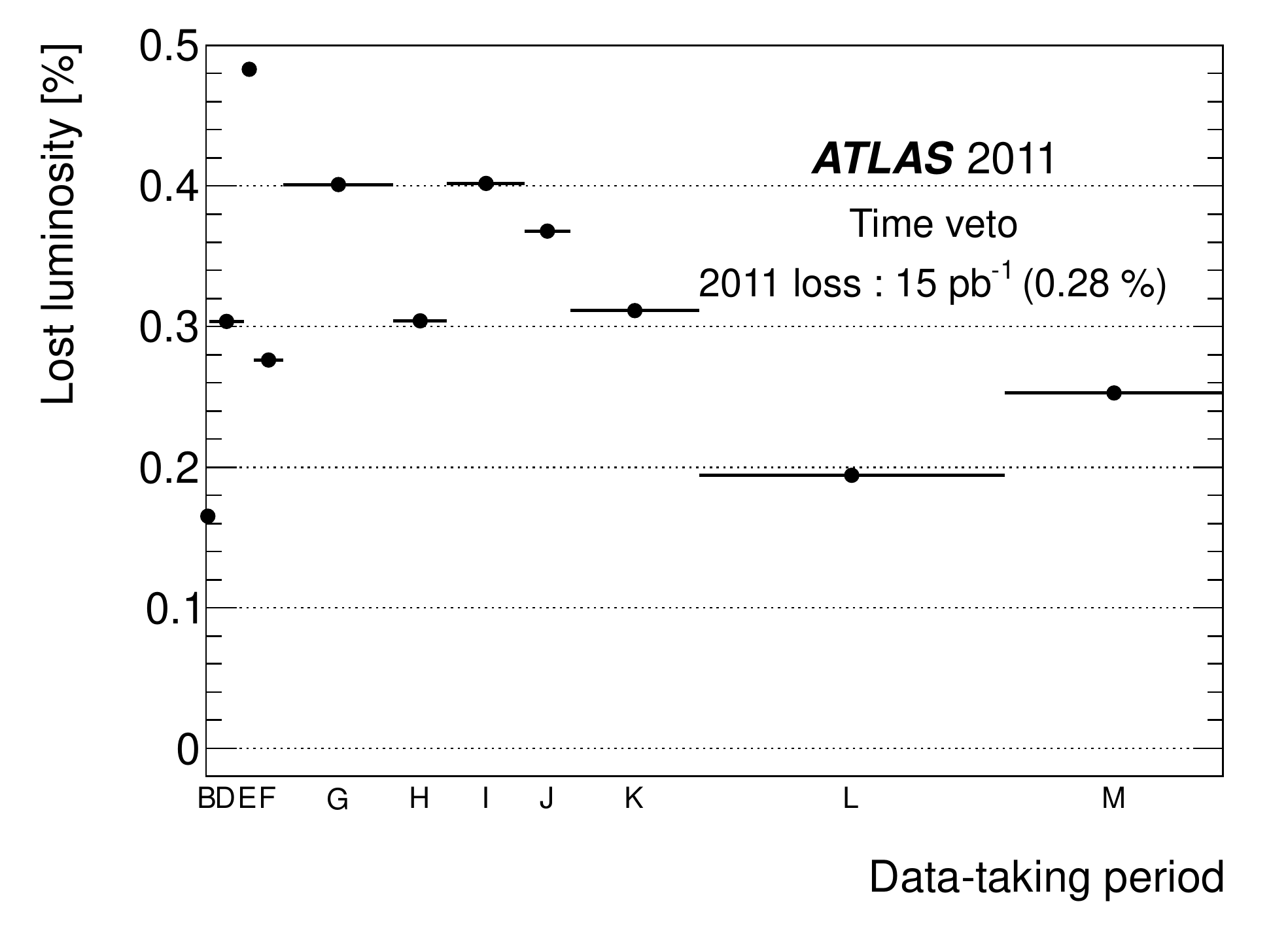}}
    \subfloat[2012 dataset]{\label{fig:veto2012}\includegraphics[width=0.48\textwidth]{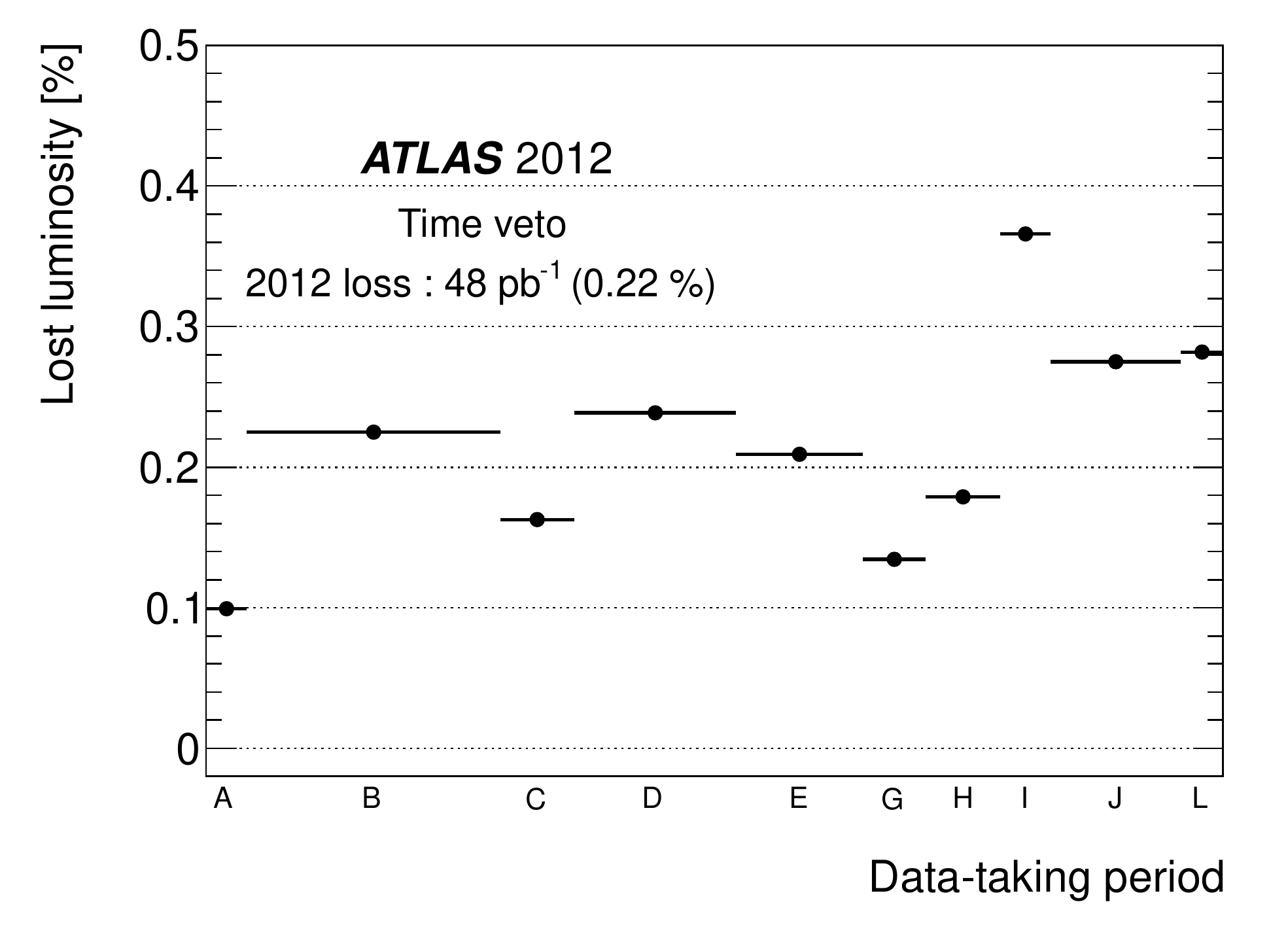}}
   \caption{Lost luminosity associated to the time-window veto as a function of the data-taking period in 2012.}
   \label{fig:vetoLoss_period}
\end{figure}

In 2011, as in 2012, the dataset was split into periods with similar data-taking conditions. The time evolution of the data rejection by defect assignment and time-window veto are displayed in figures \ref{fig:GRLLoss_period} and \ref{fig:vetoLoss_period} respectively using the datasets for proton--proton collisions collected in 2011 and 2012.

\subsection{Performance in lead--lead and lead--proton collision run (2011--2013)}

For completeness, the LAr calorimeter performance in the lead--lead and lead--proton collision runs is summarized in table \ref{tab:GRLLoss-hi}. Given the much lower peak luminosity delivered during these runs (5$\times 10^{26}~\rm{cm^{-2}s^{-1}}$ in 2011 and $10^{29}~\rm{cm^{-2}s^{-1}}$ in 2013), the impact of the phenomena correlated with the instantaneous luminosity (noise bursts and HV trips) was limited. The data rejection by the time-window veto procedure -- not shown here~-- is also negligible. In 2013 a large data rejection was observed due to a single powering problem encountered in the hadronic endcap that lasted 90~minutes. Due to the shortness of the data taking period in 2013, this caused a data loss of~1.18\%.
\begin{table}[htbp]
\caption{Summary of data rejection by defect assignment for the 2011 and 2013 lead--lead and lead--proton collision datasets.}
\begin{center}
\begin{tabular}{|p{2.15cm}|p{1.cm}||p{1.45cm}|p{2.0cm}|p{1.6cm}|p{1.4cm}|p{1.05cm}|p{1.25cm}|} 
 \hline
Year  & Total & Data corruption & Missing condition data & HV trips & Coverage & Noise bursts & Noisy\newline channels \\
  \hline
2011 (Pb--Pb) & 0.19\% & 0.16\% &  -     & 0.03\% & -  & -      & - \\
2013 (Pb--$p$)& 1.50\% & 0.05\% &  -     & 0.22\% & 1.18\%  & 0.04\% & - \\
  \hline
\end{tabular}
\end{center}
\label{tab:GRLLoss-hi}
\end{table} 

\subsection{Outlook}

The LHC is expected to restart in 2015 and to deliver collisions at the unprecedented energy and instantaneous peak luminosity of 13--14~TeV and 1--2$\times 10^{34}~\rm{cm^{-2}s^{-1}}$ respectively.

As stated in section~\ref{subs:dcs-dataRejection}, the occurrence of HV trips, currently the main source of data loss, does not depend on the absolute instantaneous luminosity, but only on its evolution over a long timescale. When the LHC running conditions are stable, the data loss remains under control. In addition, many more of the upgraded power supplies are expected to be installed on the detector before the LHC restart to further reduce this loss. 

The second largest source of data loss comes from large inefficient areas. However, out of the 0.28\% yield observed in 2012, 0.15\% were due to special runs that would have been rejected anyway. The remaining 0.13\% originating from the LAr calorimeter arose from two defects of the low-voltage power supply system.

Considering the full data rejection by both defect assignment and the time-window veto, the loss due to noise bursts reaches 0.26\% (0.20\%+0.06\%) in 2012. As explained in section~\ref{subs:noiseBurstLumi}, this yield should remain under control despite the regular increase in the frequency of noise bursts as a function of instantaneous luminosity. A parabolic extrapolation of the dependence curve of figure~\ref{fig:lumiDep}\subref{fig:occurenceVsLumi} indicates that the noise-burst rate could be in 2015 10--15 times higher than in 2012. However, there is still a lot of safety margin in the choice of the time window width to mitigate this rate increase.

The remaining sources of data losses measured in 2012 contribute less than 0.1\%, and there is no indication of any luminosity dependence that could worsen the situation.

Therefore, the increased instantaneous luminosity of the LHC in 2015 is not expected to seriously degrade the data quality performance. However, two unknowns remain. First, the evolution of the sporadic noise with the instantaneous luminosity is still poorly known as is the robustness of the adopted solution (HV settings tuning). Second, it cannot be excluded that the almost doubled centre-of-mass energy may induce new problems or affect the magnitude of the already known ones. If these two risks are properly addressed, a similar efficiency around 98--99\% can be considered as a realistic objective for the LHC restart in 2015.

\acknowledgments

We thank CERN for the very successful operation of the LHC, as well as the
support staff from our institutions without whom ATLAS could not be
operated efficiently.

We acknowledge the support of ANPCyT, Argentina; YerPhI, Armenia; ARC,
Australia; BMWF and FWF, Austria; ANAS, Azerbaijan; SSTC, Belarus; CNPq and FAPESP,
Brazil; NSERC, NRC and CFI, Canada; CERN; CONICYT, Chile; CAS, MOST and NSFC,
China; COLCIENCIAS, Colombia; MSMT CR, MPO CR and VSC CR, Czech Republic;
DNRF, DNSRC and Lundbeck Foundation, Denmark; EPLANET, ERC and NSRF, European Union;
IN2P3-CNRS, CEA-DSM/IRFU, France; GNSF, Georgia; BMBF, DFG, HGF, MPG and AvH
Foundation, Germany; GSRT and NSRF, Greece; ISF, MINERVA, GIF, I-CORE and Benoziyo Center,
Israel; INFN, Italy; MEXT and JSPS, Japan; CNRST, Morocco; FOM and NWO,
Netherlands; BRF and RCN, Norway; MNiSW and NCN, Poland; GRICES and FCT, Portugal; MNE/IFA, Romania; MES of Russia and ROSATOM, 
Russian Federation; JINR; MSTD,
Serbia; MSSR, Slovakia; ARRS and MIZ\v{S}, Slovenia; DST/NRF, South Africa;
MINECO, Spain; SRC and Wallenberg Foundation, Sweden; SER, SNSF and Cantons of
Bern and Geneva, Switzerland; NSC, Taiwan; TAEK, Turkey; STFC, the Royal
Society and Leverhulme Trust, United Kingdom; DOE and NSF, United States of
America.

The crucial computing support from all WLCG partners is acknowledged
gratefully, in particular from CERN and the ATLAS Tier-1 facilities at
TRIUMF (Canada), NDGF (Denmark, Norway, Sweden), CC-IN2P3 (France),
KIT/GridKA (Germany), INFN-CNAF (Italy), NL-T1 (Netherlands), PIC (Spain),
ASGC (Taiwan), RAL (UK) and BNL (USA) and in the Tier-2 facilities
worldwide.

\bibliographystyle{JHEP}
\providecommand{\href}[2]{#2}\begingroup\raggedright\endgroup

\onecolumn
\clearpage
\begin{flushleft}
{\Large The ATLAS Collaboration}

\bigskip

G.~Aad$^{\rm 84}$,
T.~Abajyan$^{\rm 21}$,
B.~Abbott$^{\rm 112}$,
J.~Abdallah$^{\rm 152}$,
S.~Abdel~Khalek$^{\rm 116}$,
O.~Abdinov$^{\rm 11}$,
R.~Aben$^{\rm 106}$,
B.~Abi$^{\rm 113}$,
M.~Abolins$^{\rm 89}$,
O.S.~AbouZeid$^{\rm 159}$
% ...
\bigskip
\\
% ...
$^{11}$ Institute of Physics, Azerbaijan Academy of Sciences, Baku, Azerbaijan\\
% ...
$^{21}$ Physikalisches Institut, University of Bonn, Bonn, Germany\\
% ...
$^{84}$ CPPM, Aix-Marseille Universit{\'e} and CNRS/IN2P3, Marseille, France\\
% ...
$^{89}$ Department of Physics and Astronomy, Michigan State University, East Lansing MI, United States of America\\
% ...
$^{106}$ Nikhef National Institute for Subatomic Physics and University of Amsterdam, Amsterdam, Netherlands\\
% ...
$^{112}$ Homer L. Dodge Department of Physics and Astronomy, University of Oklahoma, Norman OK, United States of America\\
$^{113}$ Department of Physics, Oklahoma State University, Stillwater OK, United States of America\\
% ...
$^{116}$ LAL, Universit{\'e} Paris-Sud and CNRS/IN2P3, Orsay, France\\
% ...
$^{152}$ Institute of Physics, Academia Sinica, Taipei, Taiwan\\
% ...
$^{159}$ Department of Physics, University of Toronto, Toronto ON, Canada\\
% ...
\end{flushleft}


\end{document}